\documentclass[iop]{emulateapj}

\usepackage{amsfonts}
\usepackage{footnote}
\usepackage{verbatim}
\usepackage{textcomp}
\usepackage{natbib,hyperref}
\usepackage{gensymb}
\shorttitle{SWIFT OBSERVATIONS OF MRK\,421 IN
2015\,December--2018\,April} \shortauthors{Kapanadze et al.}

\begin{document}

\title{SWIFT OBSERVATIONS OF MRK\,421 IN SELECTED EPOCHS. III. EXTREME X-RAY TIMING/SPECTRAL PROPERTIES AND MULTIWAVELENGTH LOGNORMALITY IN 2015\,DECEMBER--2018\,APRIL}

\author{B. Kapanadze\altaffilmark{1,2,3}, A. Gurchumelia\altaffilmark{2}, D. Dorner\altaffilmark{4}, S. Vercellone\altaffilmark{3}, P. Romano\altaffilmark{3}, P. Hughes$^{5}$,  M. Aller$^{5}$, H. Aller$^{5}$,
 O. Kharshiladze$^{6}$}

\altaffiltext{1}{Ilia State University, Colokashvili Av. 3/5,
Tbilisi, Georgia, 0162.} \altaffiltext{1}{E. Kharadze National
Astrophysical Observatory, Mt. Kanobili, Abastumani, Georgia, 0803}
\altaffiltext{3}{INAF, Osservatorio Astronomico di Brera, Via E.
Bianchi 46, 23807 Merate, Italy.} \altaffiltext{4} {Universit\"{a}t
W\"{u}rzburg, Institute for Theoretical Physics and Astrophysics,
Emil-Fischer-Str. 31, 97074 W\"{u}rzburg, Germany}
\altaffiltext{5}{Astronomy Department, University of Michigan, Ann
Arbor, MI 48109-1107, USA} \altaffiltext{6}{I. Javakhishvili State
University, Chavchavadze Av. 3, Tbilisi 0128, Republic of Georgia}

\begin{abstract}
We present the results from the timing and spectral study of
Mrk\,421 based mainly on the \emph{Swift} data in the X-ray energy
range obtained during the time interval 2015\,December--2018\,April.
The most extreme X-ray flaring activity on the long-term, daily and
intraday timescales was observed during the 2-month period which
started in 2017\,December when the 0.3--10\,keV flux exceeded a
level of 5$\times$10$^{-9}$erg\,cm$^{-2}$s$^{-1}$, recorded only
twice previously. While the TeV-band  and X-ray variabilities mostly
were correlated, the source often varied in a complex manner in the
MeV--GeV and radio--UV energy ranges, indicating that the
multifrequency emission of Mrk\,421 could not be always generated in
a single zone. The longer-term flares at X-rays and $\gamma$-rays
showed a lognormal character, possibly indicating a variability
imprint of the accretion disk onto the jet. A vast majority of the
0.3--10\,keV spectra were consistent with the log-parabolic model,
showing relatively low spectral curvature and correlations between
the different spectral parameters, predicted in the case of the
first and second-order Fermi processes. The position of the
synchrotron spectral energy distribution (SED) peak showed an
extreme variability on diverse timescales between the energies
$E_{\rm p}$$<$0.1\,keV and $E_{\rm p}$$>$15\,keV, with 15\% of the
spectra peaking at hard X-rays and was related to the peak height as
$S_{\rm p}$$\varpropto$$E^{\alpha}_{\rm p}$ with $\alpha$$\sim$0.6,
which is expected for the transition from Kraichnan-type turbulence
into the \textquotedblleft hard-sphere\textquotedblright~ one. The
0.3--300\,GeV spectra showed the features of the hadronic
contribution, jet-star interaction and upscatter in the
Klein-Nishina regime in different time intervals.
\end{abstract}


\section{INTRODUCTION}

Blazars (BL Lacertae objects and flat-spectrum radio quasars) form
the most violently variable class of active galactic nuclei (AGNs),
with timescales ranging from a few minutes (in the keV--TeV energy
range) to several years (radio to optical frequencies). Moreover, BL
Lacertae sources (BLLs) are characterized by  featureless spectra,
variable radio--optical polarization, compact radio-structure and
superluminal motion of some components and  very broad continuum
extending over the radio to the very high-energy (VHE,
$E$$>$100\,GeV)  $\gamma$-ray energy ranges. The bolometric
luminosity occasionally can reach a level of
10$^{48}$\,erg\,s$^{-1}$, particularly, during the strong outbursts,
by the $\gamma$-ray emission  (see \citealt{f14}). Consequently,
BLLs are the most frequently detected class of the extragalactic TeV
sources (65 out of 82, with redshifts
$z$=0.03--0.61\footnote{http://tevcat.uchicago.edu/}) and form one
of the most important constituents of LAT 4-year Point Source
Catalog (3FGL; \citealt{ace15}). It is widely agreed that the
extreme physical properties of BLLs are due to the beamed,
non-thermal emission from a relativistic jet which is closely
aligned with the observer$\textquotesingle$s direction (estimated
viewing angles $\theta$$<$10\textdegree) and characterized by the
bulk Lorentz factor $\Gamma $$\sim$10, which occasionally attains
values as high as $\Gamma$$ \sim$50 \citep{b08}.

In the $\log\nu$--$\log\nu F_{\rm \nu}$ plane, BLLs generally
demonstrate a double-humped, broadband SED. There is a consensus
that the lower-energy component (extended over radio to UV--X-Ray
frequencies) is produced by synchrotron emission of
ultra-relativistic electrons \citep{c08}. A sub-class of the
high-energy-peaked BLLs (HBLs, peaking at UV--X-ray frequencies;
\citealt{p95} and references therein), are particularly important
due to the disputed particle acceleration and cooling processes:
their X-ray \textquotedblleft budget\textquotedblright~ should be
filled by synchrotron photons from the highest-energy leptons
(electrons and, possibly, positrons), while the radiative lifetime
at these energies are very short \citep{m04}. Consequently, a
detailed study of the timing and spectral behaviour of these sources
on diverse timescales reveals the most plausible acceleration
mechanisms and allows us to draw conclusions about the physical
properties of the jet emission region. Moreover, since the
synchrotron and inverse Compton (IC) cooling are expected to be
extremely important at these energies, the intense X-ray
timing/spectral study of the nearby, bright HBLs may provide us with
very important clues about the injection and radiative evolution of
the freshly-accelerated particles.

In this regard, the X-ray Telescope onboard the satellite
\emph{Swift} (\emph{Swift}-XRT; \citealt{b05}) makes an outstanding
contribution by performing a regular monitoring of selected BLLs in
their \textquotedblleft visibility\textquotedblright~ periods,
particularly, during the densely-sampled Target of Opportunity (TOO)
observations\footnote{https://www.swift.psu.edu/toop/too.php}. Owing
to the excellent instrumental characteristics, good photon
statistics and low background counts of \emph{Swift}-XRT, we are
able to search for flux and spectral variability on diverse
time-scales (minutes to years), obtain high-quality spectra and
derive different spectral parameters for bright HBLs even for
exposures lasting a few hundred seconds.

\begin{table}
\hspace{-0.5cm} \vspace{-0.1cm} \tabcolsep 4pt
 \begin{minipage}{85mm}
 \centering
  \caption{\label{perdiv} The intervals and sub-intervals referred throughout the paper.} \vspace{-0.2cm}
  \begin{tabular}{cccc}
  \hline
Period & Dates & MJD \\
 \hline
1 & 2015\,December\,8 to 2016\,June\,16 & 57364--57555 \\
1a & 2015\,December\,8 to 2016\,February\,4  &57364--57422 \\
1b & 2016\,February\,6 to 2016\,June\,16 &57424--57555 \\
2 & 2016\,November\,25 to 2017\,June\,27 &57717--57931  \\
2a& 2016\,November\,25 to 2017\,January\,29  & 57717--57782 \\
2b& 2017\,January\,31 to 2017\,June\,27  & 57784--57931\\
3& 2017\,December\,3 to 2018\,April\,8 & 58090--58216 \\
3a& 2017\,December\,16 to 2018\,February\,19  &58103--58168 \\
3b& 2018\,February\,21 to 2018\,April\,8  &58170--58216 \\
 \hline
\end{tabular}
\end{minipage}
\vspace{0.2cm}
\end{table}

\begin{table*}[ht]
 \small
 \vspace{-0.4cm}
 \tabcolsep 1.5pt
   \hspace{-0.9cm}
   \begin{minipage}{189mm}
  \caption{\label{xrt} Summary of the XRT and UVOT observations in the time interval 2015\,December--2018\,April (extract; see the corresponding machine-readable table for the entire content). The columns are as follows:
  (1) -- observation ID; (2) -- observation start--end; (3) - Modified Julian date corresponding to the observation start; (4)  exposure (in seconds); (5) observation-binned 0.3--10\,keV count rate and the associated
  uncertainty shown within parentheses; (6) - (11) - de-reddened UVOT magnitudes
and corresponding fluxes (in mJy).}
 \vspace{-0.2cm}
  \begin{tabular}{ccccccccccc}
       \hline
 ObsID & Obs. Start -- End  & MJD & Exp.  & CR & UVW1 & UVW1 & UVM2 & UVM2 & UVW2 & UVW2 \\
 & (UTC) & & (s) & (cts\,s$^{-1}$) &  (mag) & (mJy)& (mag) & (mJy) & (mag) & (mJy)  \\
(1) & (2) & (3) & (4) & (5) & (6) & (7) & (8) & (9) & (10)& (11) \\
\hline
35014240  &  2015-12-08 10:15:58--11:22:39& 57364.431& 1015 & 7.11(0.09)&  11.53(0.11)& 21.79(0.79)& 11.72(0.04)& 15.65(0.28)&11.51(0.10)& 18.31(0.46)\\
35014241 &   2015-12-11 11:38:58--12:43:43 & 57367.487 &  959&21.14(0.15)& 11.78(0.10)& 17.29(0.63)& 11.90(0.04)& 13.25(0.28)&11.80(0.03)& 14.00(0.28)\\
35014242 &    2015-12-14 09:56:58--11:02:04&  57370.416 &   1069&49.04(0.22)&  11.47(0.11) & 23.03(0.79)&  11.67(0.04)&  16.39(0.28)&11.47(0.10)&  18.99(0.46)\\
35014243 &   2015-12-17 11:20:58--12:26:26& 57373.475 &  1015&43.58(0.21)& 11.54(0.11)& 21.59(0.79) &11.76(0.04)& 15.09(0.28)&11.49(0.10)& 18.65(0.46)\\
35014245 &   2015-12-18 04:50:58--23:29:49& 57374.204 &  5892&29.34(0.14)& 11.49(0.10)& 22.61(0.74)& 11.66(0.10)& 16.55(0.38)&11.50(0.10)& 18.48(0.41)\\
 \hline
\end{tabular}
\end{minipage}
\end{table*}

The nearby (\emph{z}=0.031), TeV-detected HBL source Mrk\,421
provides an unique X-ray space laboratory due to the features as
follows \citep{b16,k16,k18a,k18b}): (i) high brightness (with the
\emph{Swift}-XRT 0.3--10\,keV count rates CR$>$100\,cts\,s$^{-1}$
during strong flares, corresponding to de-absorbed fluxes $F_{\rm
0.3-10 keV}\gtrsim$2.5$\times$10$^{-9}$\,erg\,cm$^{-2}$s$^{-1}$; ;
(ii) exceptionally strong outbursts (e.g., in 2013\,April;
\citealt{p14,k16}); (iii) very large and fast timing/spectral
variability on timescales down to a few hundred seconds; (iv)
extremely hard spectra during strong flares with the photon index
smaller than 1.6 and synchrotron SED peak shifting beyond 10\,keV
etc. Moreover, the source is also bright  in other spectral ranges,
making it a frequent target of densely-sampled MWL campaigns
(\citealt{m95}, \citealt{a12,a15a,a15b}, \citealt{b16} etc.), which
are crucial for checking the viability of models explaining the
origin of the higher-energy SED component via the inter-band
correlation study: (1) synchrotron self-Compton (SSC, scattering of
synchrotron photons by their \textquotedblleft
parent\textquotedblright~ lepton population; \citealt{m85}); (2)
external Compton (EC), with the low-energy photons from the
accretion disc (AD), dust torus, narrow/broad line clouds
upscattered by the jet ultra-relativistic particles \citep{d92}; (3)
hadronic models, which embody a generation of the keV--TeV emission
by relativistic protons, either directly (synchrotron-proton
scenario; \citealt{ab11}) or indirectly (e.g., synchrotron radiation
by the electron population, produced by a cascade induced by the
interaction of high-energy protons with the ambient photons;
\citealt{m93}). Despite the large number of publications related to
the aforementioned observations, the details of the physical
processes underlying the higher-energy SED component remain mainly
unknown  owing to (i) sparse multiwavelength (MWL) data during long
periods; (ii) moderate or low sensitivity in the hard X-ray and
$\gamma$-ray energy ranges in the past;  (iii) many previous MWL
campaigns were triggered in the epochs of enhanced X-ray and
$\gamma$-ray activity, and, consequently, these studies are biased
towards the high states of the source, while the distinct physical
processes may play a dominant role during the moderate and\emph{}
lower brightness states.

\begin{figure*}[ht!]
\vspace{-0.1cm}
  \includegraphics[trim=7.3cm 0.2cm 0cm 0cm, clip=true, scale=0.825]{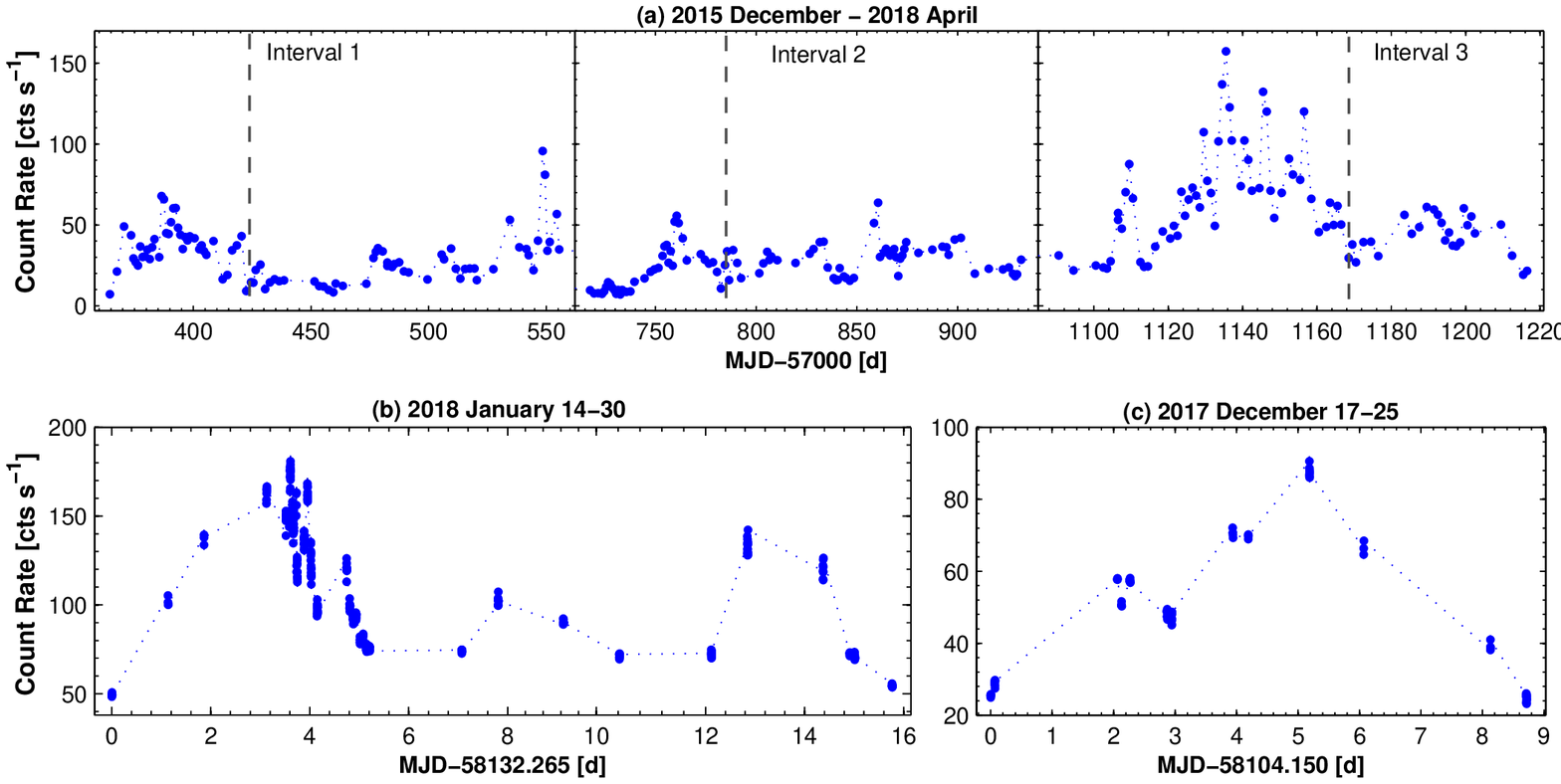}
\includegraphics[trim=7.3cm 0cm 0cm 0cm, clip=true, scale=0.825]{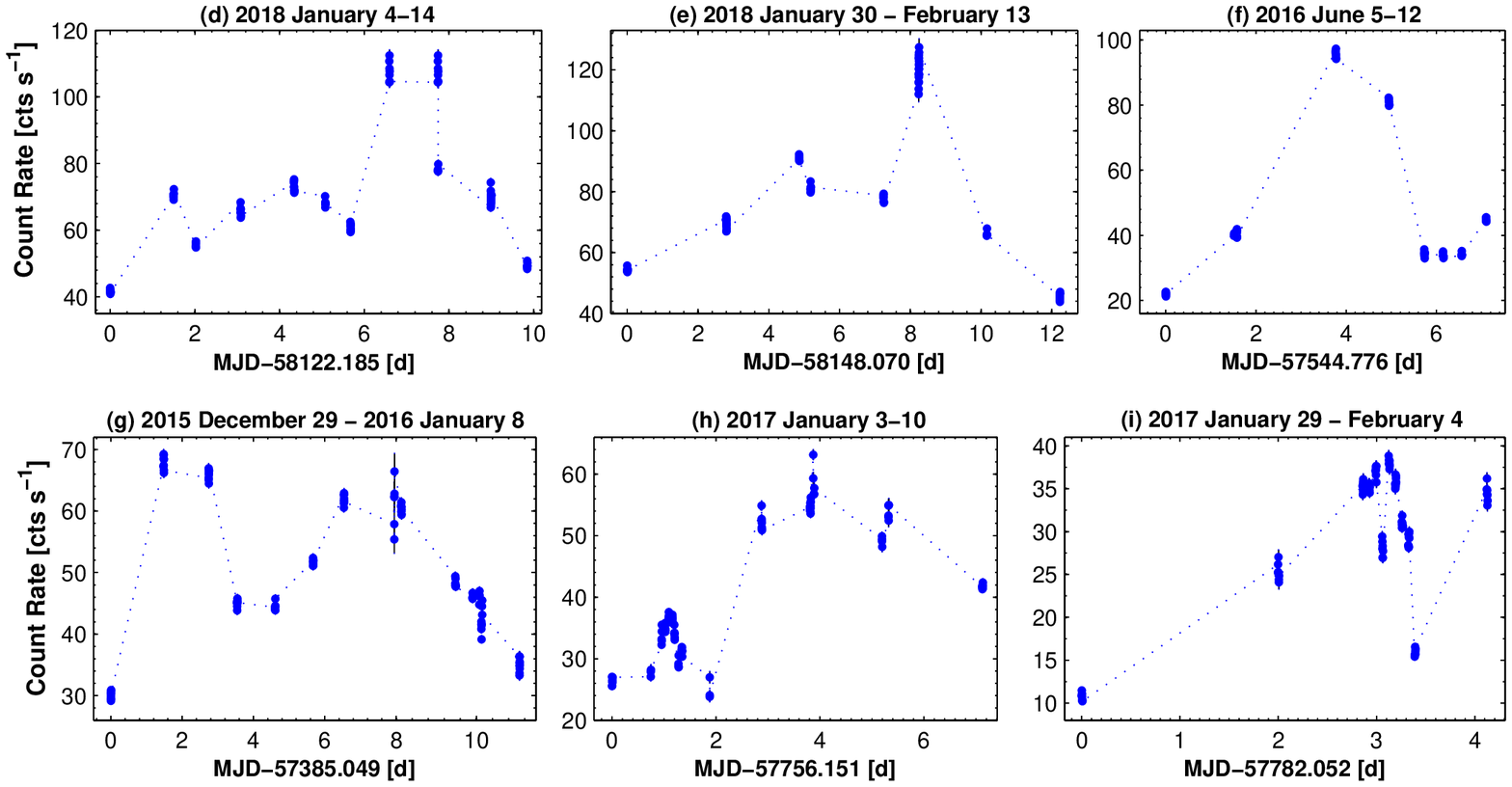}
\vspace{-0.5cm}
 \caption{\label{flares} Panel\,(a): X-ray flaring activity of Mrk\,421 in the 0.3--10\,keV energy range during different periods. In each plot, the vertical
 dashed line denotes a boundary between the different sub-intervals; panels (b)--(i): the strongest
 X-ray flares during Intervals\,1--3.}
 \end{figure*}

For the aforementioned reasons, we performed a detailed study of
X-ray spectral and flux variability in the MWL context, focussed on
the rich archival data obtained with XRT during
2005\,March--2015\,June \citep{k16,k17a,k18a,k18b}. This campaign
revealed extreme X-ray flares by a factor of 3--20 on the timescales
of a few days--weeks between the lowest historical state and that
corresponding to the \emph{Swift}-XRT rate higher than
200\,cts\,s$^{-1}$. The source was characterized by extreme spectral
and intraday flux variability, particularly during the strong
flares. In the latter case, we were able to extract the 0.3--10\,keV
spectra for the time intervals of 50--100 seconds and explore the
observational features predicted in the framework of the various
particle acceleration and emission scenarios. The distribution of
different spectral parameters and their cross-correlations hinted at
the importance of the first and second-order Fermi accelerations,
changes in the turbulence type, possible hadronic contribution to
the MeV--GeV emission etc.

In this paper, we present the results of our detailed study of the
timing/spectral behaviour of Mrk\,421 during the period
2015\,December--2018\,April, which was characterized by a very
strong X-ray flare in the time interval
2017\,December--2018\,February, and showed an intensive flaring
activity also in other parts of the here-presented period, revealed
by the densely-sampled \emph{Swift}-XRT observations. Using XRT
observations, we checked correlations between the 0.3--10\,keV flux
variability and those observed with different instruments: the
Ultraviolet-Optical Telescope (UVOT; \citealt{r05}) and the Burst
Alert Telescope (BAT; \citealt{ba05}) onboard \emph{Swift}, the
Large Are Telescope (LAT) onboard \emph{Fermi} \citep{at09}, MAXI
\citep{m09}, the First G-APD Cherenkov Telescope (FACT;
\citealt{a13}), the 40-m telescope of Owens Valley Radio Observatory
(OVRO; \citealt{r11}), the optical telescopes of Steward Observatory
\citep{s09}.

The paper is organized as follows: Section\,2 encompasses the
description of the data processing and analyzing procedures. The
results of the X-ray and MWL timing study, as well as those from the
X-ray spectral analysis are presented in Section\,3. We provide a
discussion, based on our results, and the corresponding conclusions
in Section\,4. Finally, the summary of our study is given in
Section\,5.

\begin{table*}
 \tabcolsep 3pt
 \hspace{-0.4cm}
 \begin{minipage}{185mm}
  \caption{\label{persum}  Summary of the XRT and  UVOT observations in different intervals.  Cols\,(2)--(5): maximum 0.3--10\,keV flux
(in cts\,s$^{-1}$), maximum-to-minimum flux ratio, mean flux and
fractional amplitude, respectively;  maximum de-absorbed flux (in
10$^{-10}$erg\,cm$^{-2}$s$^{-1}$), maximum-to-minimum flux ratio,
mean flux and fractional amplitude in the 0.3--2\,keV
(Cols\,(6(-(9)) and 2--10keV (Cols\,(10)-(13)) bands; maximum
unabsorbed flux (in mJys), maximum-to-minimum flux ratio and
fractional amplitude in the bands UVW1 (Cols\,(15)--(18)), UVM2
(Cols\,(19)--(22)) and UVW2 (Cols\,(23)--(26)).} \vspace{-0.2cm}
  \begin{tabular}{cccccccccccccccccc}
    \hline
   & \multicolumn{12}{c}{XRT}    \\
  \hline
    & \multicolumn{4}{c}{0.3--10\,keV}& \multicolumn{4}{c}{0.3--2\,keV}& \multicolumn{4}{c}{2--10\,keV}\\
    \hline
Per. & $CR_{\rm max}$  &$\Re$ & Mean& $100\times F_{\rm var}$&$F^{\rm max}$& $\Re$ &Mean&$100\times F_{\rm var}$ &$F_{\rm max}$ & $\Re$ & Mean & $100\times F_{\rm var}$    \\
(1) & (2) & (3) & (4) & (5) & (6) & (7) & (8) & (9) & (10) & (11) & (12) & (13)  \\
 \hline
1 & 95.72(0.32) & 13.5& 31.53(0.02) &50.5(0.1)& 15.85(0.18)&12.3 &7.19(0.01) & 41.6(0.1) &13.93(0.38) &59.5 & 3.55(0.01) & 75.0(0.2) \\
1a & 67.83(0.28)&9.5&37.98(0.03) &35.4(0.1)&12.76(0.26)&9.9 &7.60(0.01) &30.3(0.1)  &7.96(0.24) &34.0 &3.31(0.01) & 44.5(0.3)   \\
1b &95.72(0.32) & 11.4&26.56(0.03) & 60.3(0.1)& 15.85(0.18) &8.3 &6.62(0.01) &55.9(0.1) &13.93(0.38) & 46.3&38.90(0.01) & 95.9(0.4)  \\
2 &63.75(0.30) & 9.0 & 25.81(0.02)& 44.5(0.1)&12.02(0.14) & 7.9&5.54(0.01) &34.2(0.1)&12.25(0.33) &47.5 & 38.24(0.01)&58.8(0.2)  \\
2a &55.70(0.18) &7.9 &21.40(0.03) &62.4(0.1)& 9.66(0.15) &6.4 &5.45(0.01) &39.7(0.1)  &9.33(0.25) &36.2 &37.70(0.01) &65.8(0.4)  \\
2b & 63.75(0.30)&4.2 & 28.43(0.03)& 33.5(0.1)& 12.02(0.14) &4.2 &5.61(0.01) &30.1(0.1) & 12.25(0.33))& 12.7&38.62(0.01) &53.9(0.3))\\
3 & 162.88(0.70) & 8.5 &60.89(0.03) &51.0(0.1)&26.85(0.30)  & 13.2& 14.29(0.01)&38.8(0.01) &29.04(0.79) & 31.2  &11.80(0.02) &63.4(0.2)    \\
\textbf{3a} &162.88(0.70) & 7.1& 71.26(0.04) &45.9(0.1)& 26.85(0.30) & 6.1& 15.60(0.01)&33.5(0.01) &29.04(0.79) &22.1& 13.30(0.02) &55.6(0.2)  \\
3b & 61.10(0.26)&3.2 &43.33(0.05)  & 27.2(0.1)&19.45(0.27)&9.6 &8.52(0.01) & 38.5(0.02)&10.38(0.32) & 11.1&5.15(0.02)&  40.4(0.4)   \\
 \hline
  &  \multicolumn{12}{c}{UVOT}     \\
  \hline
    & \multicolumn{4}{c}{UVW1}& \multicolumn{4}{c}{UVM2}& \multicolumn{4}{c}{UVW2}\\
  \hline
Per. & $F_{\rm max}$  &$\Re$ & Mean& $100\times F_{\rm var}$&$F_{\rm max}$& $\Re$ &Mean&$100\times F_{\rm var}$ &$F_{\rm max}$ & $\Re$ & Mean & $100\times F_{\rm var}$\\
(14) & (15) & (16) & (17) & (18) & (19) & (20) & (21) & (22) & (23) & (24) & (25) & (26)  \\
 \hline
1 &29.83(1.05) &2.4& 22.8(0.4)& 20.36(0.07)  & 21.22(0.41) & 2.1 &15.19(0.03)  &21.3(0.2)  & 26.24(0.74) & 2.4 &18.10(0.05)  & 23.0(0.3)\\
1a &28.75(0.97) &1.7 &22.86(0.13)  &12.5(0.6)  & 20.64(0.28) &1.7  & 15.88(0.04) &16.5(0.3)  & 26.24(0.74) & 1.9 &20.09(0.07)  & 15.1(0.4)\\
1b &29.83(1.05) &2.4 &18.15(0.09) &26.5(0.5)  & 21.22(0.41) & 2.1 & 14.55(0.03) &25.0(0.3)  &25.06(0.64)  &2.3  &16.16(0.06)  &26.1(0.4) \\
2 &18.11(0.63)& 2.1 & 11.33(0.04) & 18.4(0.3)&11.11(0.19)  &2.7  & 7.73(0.02) & 20.0(0.2) &14.13(0.28)  & 1.9 &9.90(0.02)  &14.9(0.2) \\
2a &18.11(0.63) &2.1 &12.8(0.07) & 19.9(0.6) & 11.11(0.19) & 2.7 &7.56(0.03)  &28.2(0.4)  & 14.13(0.28) &1.9  & 10.65(0.04) & 17.1(0.4)\\
2b &13.71(0.50) &1.5 &10.38(0.04) &7.9(0.04)  &10.51(0.19)  & 1.9 &7.84(0.02)  & 13.1(0.3) &13.00(0.28)  &1.7  &9.42(0.03)  &10.1(0.3) \\
3 &25.50(0.64) & 2.1 &19.40(0.07) & 20.1(0.4) & 18.82(0.28) & 2.0 & 13.6(0.03) & 17.2(0.2) & 22.43(0.56) &2.0  & 17.16(0.04) & 19.0(0.3)\\
3a &25.50(0.64) & 1.4 &21.82(0.10)& 7.2(0.5) &18.82(0.28)  & 1.6 & 14.92(0.04) & 10.8(0.3) &22.43(0.56)  & 1.4 &19.29(0.06)  &7.1(0.3) \\
3b & 19.14(0.63)&1.6  &14.10(0.10) & 14.8(0.7) & 15.08(0.28) & 1.6 & 11.17(0.04) & 12.1(0.4) & 17.64(0.41) & 1.6 & 12.99(0.06) & 13.1(0.5)\\
 \hline
\end{tabular}
\end{minipage}
\end{table*}

\section{DATA SETS, REDUCTION and ANALYSIS }

\subsection{X-Ray Data}
We retrieved the raw \emph{Swift}-XRT data from \emph{NASA's Archive
of Data on Energetic
Phenomena}\footnote{https://heasarc.gsfc.nasa.gov/} (HEASARC). The
Level\,1 event files were reduced, calibrated and cleaned via the
\texttt{XRTPIPELINE} script (included in the package
\texttt{HEASOFT} v.\,6.26) by applying the standard filtering
criteria and the latest calibration files of XRT CALDB v.20190412.
The events with 0–-2 grades are selected for those observations
performed in the Windowed Timing (WT) mode. The selection of the
source and background extraction regions was performed with
\texttt{XSELECT}, using the circular area with radii of 25--50
pixels depending on the source brightness and position in the XRT
field-of view (FOV), as well as on the exposure length. We produced
a pile-up correction for the count rate
CR$\gtrsim$100\,cts\,s$^{-1}$ by excluding the central area with
radii of 1--3 pixels from the source extraction region, following
the recipe provided by \cite{ro06}. Afterwards, the light curves
were corrected using the task \texttt{XRTLCCORR} for the resultant
loss of the effective area, bad/hot pixels, pile-up, and vignetting.
Moreover, the corrections on the point-spread function losses,
different extraction regions, vignetting and CCD defects were done
by generating the ancillary response files (ARFs) using the
\texttt{XRTMKARF} task.

Due to the high X-ray brightness of the source, it generally was not
observed in the Photon Counting (PC) regime. However, Mrk\,421 was
accidentally targeted in this regime three times in the
here-presented period (ObsIDs\,35014255\footnote{The three leading
zeroes of each ObsID are omitted throughout the paper.}, 34228023
and 34228026). In those cases, we used the events with 0-–12 grades
for our analysis. The pile-up correction was done according to the
prescription of \citealt{m05}. The radius, below which the model
overproduced the data, was accepted as a region affected by pile-up
(11--15 pixels for the particular observation). The source events
were extracted from an annular region with the inner radius
encircling the pileup area and the outer radius of 50--60 pixels.
The loss of counts caused by the inner hole in the source region,
vignetting and bad pixels were corrected by generating the
corresponding ARF-file. The background counts were extracted from a
surrounding annulus with radii of 80 and 120 pixels.

\begin{figure*}
\vspace{-0.1cm}
\includegraphics[trim=7.3cm 0.8cm -1cm 0cm, clip=true, scale=0.82]{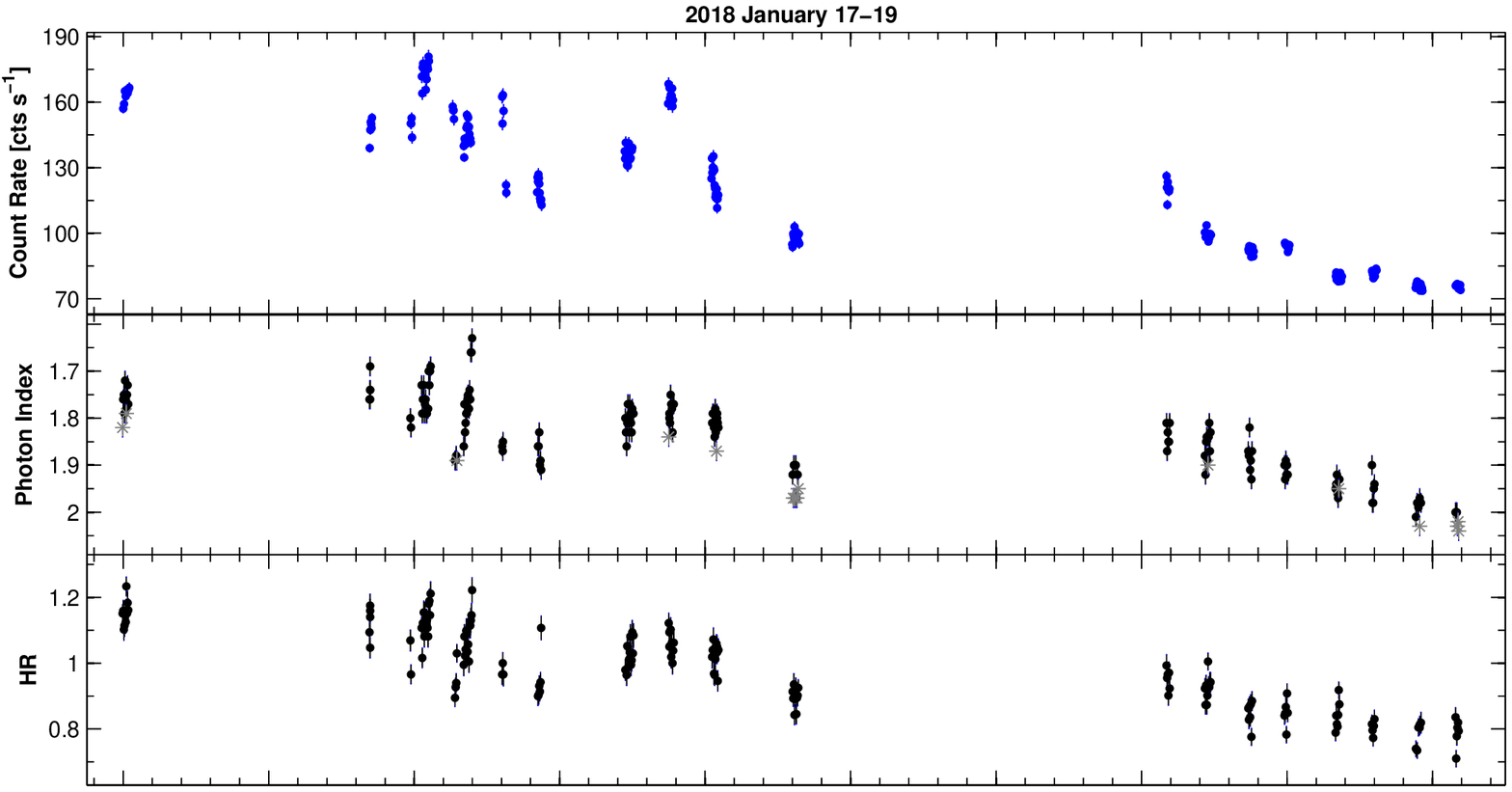}
\includegraphics[trim=7.3cm 3.8cm -1cm 0cm, clip=true, scale=0.82]{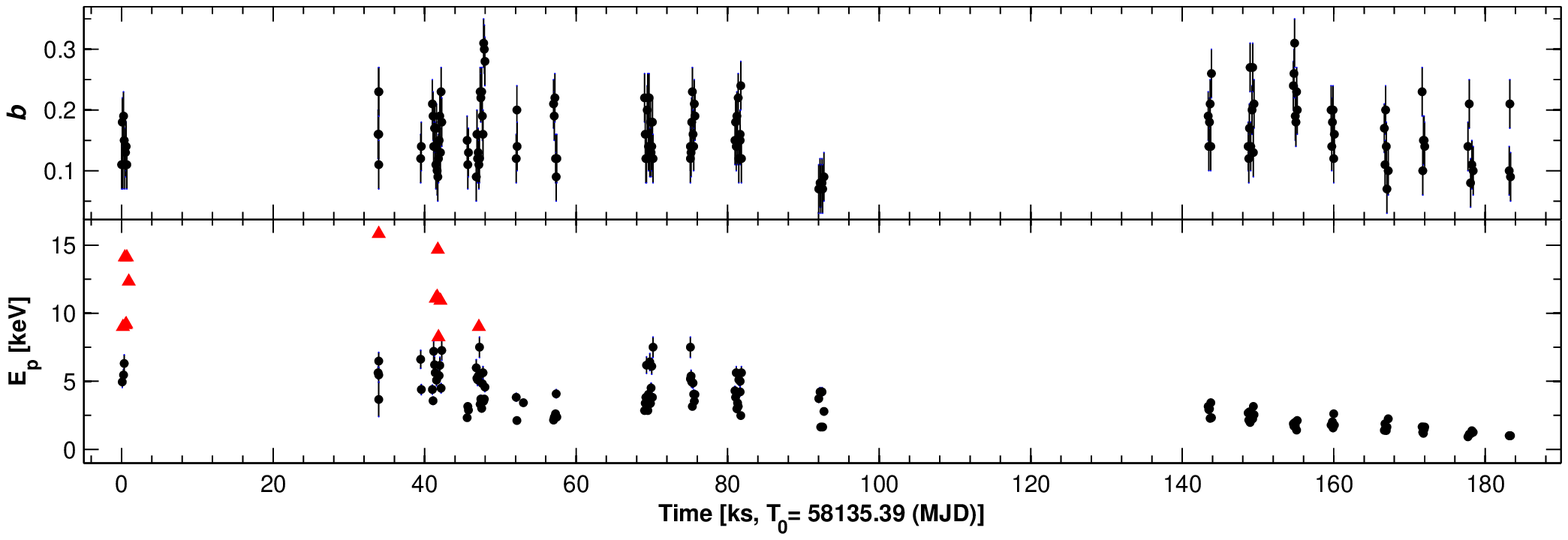}
\vspace{-0.6cm}
 \caption{\label{idvext1} Top panel: intraday 0.3--10\,keV variability during the strongest X-ray flaring activity in 2018\,January 17--19. The subsequent panels:
 timing behaviour of different spectral parameters. In the second panel, the black points and the gray asterisks stand for the photon
 indices \emph{a} and $\Gamma$, respectively. Red triangles in the bottom panel stand for the lower limits to the intrinsic position of the synchrotron SED
 peak.}
  \end{figure*}

From the publicly available, daily-binned
BAT\footnote{http://swift.gsfc.nasa.gov/results/transients/weak/Mrk421/}
and MAXI\footnote{http://maxi.riken.jp/} data, we used only those
corresponding to the target's detection with a minimum significance
of 5\,$\sigma$ to study a variability of the 15--150\,keV and
2--20\,keV fluxes, respectively.

 \begin{figure*}
\vspace{-0.1cm}
\includegraphics[trim=7.3cm 2.6cm -1cm 0cm, clip=true, scale=0.83]{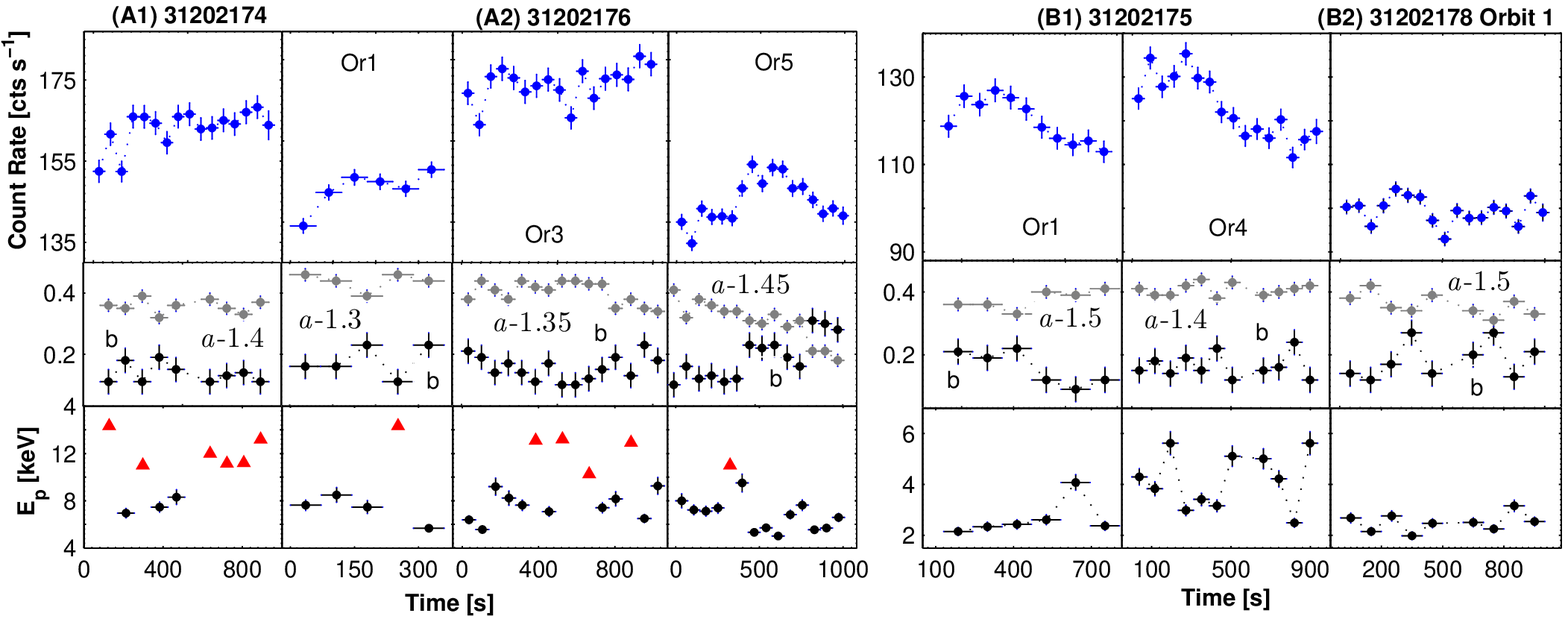}
\includegraphics[trim=7.3cm 2.5cm -1cm 0cm, clip=true, scale=0.83]{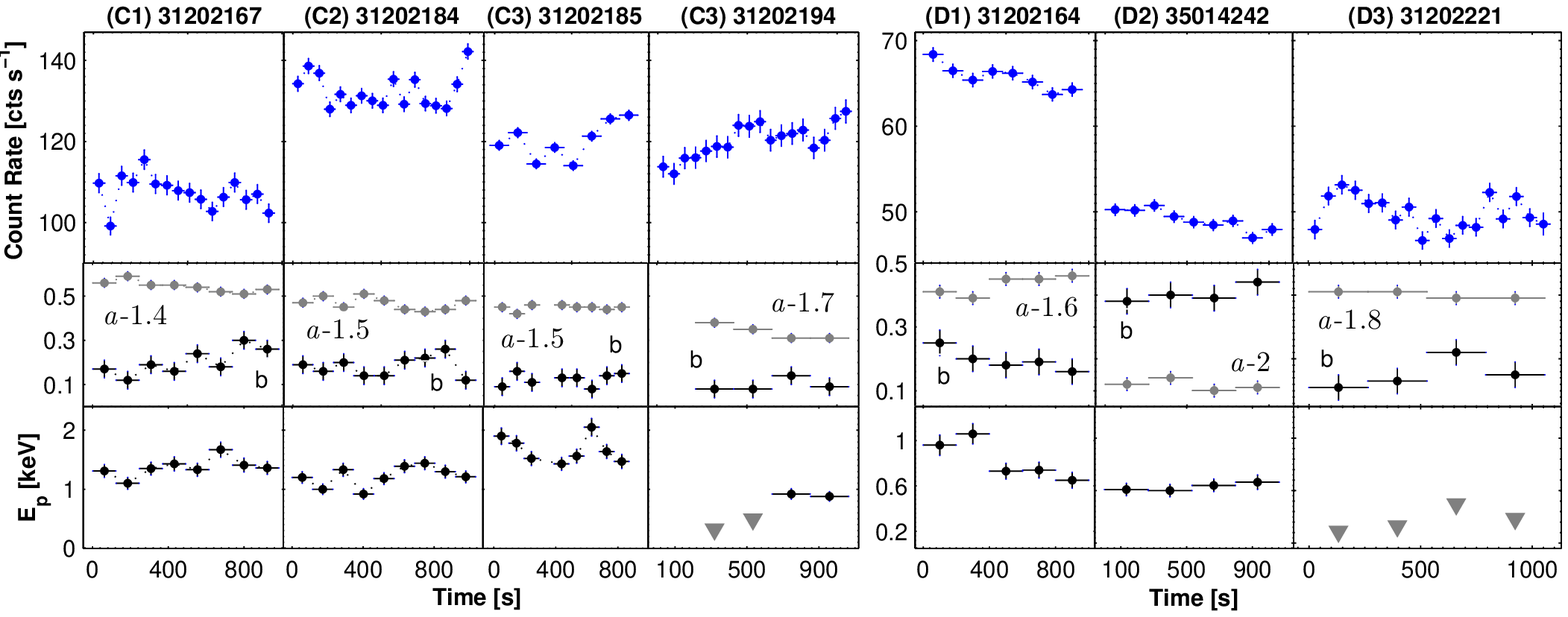}
\includegraphics[trim=7.3cm 6.6cm -1cm 0cm, clip=true, scale=0.83]{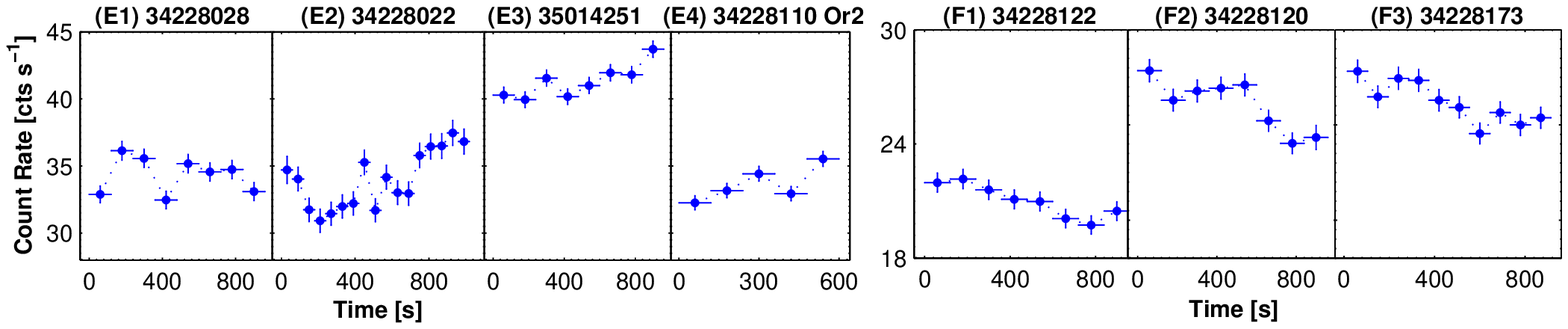}
\vspace{-0.4cm}
 \caption{\label{idv1ks} The fastest 0.3--10\,keV IDVs. In the
 second panel of Figures\,A1--D3, the gray points correspond to the
 photon index $a$,  arbitrarily shifted for the better resolution.
 The red and gray triangles in the bottom panels of (A1)--(A2), (C3) and (D3) plots stand for the lower and
 upper limits to the intrinsic position of the synchrotron SED peak, respectively.}
 \end{figure*}

 \subsection{$\gamma$-ray Observations}
The reduction of the \emph{Fermi}--LAT data was performed  with
\texttt{ScienceTools} (version v11r5p3), adopting the instrument
response function \texttt{P8R3\_SOURCE\_V2} and the unbinned maximum
likelihood method \texttt{GTLIKE}. We selected the 0.3--300\,GeV
energy  range for extraction of the photon flux and spectral
information, since the effective area of the instrument is larger
($>$0.5m$^2$) and the angular resolution relatively good (the 68\%
containment angle smaller than 2\,deg) in that case \citep{at09}.
Consequently, we obtain smaller systematic errors and the spectral
fit is less sensitive to possible contamination from unaccounted,
transient neighbouring sources \citep{ab11}. The events of the
diffuse class (\texttt{evclass}=128, \texttt{evtype}=3), i.e, those
with the highest probability of being photons,  from a region of
interest (ROI) with the 10-deg radius centered at the location of
Mrk\,421 were included in our analysis. Moreover, we discarded the
events at zenith angles $>$100\,deg (to avoid a contamination from
the Earth-albedo photons, generated by cosmic rays interacting with
the upper atmosphere) and those recorded when the spacecraft rocking
angle was larger than 52\,deg (greatly reducing the contamination
from Earth-limb photons).

 \begin{figure*}
\vspace{-0.1cm}
\includegraphics[trim=7.2cm 0.7cm -1cm 0cm, clip=true, scale=0.82]{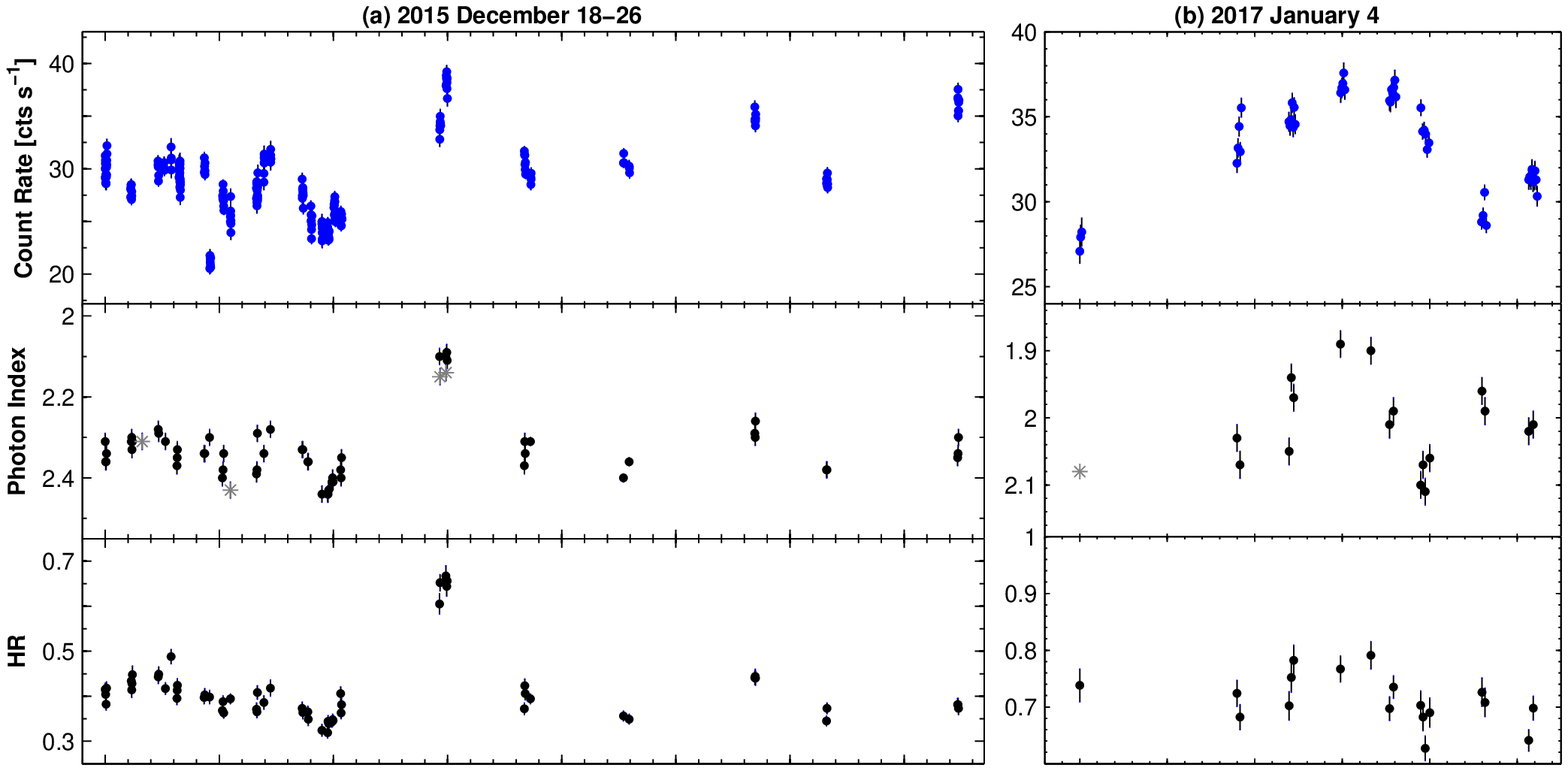}
\includegraphics[trim=7.2cm 3.9cm -1cm 0cm, clip=true, scale=0.82]{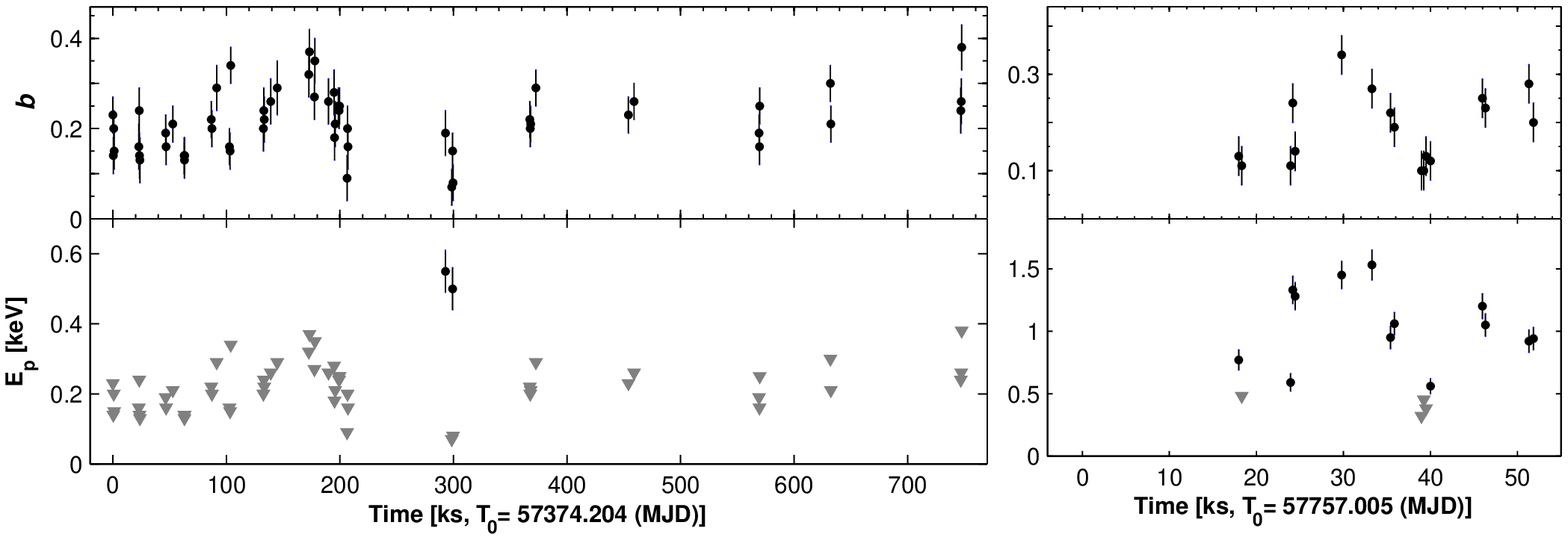}
\vspace{-0.5cm}
 \caption{\label{idvext2} Two examples of the  densely sampled 0.3--10\,keV light curves with a weaker intraday variability. In the second panel, the black points and the gray asterisks stand for the photon
 indices \emph{a} and $\Gamma$, respectively. Gray triangles in the bottom panel
 stand for the upper limits to the intrinsic position of the synchrotron SED peak.}
 \end{figure*}

The background model  \texttt{gll\_iem\_v07.fits} was created to
extract the $\gamma$-ray signal from (i) the Galactic
diffuse-emission component; (ii) an isotropic component, which is
the sum of the extragalactic diffuse emission and the residual
charged particle background (parameterized by the file
\texttt{iso\_P8R3\_SOURCE\_V2\_v1.txt}); (iii) all $\gamma$-ray
sources from the \emph{Fermi}-LAT 4-yr Point Source Catalog (3FGL,
\citealt{ace15}) within 20\,deg of Mrk\,421. For the spectral
modelling of our target, we adopted a simple power law, similar to
the 3FGL catalogue. The spectral parameters of the sources within
the ROI were left free during the minimization process while those
outside of this range were held fixed to the 3FGL catalog values.
The normalizations of both components (i)--(ii) in the background
model were allowed to vary freely during the spectral fit. The
photon flux and spectral parameters were estimated using the
unbinned maximum-likelihood technique \citep{m96}.

When the target's detection significance is less than 5$\sigma$
(i.e., the corresponding test-statistics TS$<$25) and/or the  number
of the model-predicted counts $N_{\rm pred}$$\lesssim$10, such
detections are not robust. For example, even a small change in the
time-bin width can result in significantly different values of the
photon flux and spectral parameters. In such cases, we calculated
the upper limit to the photon flux\footnote{See
fermi.gsfc.nasa.gov/ssc/data/analysis/scitools/upper\_limits.html}.

We used the user-contributed tool
\texttt{likeSED}\footnote{https://fermi.gsfc.nasa.gov/ssc/data/analysis/user/}
to construct the 300\,MeV--300\,GeV SED of Mrk\,421. In that case,
the photon indices of the sources were frozen to the best-fit values
obtained from the full spectral analysis when performing unbinned
likelihood fits in differential energy bins (following the recipe
provided in \citealt{a15a}).

The source was observed with FACT at VHE energies during 363 nights
for a total of 1408\,hr in the period
2015\,December\,8--2018\,April\,8. For the timing study, we have
used only the nightly-binned TeV excess rates corresponding to
detection significances higher than 3$\sigma$\footnote{ See
http://www.fact-project.org/monitoring/}, since more than 98\% of
these data are taken with a zenith distance small enough to not
influence significantly the energy threshold of the analysis (see
\citealt{d15} for the data reduction and analysis details). More
than 84\% of the same data are taken under light conditions not
increasing the analysis threshold. This results in 190 nights for
which the nightly observation time ranges of 0.66--7.32\,hr. In the
case of the 20-min binned data, the source was detected 456-times in
the here-presented period, and the corresponding rates were used in
searching for the intraday brightness variability.

\subsection{UV, Optical and radio data}
The source was targeted with \emph{Swift}-UVOT in the ultraviolet
bands \textit{UVW1}, \textit{UVM2}, and \textit{UVW2} simultaneously
with XRT. Generally, Mrk\,421 was not observed with UVOT in the
optical \emph{V--U} bands due to the presence of very bright stars
in the telescope's FOV. The absolute photometry for the
sky-corrected images was performed by means of the
\texttt{UVOTSOURCE} tool (distributed within \texttt{HEASOFT}) and
the calibration files included in the CALDB v.20170922. The
measurements were done using a 20\,arcsec radius due to the target's
high UV-brightness. When the source was brighter than 12\,mag, a
pile-up was estimated and the corresponding correction was performed
using the recipe provided in \cite{pa13}. According to the latter, a
systematic uncertainty of $\pm$0.1\,mag should be added to the
measurements. The magnitudes were then corrected for the Galactic
absorption adopting $E(B-V)$=0.028\,mag (see \citealt{k18a}), and
the $A_\lambda/E(B-V)$ values derived from the interstellar
extinction curves \citep{fi07}. For this purpose, we used the
effective wavelength of each filter adopted from \cite{p08}.
Finally, the magnitudes were converted into milli-Janskys by
adopting the latest photometric zero-points for each band provided
in \cite{b11}, and the host contribution was removed by subtracting
the values of 0.09\,mJy, 0.05\,mJy and 0.06\,mJy for the \emph{UVW1,
UVM2, UVW2} bands, respectively \citep{c12}.

The publicly available  \emph{V} and \emph{R}-band magnitudes,
obtained with the 2.3\,m Bock and 1.54\,m Kuiper telescopes of
Steward observatory\footnote{See
http://james.as.arizona.edu/~psmith/Fermi/} (see \citealt{s09} for
details), were de-reddened and converted into milli-Janskys
according to \cite{b79}. In both bands, the host contribution was
subtracted following \cite{n07} and \cite{f95}.

The 15\,GHz radio fluxes, obtained with the OVRO 40-m telescope,
were retrieved from the corresponding website\footnote{see
http://www.astro.caltech.edu/ovroblazars/} (see \citealt{r11} for
the data reduction and calibration steps). The sharp spikes or drops
in the light curve, associated with less-favourable observing
conditions, were not included in our analysis.

 \begin{table*}
 \tabcolsep 1.3pt \hspace{-0.9cm}
     \begin{minipage}{189mm}
  \caption{\label{idvtable}  The 0.3--10\,keV IDVs during 2015\,December--2018\,April (extract; see the corresponding machine-readable
table for the full version). Col.\,(1) contains the ObsIDs of those
\emph{Swift}-XRT pointings to the source during which the given
event was recorded. Col.\,(2)  presents the MJD of the observation
start and the total length of the particular observation (including
the intervals between the separate orbits); Col.\,(3): reduced
chi-squared and the corresponding degrees-of-freedom for the given
observations, along with the time bin used for the variability
search; Col.\,(4): fractional variability amplitude and the
associated error in the parenthesis; Cols\,(5)--(8): the ranges of
different spectral parameters, obtained via the log-parabolic (LP)
or power-law (PL) fits with the spectra extracted from  the separate
orbits (or segments) of  the corresponding  XRT observation. The
last column provides a remark related to the variable spectral
parameter  making the basic contribution in the observed IDV:  1 --
photon index; 2 -- curvature; 3 -- $E_{\rm p}$; 4 -- $S_{\rm p}$;  5
-- change in the particles' energy distribution from the
log-parabolic  functional shape into the power-law one or vice
versa.}
   \vspace{-0.2cm}
   \hspace{-0.5cm}
  \begin{tabular}{cccccccccc}
  \hline
 ObsID(s)& MJD/$\Delta$T(hr) & $\chi^2_{\rm r}$/dof/bin &  $100\times F_{\rm var}$  &
  $ a$ or $\Gamma$& $ b $ & $ E_{\rm p}$ (keV) & \emph{HR}& Remark \\
   (1)& (2) & (3)  & (4)  & (5) & (6)  & (7)  & (8) & (9) \\
\hline
35014242 &   57370.416/0.30 & 3.41/8/120\,s &  2.1(0.5)&2.20(0.02)$-$2.24(0.02) &  0.38(0.04)$-$0.44(0.04)&0.50(0.06)$-$0.58(0.07)&0.41(0.01)$-$0.44(0.02) &  1\\
34014243$-$ &  57373.475/17.98 &2925/1/Or & 26.3(0.3)&2.17(0.02)$-$2.31(0.02)&   0.20(0.05)$-$0.27(0.05)&0.21(0.04)$-$0.48(0.06)& 0.38(0.01)$-$0.51(0.02) &  1,3\\
31202245\,Or1& \\
35014245 &   57374.204/18.66& 28.63/5/Or & 3.4(0.3)&2.28(0.02)$-$2.36(0.02)\,LP& 0.13(0.04)$-$0.23(0.04)&0.05(0.02)$-$0.24(0.05)& 0.38(0.01)$-$0.45(0.02) &  1,2,4,5\\
&&&&2.31(0.02)\,PL \\
35014245\,Or1 &   57374.204/0.48 & 2.82/18/90\,s &   2.7(0.5)&2.31(0.02)$-$2.36(0.02)&   0.14(0.04)$-$0.23(0.04)&0.05(0.02)$-$0.21(0.04)&0.38(0.01)$-$0.42(0.02)  & 1,2,4\\
 \hline
\end{tabular}
\end{minipage}
\end{table*}

\begin{figure*}[ht!]
\vspace{-0.1cm}
\includegraphics[trim=7.3cm 0.9cm 0cm 0cm, clip=true, scale=0.86]{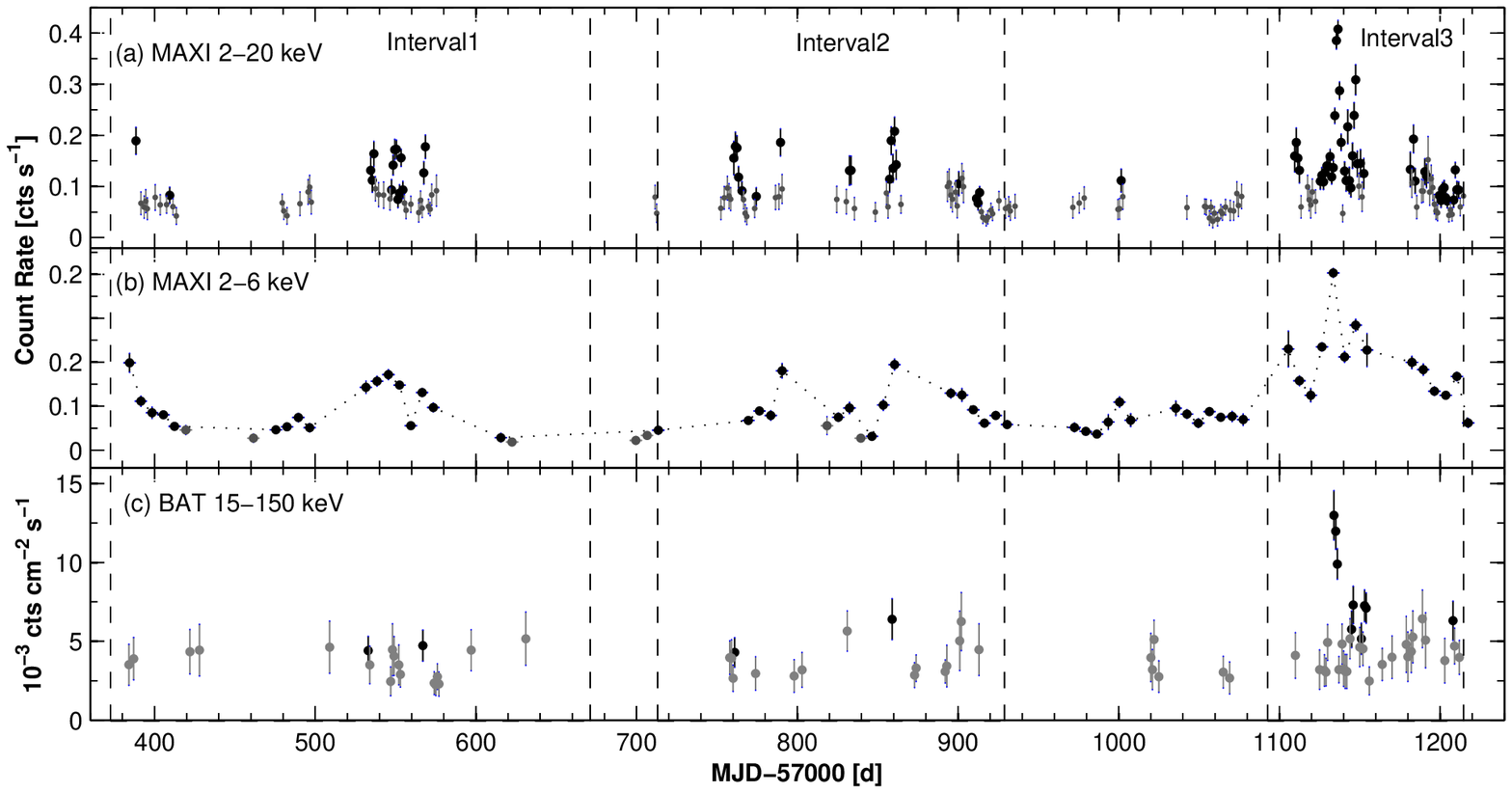}
\vspace{-0.4cm}
 \caption{\label{maxi} Long-term behaviour of Mrk\,421 in the MAXI 2--20\,keV and 2--6\,keV bands (panels (a) and (b), respectively), as well as in the
 BAT energy range (bottom panel). The daily-binned data are used to construct light curves. The black and gray points correspond to the detections with 5$\sigma$ and 3$\sigma$ significances, respectively.} \vspace{0.2cm}
 \end{figure*}

 \subsection{Analysis Methods}
The 0.3--10\,keV spectra, extracted from the XRT observations and
corrected for the different effects (see Section\,2.1), were further
reduced as follows: using the \texttt{GRPPHA} task, we combined the
instrumental channels to include at least 20\,photons per bin,
making a spectrum valid for the $\chi^2$-statistics. The reduced
spectra were fitted with three different models, generally adopted
for the blazar X-ray spectra (by fixing the Hydrogen column density
to the Galactic value $N_{\rm H}=1.90\times10^{20}$ cm$^{-2}$,
obtained within the Leiden/Argentine/Bonn (LAB) survey;
\citealt{k05}): (1) log-parabolic model \citep{m04}
 \vspace{-0.2cm}
\begin{equation} F(E)=K(E/E_{\rm 1})^{-(a+b
log(E/E_{\rm 1}))},
 \vspace{-0.1cm}
\end{equation}
with $E_{\rm 1}$ fixed to 1\,keV; $a$, the photon index at the
energy $E_{\rm 1}$; $b$, the curvature parameter; $K$, the
normalization factor. The position of the synchrotron SED peak was
calculated as $E_{\rm p}$=10$^{(2-a)/2b}$\,keV; (2) simple power-law
$ F(E)=KE^{-\Gamma}$, with $\Gamma$, the photon index throughout the
entire 0.3--10\,keV energy range; (3) broken power-law
 \vspace{-0.2cm}
\begin{equation}
\begin{array}{c l}
F(E)=KE^{-\Gamma_1}, ~ E\leq E_{\rm br}  \\
F(E)=KE^{\Gamma_2-\Gamma_1}_{\rm br} (E/1keV)^{-\Gamma_2}, ~
E>E_{\rm br} ,
 \end{array}
 \vspace{-0.1cm}
\end{equation}
with $E_{\rm br}$: break point for the energy in keV, $\Gamma_1$:
photon index for $E\leq E_{\rm br}$, $\Gamma_2$: photon index for
$E\leq E_{\rm br}$. The model validity was determined using the
reduced chi-squared ($\chi^2_{\rm r}$), distribution of the
residuals, and F-test. The high X-ray brightness of Mrk\,421 allowed
us to extract the spectra from separate orbits of the particular
ObsID (especially important when it is impossible to use the same
source and/or background extraction regions for all orbits, or the
source is variable), or even from the separated segments of a single
orbit in the medium and higher brightness states. The unabsorbed
0.3--2\,keV, 2--10\,keV and 0.3--10\,keV fluxes and their errors (in
logarithmic units) were derived using the task \texttt{EDITMOD}.

 The hardness ratio (HR) was determined as HR=$F_{\rm 2-10\,keV}/F_{\rm 0.3-2\,keV}$ where the
symbols $F_{\rm 2-10\,keV}$ and $F_{\rm 0.3-2\,keV}$ stand for the
de-absorbed 2--10\,keV to 0.3--2\,keV fluxes, respectively.

In order to study the statistical properties of different spectral
parameters, we constructed a histogram and normalized cumulative
distribution for the values of each parameter. A Kolmogorov-Smirnov
(K-S) test was adopted to compare the distributions of the
particular parameter in different time intervals, defined in
Table\,1, and to measure the distance $D_{\rm K-S}$ between the
normalized cumulative distributions of parameters corresponding to
the two different periods, following the recipe provided by
\cite{m11a}.  Since our samples are not statistically complete, a
comparison of the corresponding distributions can be affected by
biases and there is a risk of obtaining a large $D_{\rm K-S}$ value
between the selected cumulative distributions indicating that they
are different, simply due to the lack/absense of the data in some
particular bins of each histogram.  To check the significance of the
results provided by the K-S test, we adopted a method based on the
Monte Carlo simulations (developed by \citealt{m11a}) to account for
this effect and estimate its relevance: firstly, we performed the
K--S test and derived the $D_{\rm K-S}$ quantity for two normalized
cumulative distributions. Afterwards, we randomly simulated two
distributions for both data sets with the same number of components,
adopting two different shapes for the simulated distributions:
log-uniform and lognormal - the former having simply the same
maximum and minimum values of the observed distribution, while the
latter with the same variance, the same median of the observed
distribution and spanning the same range of values. We measured the
$D_{\rm K-S,simul}$ distance between the simulated distributions,
repeated simulations at least 30000 times and built a distribution
of the obtained $D_{\rm K-S,simul}$. Finally, we estimated the
probability of obtaining the observed $D_{\rm K-S}$ randomly, that
provides the confidence level of our K-S test.

\begin{figure*}[ht!]
\includegraphics[trim=6.9cm 1.65cm 0cm 0cm, clip=true, scale=0.78]{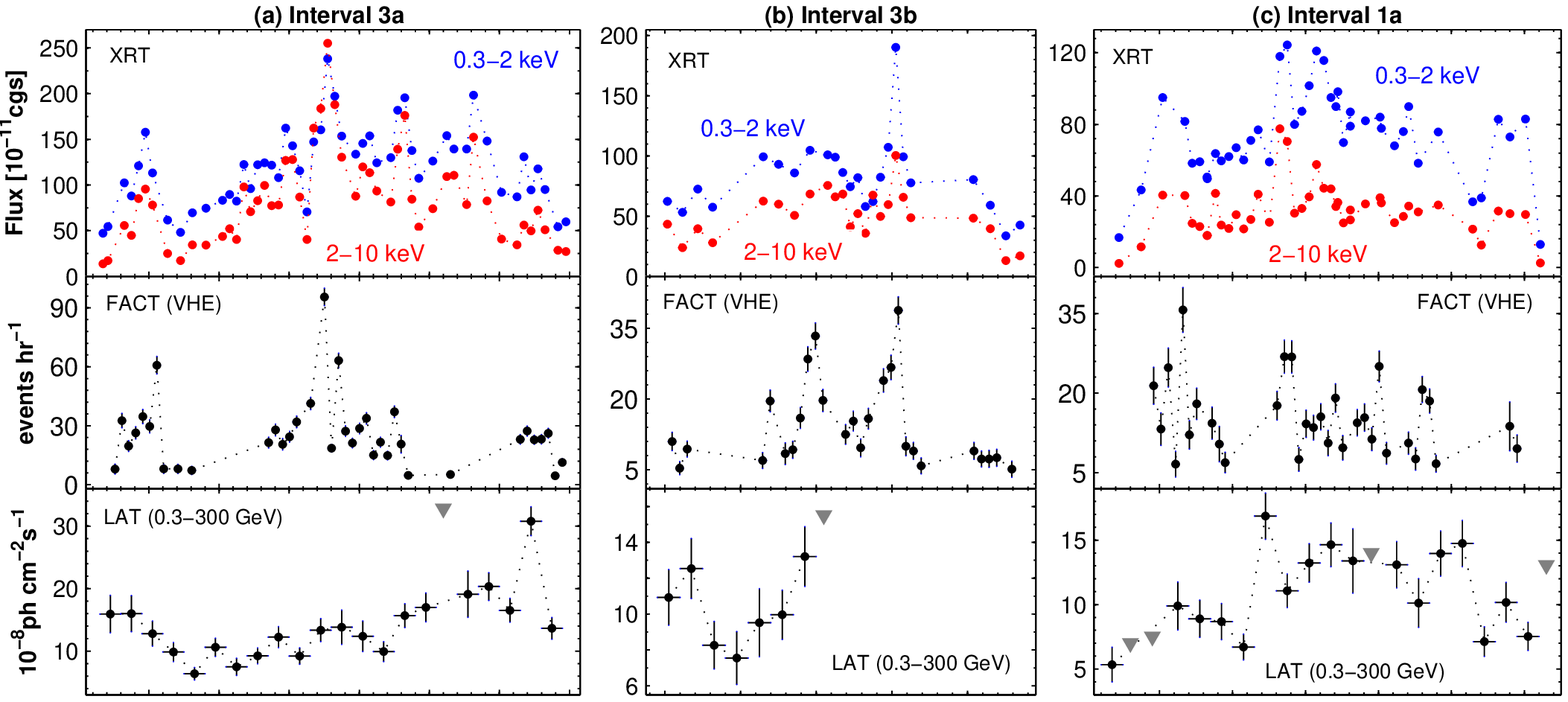}
\includegraphics[trim=6.9cm 4.4cm 0cm 0cm, clip=true, scale=0.78]{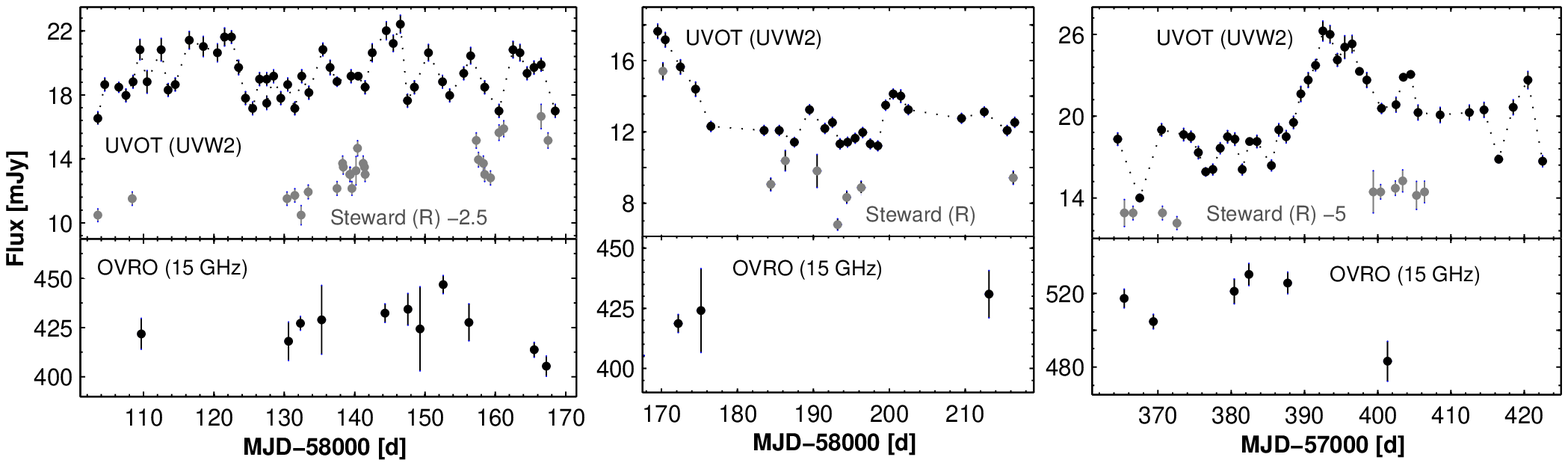}
\includegraphics[trim=6.9cm 1.6cm 0cm 0cm, clip=true, scale=0.78]{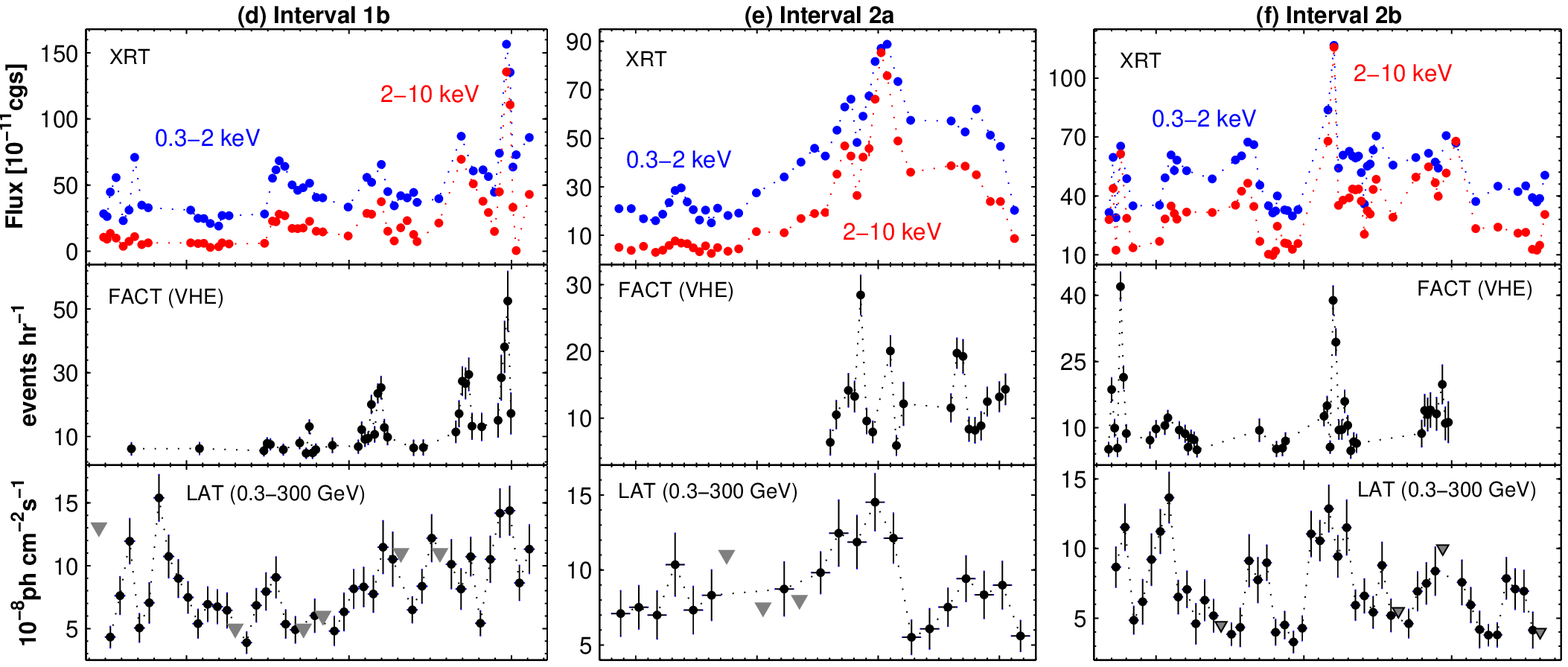}
\includegraphics[trim=6.9cm 4.8cm 0cm 0cm, clip=true, scale=0.78]{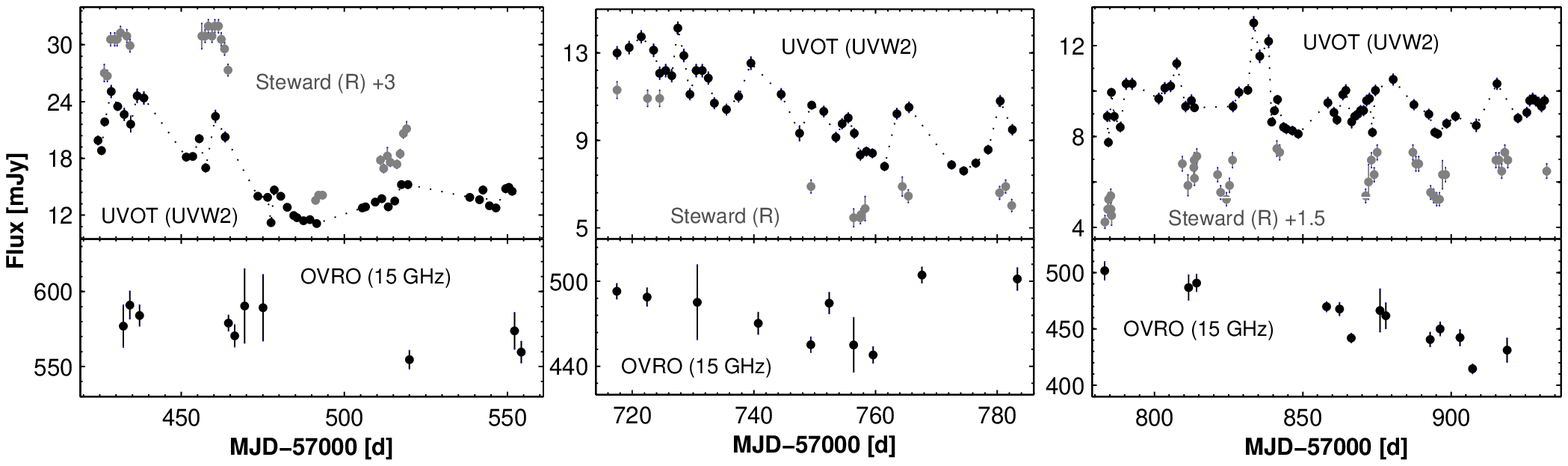}
\vspace{-0.4cm}
 \caption{\label{subper}  The MWL variability of Mrk\,421 in different sub-intervals. The daily bins are used for the XRT,
FACT, UVOT, Steward, OVRO light curves, while 3-d bins are used for
the LAT 0.3-300 GeV data. The triangles in the LAT-related plots
stand for the 2$\sigma$ upper limits to the 0.3--300\,GeV flux when
the source was detected below the 5\,sigma significance and/or
showing $N_{\rm pred}<$10. The acronym \textquotedblleft
cgs\textquotedblright stands for erg\,cm$^{-2}$s$^{-1}$.}
\vspace{0.2cm}
 \end{figure*}

 For each variability instance,  we calculated the fractional variability amplitude and its error according to \cite{v03}

\begin{equation}
\begin{array}{c l}
 F_{\rm var}=(S^2-\overline{\sigma^2_{\rm err}})^{1/2} /{\overline{F}} \\
 err(F_{\rm var})=\Biggl\lbrace \left ( \sqrt{1\over 2N} {\overline{\sigma^2_{\rm err}} \over {\overline{x}}^2 F_{\rm var}} \right )^2+\left ({\sqrt{\overline{\sigma^2_{\rm err}} \over N}} {1 \over \overline {x}} \right
 )^2 \Biggr\rbrace ^{1/2} ,
 \end{array}
\end{equation}
with  $S^2$, the sample variance; $\overline{\sigma^2_{\rm err}}$,
the mean square error; $\overline{F}$, the mean flux.

In order to investigate the possible quasi-periodical behavior of
the flux variations, we first constructed the Lomb-Scragle (LSP)
periodogram \citep{l76,s82}, which is an improved Fourier-based
technique suitable for unevenly-sampled time series $g_n$ without
interpolation for the data gaps \citep{v18}: \vspace{-0.2cm}
\begin{equation}
\begin{array}{c l}
 P(f)={A^2\over 2}\left(\sum\limits_{n}{g_n cos(2\pi f[t_n-\tau]} \right)^2+\\
 {A^2\over 2}\left(\sum\limits_{n}{g_n sin(2\pi f[t_n-\tau]} \right)^2,
 \end{array}
 \vspace{-0.2cm}
\end{equation}
where \emph{A}, \emph{B}, and $\tau$ are arbitrary functions of the
frequency \emph{f} and observing times $\{t_i\}$.  The LSP yields
the most significant spectral power peak, and estimates its
significance level by testing the false alarm probability of the
null hypothesis.

   \begin{table}
 \tabcolsep 3pt
 \vspace{-0.2cm}
    \begin{minipage}{89mm}
  \caption{\label{idvuvottable}  The optical--UV and $\gamma$-ray IDVs during 2015\,December--2018\,April.}
   \vspace{-0.2cm}
     \begin{tabular}{ccccc}
  \hline
 MJD & Band & $\Delta$T(hr) & $\chi^2$/dof &  $100\times F_{\rm var}$  \\
   (1)& (2) & (3)  & (4)  & (5)  \\
\hline
57380.79$-$57381.56 &   \emph{UVW2} &   17.47 &   12.46/1 & 8.5(1.8)\\
57391.56$-$57392.21 &  \emph{UVW1} &   12.26  & 7.77/2 & 10.3(2.3)\\
57425.25$-$57426.23 &  \emph{UVM2} &   22.37 &  23.24/1 &7.2(1.1)\\
57426.18$-$57426.49 &  \emph{UVW2} &   6.38  &  12.38/1 &8.0(1.7)\\
57430.17$-$57430.35  & \emph{UVM2}  &  3.29  &  14.88/1 &4.6(0.9)\\
57477.97$-$57478.21 &  \emph{UVM2}  &  4.61  &  14.67/1 &7.0(1.4)\\
57477.97$-$57478.21  &  \emph{UVW2}   &  4.58   &  49.42/1  &18.5(1.9)\\
57724.44$-$57725.29   &  \emph{UVM2}   &  19.39   & 12.31/1  &5.4(1.2)\\
57726.44$-$57727.29   &  \emph{UVW1 }  &  19.44  &  29.94/1  &19.1(2.5)\\
57726.44$-$57727.29  & \emph{UVW2}  &  19.44 &  27.79/1 &11.0(1.5)\\
57728.49$-$57729.28 &  \emph{UVW1}  &  17.88 &  19.70/1 &14.4(2.4)\\
57728.49$-$57729.28  &\emph{ UVM2 }  & 17.98  & 78.32/1& 14.4(1.2)\\
57728.49$-$57729.28  & \emph{UVW2}  &  17.81 &  24.44/1 &10.6(1.6)\\
57755.22$-$57756.20  & \emph{UVM2} &   22.32  & 22.63/1 &9.0(1.4)\\
57784.17$-$57785.08  & \emph{UVM2 } &  20.66 &  53.39/1& 13.6(1.3)\\
57784.17$-$57785.08 & \emph{ UVW2}  &  20.69 &  16.47/1& 9.5(1.7)\\
57785.03$-$57785.54 &  \emph{UVW1}  &  11.14  & 4.17/7 & 4.1(1.2)\\
57785.03$-$57785.54 &  \emph{UVM2}  &  11.12 &  13.94/7 &6.9(0.7)\\
57785.03$-$57785.54 & \emph{ UVW2}   & 11.11 &  15.95/7& 7.7(0.8)\\
57785.37$-$57786.27  & \emph{UVW1}  &  20.76 &  6.04/3 & 7.1(1.7)\\
57785.37$-$57786.27  &  \emph{UVM2}  &  20.74 &  5.56/1 & 3.5(0.9)\\
57785.37$-$57786.27  &  \emph{UVW2}  &  20.72 &  5.67/1 & 4.4(1.1)\\
57839.93$-$57840.37  & \emph{UVM2} &   9.50 &   10.74/5 &5.3(0.8)\\
57839.93$-$57840.37  & \emph{UVW2}  &  9.53 &   6.48/5&  5.3(0.9)\\
57843.91$-$57844.41 &  \emph{UVM2}  &  9.51 &   21.80/1 &8.3(1.3)\\
57873.87$-$57874.12 &  \emph{UVM2 } &  4.85 &   12.68/1& 6.1(1.3)\\
57873.87$-$57874.12 &  \emph{UVW2 } &  4.86  &  40.44/1 &14.3(1.6)\\
57926.42$-$57927.40 &  \emph{UVM2}  &  22.27 &  14.22/1& 6.191.2)\\
58103.44$-$58104.34 &  \emph{UVM2}  &  20.66 &  28.82/1& 8.9(1.2)\\
58103.44$-$58104.34  & \emph{UVW2}  &  20.64 &  12.50/1 &7.8(1.6)\\
58112.36$-$58113.35 &   \emph{UVM2}  &   8.28 &    38.20/1&  13.3(1.5)\\
58123.91$-$58124.48 &   \emph{UVM2} &    12.65&    26.45/1&  9.3(1.3)\\
58123.91$-$58124.48 &   \emph{UVW2}  &   12.64 &   12.46/1&  7.7(1.6)\\
58137.05$-$58137.55 &  \emph{UVW1}  &  11.05  & 13.41/1& 11.7(2.4)\\
58146.68$-$58147.32 &  \emph{UVW1 }&   14.50 &  29.12/1& 18.2(2.4)\\
58146.68$-$58147.32 &  \emph{UVW2}  &  14.49 &  46.25/1 &16.2(1.7)\\
58147.217$-$58148.12 &  \emph{UVM2}  &  19.03 &  14.88/1 &6.2(1.2)\\
58199.81$-$58200.06  & \emph{UVW1} &   4.73 &   19.56/1 &13.5(2.2)\\
58139.60$-$58140.40 &  \emph{V} &  19.20  & 14.19/2 &7.2(1.7)\\
58139.60$-$58140.40 &  \emph{R} &  19.21 &  7.59/2 & 7.2(2.2)\\
58158.00$-$58158.98&0.3$-$300 GeV &23.52 &  11.62/1& 57.6(12.8)\\
58135.059$-$58136.058& VHE & 23.58&2.82/16& 23.4(5.4)\\
58135.236$-$56136.073& VHE &20.09 &  4.44/5 & 41.1(10.1)\\
58136.250$-$58137.141& VHE &21.38 &  4.22/9 & 32.3(7.8)\\
 \hline
\end{tabular}
\end{minipage}
\end{table}

\cite{f96} introduced the weighted wavelet Z-transform (WWZ) method,
which is a periodicity analysis technique in both the time and
frequency domains. Note that WWZ is suited for discovering
variability timescales and is robust against missing data. It is
defined as follows \vspace{-0.1cm}
\begin{equation}
WWZ={(N_{\rm eff}-3)V_y\over{2(V_{\rm x}-V_{\rm y})}},
\vspace{-0.1cm}
\end{equation}
with $N_{\rm eff}$, the so-called effective number of data points;
$V_{\rm x}$ and $V_{\rm y}$, the weighted variation of the data
$x(t)$ and model function $y(t)$, respectively. WWZ is based on the
Morlet wavelet \citep{g84} $ f(z)=e^{-cz^2}(e^{iz}-e^{-1/4c})$ where
the constant $e^{-1/4c}$ is inserted so that the wavelet's mean
value is zero.

Throughout the paper, the errors are quoted at the 90\% confidence
level for the one parameter of interest, unless otherwise stated.

\begin{figure*}[ht!]
\vspace{-0.1cm}
  \includegraphics[trim=7.2cm 3.1cm 0cm 0cm, clip=true, scale=0.81]{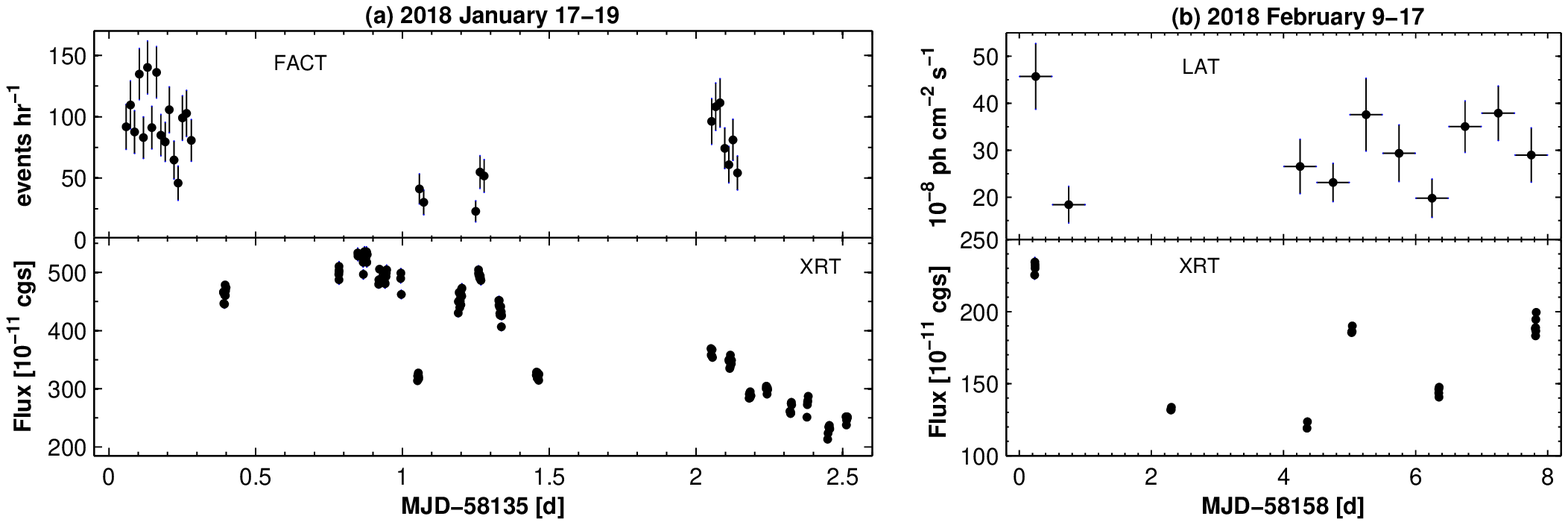}
\includegraphics[trim=7.2cm 3cm 0cm 0cm, clip=true, scale=0.81]{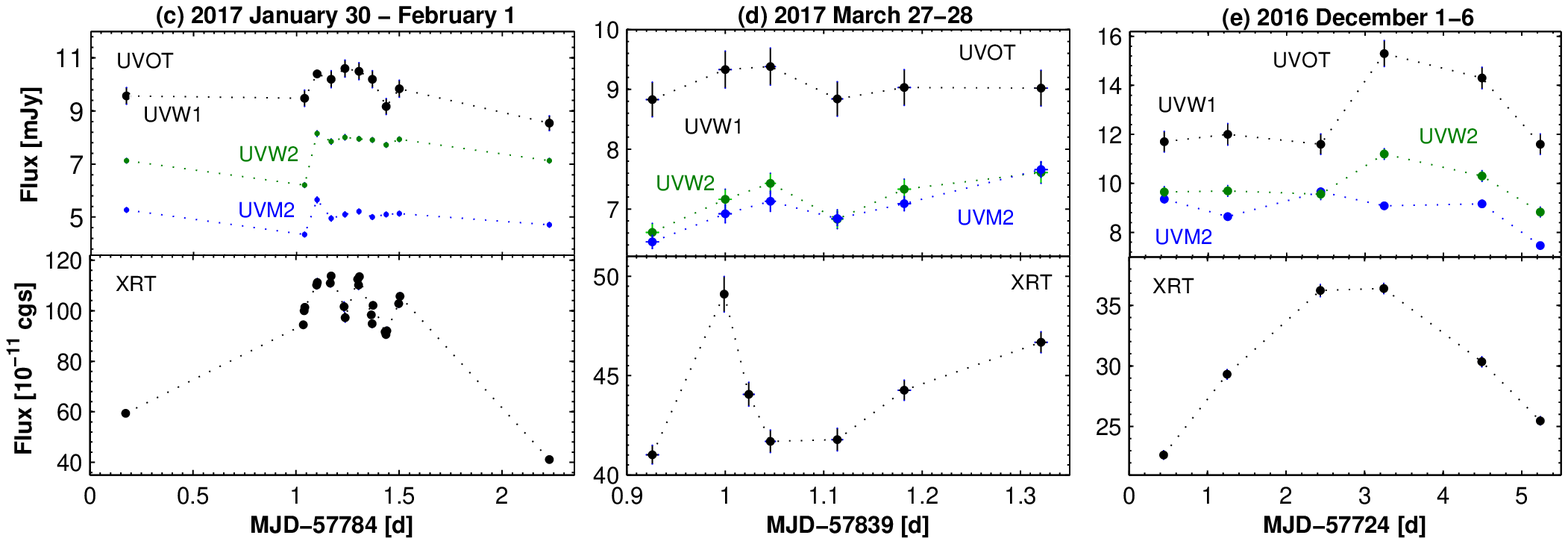}
\vspace{-0.6cm}
 \caption{\label{idvvhe} Intraday variability (IDV) of Mrk\,421 in the FACT (panel\,(a)), LAT (panel\,(b)) and UVOT (panels\,(c)--(e)) bands. }
 \vspace{0.2cm}
 \end{figure*}

\vspace{0.3cm}
\section{RESULTS}
\subsection{X-Ray Variability}
Table\,\ref{xrt} provides a summary of  the \emph{Swift}-XRT and
UVOT observations in 2015\,December\,8--2018\,April\,8. The source
was targeted 299-times, with the net exposure time (sum of the Good
Time Intervals, GTIs) of 314.5\,ks. Based on the target's
\textquotedblleft visibility\textquotedblright ~for \emph{Swift},
the observations were performed during 2015\,December--2016\,June,
2016\,December--2017\,June and 2017\,December--2018\,April, denoted
as Interval\,1, Interval\,2 and Interval\,3, respectively. Each
interval was split into two sub-periods (e.g., intervals 1a and 1b),
according to the flaring activity of Mrk\,421 in the
\emph{Swift}-XRT band.

The source was highly variable during all Intervals\,1--3 (see
Table\,\ref{persum} for the corresponding $F_{\rm var}$ and
maximum-to-minimum flux ratios in different bands), although it
showed significantly stronger X-ray flaring activity in Interval\,3
compared to the previous ones (Figure\,\ref{flares}a): the mean
0.3--10\,keV count rate was a factor $\sim$2-2.5 higher; during the
eleven XRT pointings, the brightness exceeded a level of
100\,cts\,s$^{-1}$ (see Table\,2), which never was recorded in
Intervals\,1--2; a generally higher state was superimposed by fast,
strong flares by a factor of 2.2--3.8 (lasting 4--12\,d) which were
considerably fewer in the previous periods.

In Interval\,3, the most extreme flare occurred during
2018\,January\,14--30 with a 2-min binned count rate of
180.9$\pm$2.8\,cts\,s$^{-1}$ (to the level observed on
2008\,June\,12; see Figure\,\ref{flares}b) and the highest levels
were recorded only during the giant X-ray outburst in 2013 April
10--17 \citep{k16,k18a}. The source underwent fast brightness
variations by 30\%--45\% within 13--35\,ks during the declining
phase of the flare (Figure\,\ref{idvext1}, top panel). Note that the
flux doubling time ($\tau_{\rm d}$$\approx$92\,ks, defined as
$\tau_{\rm d}$=$\Delta t \times ln(2)/ln(F_2/F_1)$; \citealt{sa13})
was faster than the halving one ($\tau_{\rm d}$$\approx$113\,ks),
and a similar situation was seen during the next flare when an
intraday flux doubling with $\tau_{\rm d}\approx$64\,ks was recorded
(Figure\,\ref{flares}b).

These events were preceded and followed by strong 0.3--10\,keV
flares by a factor of $\sim$2.5 and peak fluxes exceeding a level of
100\,cts\,s$^{-1}$ (Figures\,\ref{flares}c--\ref{flares}c). They
were parts of a well-defined long-term flare lasting more than
2\,months (MJD\,58103--58170). Other strong XRT-band flares are
presented in Figures\,\ref{flares}f--\ref{flares}i, with the
intraday flux-doubling/halving events within 4.8--18.9\,hr. Among
these flares, the 2015\,December\,29--2016\,January\,8 event was
characterized by a two-peak maximum, possibly related to the
propagation of forward and reverse shocks after the collision of two
\textquotedblleft blobs\textquotedblright ~ in the blazar jet (see
\citealt{b10}).

\begin{table*}
\vspace{-0.3cm} \tabcolsep 3.5pt \centering
\begin{minipage}{180mm}
\caption{\label{lp} The results of the \emph{Swift}-XRT spectral
analysis with the log-parabolic model (extract; see the
corresponding machine-readable table for the full version).
Col.\,(1) gives the ObsID of the particular observation, along with
the duration of the separate segments (in seconds) used to extract a
spectrum  in parentheses. The abbreviations \textquotedblleft
Or\textquotedblright~ and  \textquotedblleft S\textquotedblright
~stand for \textquotedblleft orbit\textquotedblright~ and
\textquotedblleft segment\textquotedblright, respectively;
Cols\,(2)--(3) present the values of the photon index at 1\,keV and
the curvature parameter, respectively.  The $E_{\rm p}$ values
(Col.\,(4)) are given in keV; Cols 5 and 6 present the norm and the
reduced Chi-squared, respectively; de-absorbed 0.3--2\,keV,
2--10\,keV and 0.3--10\,keV fluxes (Col.\,(7)--(9)) - in
erg\,cm$^{-2}$s$^{-1}$. The hardness ratio is presented in the last
column.} \vspace{-0.1cm}
  \begin{tabular}{cccccccccc}
    \hline
ObsId & $a$ & $b$ &$E_{\rm p}$  & 10$\times K$ & $\chi^2/d.o.f.$ & $F_{\rm 0.3-2\,keV}$ & $ F_{\rm 2-10\,keV}$ & $ F_{\rm 0.3-10\,keV}$ & HR \\
(1) & (2) & (3) & (4) & (5) & (6) & (7) & (8) & (9) & (10) \\
\hline
35014240  &  2.92(0.02)&  0.27(0.08)&  0.02(0.01) &0.41(0.01)&1.04/145  &  16.71(0.27)& 2.34(0.14) & 19.05(0.35)& 0.14(0.01)\\
 35014241 &  2.49(0.01) & 0.48(0.04)  &0.31(0.05) & 1.31(0.01) & 1.10/244  &  43.35(0.40)& 11.59(0.32)& 54.95(0.50)& 0.27(0.01)\\
 35014242\,S1\,(267\,s)& 2.22(0.02) & 0.38(0.04) & 0.51(0.06)&  3.12(0.04) & 1.07/230  &  94.62(1.08)& 40.46(1.19)& 135.21(1.55)& 0.43(0.01)\\
 35014242\,S2\,(267\,s) &2.24(0.02) & 0.40(0.04)&  0.50(0.06) & 3.19(0.04) & 1.00/236   & 97.05(1.11) &39.90(1.18)& 136.77(1.57)& 0.41(0.01)\\
 \hline
\end{tabular}
\end{minipage}
\end{table*}

The source also underwent extremely fast instances of the
0.3--10\,keV intraday variability (IDV: brightness change within a
day, detected by means of the chi-squared statistics; see
Table\,\ref{idvtable} for details) at the 99.9\% confidence level.
Namely, the brightness showed a rise by 5\%--18\% (taking into
account the associated measurement errors) in 180--600 seconds in
Figures\,\ref{idv1ks}A1 (panels 1--2),
\ref{idv1ks}C2--\ref{idv1ks}C4, \ref{idv1ks}D3, \ref{idv1ks}E2
 and \ref{idv1ks}E3--\ref{idv1ks}E4 .
Figs \ref{idv1ks}A--\ref{idv1ks}C belong to the epoch of the
strongest X-ray flaring activity in 2018\,January\,14--30. Note that
the states with CR$\gtrsim$100\,cts\,s$^{-1}$ sometimes were
associated with very fast and large drops just at the start of the
XRT orbit, which were related to instrumental effects. The source
also showed very fast drops by 6--16 per cent within 180--840
seconds (Figures\,\ref{idv1ks}B1, \ref{idv1ks}C2,
\ref{idv1ks}D1--\ref{idv1ks}D3, \ref{idv1ks}F1--\ref{idv1ks}F3).
Finally, entire brightness rising and dropping cycles with 6\%--13\%
 within 420-960 seconds were also observed
(Figures\,\ref{idv1ks}A2, \ref{idv1ks}B2, \ref{idv1ks}C1,
\ref{idv1ks}E1). On the contrary, Mrk\,421 sometimes showed a slow,
low-amplitude variability during some densely-sampled XRT
observations in lower X-ray states (see Figure\,\ref{idvext2}).
Moreover, tens of 0.3--10\,keV IDVs were detected by us, whose
details are reported in Table\,\ref{idvtable}.

MAXI detected the source  78 and 61 times with 5$\sigma$
significance  in the 2--20\,keV and 2--6\,keV energy ranges,
respectively (Figure\,\ref{maxi}a--b). Owing to lower instrumental
capabilities, a strong, long-term flare is evident only in
Interval\,3. Significantly  fewer detections with 5$\sigma$
significance are found from the daily-binned BAT data, which show a
strong flare during the highest XRT and MAXI-band states
(Figure\,\ref{maxi}c).

\begin{table}
\tabcolsep 4pt
 \centering
  \begin{minipage}{85mm}
  \caption{\label{distrtable} Distribution of spectral parameters in different periods: minimum and maximum values
(Cols (2) and (3), respectively), mean value (Col.\,(4)) and
skewness (last column). }
 \vspace{-0.1cm}
   \begin{tabular}{cccccc}
  \hline
Par. & Min. & Max.  & Mean  &Skewness \\
(1) & (2) & (3) & (4) & (5)  \\
 \hline
 &  &  2015--2018  & & \\
\hline
$b$ & 0.07(0.04) & 0.48(0.04) & 0.20(0.01)  &0.66  \\
$a$ & 1.63(0.02) & 2.92(0.02) & 2.12(0.01)&0.58  \\
$\Gamma$ & 1.79(0.02) & 2.91(0.02) & 2.12(0.02)&0.98 \\
$HR$ & 0.14(0.01) & 1.23(0.03) &  0.64(0.01)&0.28\\
$E_{\rm p}$ & 0.50(0.06) & 7.50(0.76) &  1.85(0.09) &1.65 \\
\hline
&  &  Int1  & & \\
\hline
$b$ & 0.07(0.04) & 0.48(0.04) & 0.25(0.01) &0.34  \\
$a$ & 1.77(0.02) & 2.92(0.02) & 2.29(0.01)&0.31  \\
$\Gamma$ & 1.99(0.02) & 2.91(0.02) & 2.12(0.04) &1.37 \\
$HR$ & 0.14(0.01) & 0.89(0.03)&0.45(0.01) &0.91\\
$E_{\rm p}$ & 0.50(0.06)  & 2.03(0.14)&  1.02(0.18)&0.94 \\
\hline
&  &  Int1a  & & \\
\hline
$b$ & 0.09(0.04) & 0.48(0.04) & 0.25(0.01)&0.61  \\
$a$ & 1.98(0.02) & 2.92(0.02) & 2.30(0.01) &-0.37 \\
$HR$ & 0.14(0.01) & 0.68(0.02) &0.42(0.01)&0.98\\
\hline
&  &  Int1b  & & \\
\hline
$b$ & 0.08(0.04) & 0.42(0.05) & 0.25(0.01) &0.25  \\
$a$ & 1.77(0.02) & 2.91(0.02) & 2.26(0.01)  &0.41  \\
$HR$ & 0.14(0.01) & 0.89(0.03) &0.49(0.01) &0.33\\
\hline
&  &  Int2  & & \\
\hline
$b$ & 0.07(0.04) & 0.42(0.05) & 0.20(0.01)&0.58  \\
$a$ & 1.74(0.02) & 2.82(0.02) & 2.12(0.01) &1.02  \\
$\Gamma$ & 1.92(0.02) & 2.54(0.02) & 2.13(0.03)&1.22 \\
$HR$ & 0.17(0.01) & 1.09(0.03) &  0.63(0.01) &-0.20\\
$E_{\rm p}$ & 0.51(0.06)  & 6.31(0.65) &1.46(0.22)&2.17 \\
\hline
&  &  Int2a  & & \\
\hline
$b$ & 0.10(0.04) & 0.40(0.04) & 0.22(0.01)&0.48  \\
$a$ & 1.74(0.02) & 2.82(0.02) & 2.14(0.01)  &0.84  \\
$HR$ & 0.17(0.01) & 1.09(0.03) &  0.61(0.01) &-0.23\\
\hline
&  &  Int2b  & & \\
\hline
$b$ & 0.07(0.04) & 0.40(0.04) & 0.18(0.01)&0.12  \\
$a$ & 1.80(0.02) & 2.50(0.02) & 2.10(0.01) &0.47  \\
$HR$ & 0.29(0.01) & 1.06(0.04) &  0.65(0.01) &0.14\\
\hline
&  &  Int3  & & \\
\hline
$b$ & 0.07(0.04) & 0.36(0.04) & 0.18(0.01)&0.53  \\
$a$ & 1.63(0.02) & 2.64(0.01) & 2.01(0.01)&0.43  \\
$\Gamma$ & 1.79(0.02) & 2.60(0.02) & 2.08(0.02)&0.05\\
$HR$ & 0.28(0.01) & 1.23(0.03) &  0.75(0.01)&0.19\\
$E_{\rm p}$ & 0.50 & 7.50(0.76) &  2.16(0.17)&1.30 \\
\hline
&  &  Int3a  & & \\
\hline
$b$ & 0.07(0.04) & 0.30(0.04) & 0.17(0.01)&0.34  \\
$a$ & 1.63(0.02) & 2.52(0.02) & 1.98(0.01) &0.54  \\
$HR$ & 0.28(0.01) & 1.23(0.03) &  0.79(0.01)&-0.04\\
\hline
&  &  Int3b  & & \\
\hline
$b$ & 0.08(0.04) & 0.36(0.04) & 0.21(0.01)&0.16  \\
$a$ & 1.95(0.02) & 2.39(0.02) & 2.11(0.01)&0.78  \\
$HR$ & 0.36(0.02) & 0.80(0.03) &  0.60(0.01)&-0.19\\
\hline
\end{tabular}
\end{minipage}
\end{table}

\subsection{Multiwavelength Variability on Various Timescales}
Similar to the XRT-band, the strongest FACT VHE $\gamma$-ray flaring
activity of Mrk\,421 occurred in Period\,3a (Figure\,\ref{subper}a).
Namely, two strong flares peaking at MJD\,58111 and MJD\,58135, as
well as lower-amplitude ones with the peaks at MJD\,58141, 58145 and
58164 accompanied the X-ray counterparts. The UVOT-band behaviour
was also predominantly correlated with the 0.3--10\,keV flaring
activity, while the source demonstrated different timing properties
in the LAT-band, showing slower variability and  only one peak at
the end of this sub-interval, coinciding with the X-ray and VHE
peaks.

The three subsequent peaks in the FACT-band light curve were evident
during the strongest 0.3--10\,keV activity of the source in
Interval\,3b (Figure\,\ref{subper}b). No LAT-band GTIs were obtained
during the second half of this period and a correlated behaviour
with X-ray variability was not observed in the first half. The
optical--UV light curves showed a decline at the beginning of the
sub-interval, followed by the weak variability which was not
strongly correlated with the VHE--X-ray activity. However, a
stronger correlation was observed during Period\,1a, in the three
consecutive peaks having X-ray and VHE counterparts
(Figure\,\ref{subper}c). Similarly, the LAT-band behaviour was more
correlated with those in the XRT and FACT bands, compared to that
shown in Interval\,3.

The majority of Interval\,1b also was characterized by a correlated
X-ray and VHE flaring activity (Figure\,\ref{subper}d). The peak
days of the X-ray--VHE fluxes also coincided with enhanced activity
of Mrk\,421 in the LAT-band. However, the source was not targeted
with \emph{Swift} and FACT during the strongest GeV-band flare
peaking on MJD\,57441. Initially, the source was flaring at
UV-optical frequencies (correlated with the higher-energy activity),
while it showed a declining trend and low states afterwards, during
the four consecutive keV--TeV flares. A similar situation was
observed in Interval\,2a  when the source exhibited enhanced VHE and
LAT-band activities during the long-term X-ray flare (see
Figure\,\ref{subper}e). Finally, Mrk\,421 underwent strong VHE
flares along with the X-ray ones in Interval\,2b, while the latter
showed fewer correlation with the LAT and UVOT-band fluctuations
(Figure\,\ref{subper}f). The source did not show flares at the radio
frequencies during any sub-interval(see the bottom panels of
Figure\,\ref{subper}).

Although Mrk\,421 frequently showed 8--16 detections a night with
3$\sigma$ significance \textbf{in} the 20-min binned FACT data
(particularly, in the time interval 2018\,January--February), only 3
instances of a VHE IDV at the 99.9\% confidence level were
detected\footnote{Note that the FACT results  are still provided in
the form of the excess rates, and this result should be considered
with caution.}, belonging to the epoch of the strongest X-ray
flaring activity in the period presented here (see
Figure\,\ref{idvvhe}a and Table\,\ref{idvuvottable}). While no
correlated X-ray--VHE variability was evident during
MJD\,58136.1--58136.3, the subsequent VHE data exhibit a brightness
decline similar to the XRT ones.  In the same period, the source
showed one instance of a LAT-band IDV with a brightness decline by
42\% (taking into account the associated errors) and accompanied by
similar behaviour in X-rays (Figure\,\ref{idvvhe}b). Finally, three
ultraviolet IDVs, showing a brightness increase nearly
simultaneously with that in the 0.3--10\,keV energy range, are
presented in Figures\,\ref{idvvhe}c--\ref{idvvhe}d. The details of
another 35 optical--UV IDVs are provided in
Table\,\ref{idvuvottable}.

\subsection{Spectral Variability}
\subsubsection{Curvature Parameter}
Similar to the period 2005\,March--2015\,June
\citep{k16,k17a,k18a,k18b}, a vast majority of the 0.3--10\,keV
spectra of Mrk\,421 (886 out of 980) show a significant curvature
and  are well fitted with the log-parabolic model. The corresponding
results are presented in Table\,\ref{lp}. The distribution of values
of the curvature parameter $b$ (corresponding to the curvature
detection significance of 3$\sigma$ and higher) for different
periods are provided in Figures\,\ref{figdistr}A1--\ref{figdistr}A4
and the corresponding properties (minimum, maximum and mean values,
distribution skewness) are listed in Table\,\ref{distrtable}.

Figure\,\ref{figdistr}A1 demonstrates that the source was
characterized by a relatively low curvature in the period
2015\,December--2018\,April: 98.1\%  of the values of the parameter
$b$ were smaller than $b$=0.4 (conventional threshold between the
lower and higher curvatures). Moreover, 46.2\% of the spectra shows
$b$$<$0.2 (see Section\,4.3 for the corresponding physical
implication). The lowest curvatures were observed in Interval\,3:
there were no spectra with $b$$>$0.4; 62.3\% of the values were
lower than 0.2; the mean value $\overline{b}$=0.18$\pm$0.01 is
significantly smaller than that recorded in Interval\,1 (see
Table\,\ref{distrtable}). The latter was characterized by the
majority of the spectra showing curvatures with $b$$>$0.4 (5 out of
9, including the highest value for the entire 2.3-yr period
presented here). Interval\,2 was different from both cases and
characterized by \textquotedblleft intermediate\textquotedblright~
properties of the parameter $b$ between Intervals\,1 and 3. This
situation is clearly evident from the cumulative distributions of
the curvature parameter presented in Figure\,\ref{figdistr}A3 and
Table\,\ref{ks}, providing the results of the K-S test and distances
between the corresponding distributions.

According to Figure\,\ref{figdistr}A3 and Table\,\ref{ks}, the
distribution of the parameter $b$ shows differences between the
different parts of Intervals\,2--3. The lowest curvatures are found
for Interval\,3a ($\overline{b}$=0.17$\pm$0.01, 66\% of the spectra
with $b$$<$0.2, and $b_{\rm max}$0.30$\pm$0.04), while Intervals 1a
and 1b are not significantly different from each other.

The curvature parameter showed a weak positive correlation with the
photon index at 1\,keV and an anti-correlation with the position of
the synchrotron SED peak, as observed in different sub-intervals
(see Figures\,\ref{figcor}a--\ref{figcor}b, as well as
Table\,\ref{cortable} for the corresponding values of the Spearman
correlation coefficient $\rho$). Moreover, this parameter showed an
anti-correlation with the de-absorbed 0.3--10\,keV flux in
Intervals\,1--2(Figure\,\ref{figcor}c).

 The sixth column of Table\,\ref{idvtable} demonstrates  that the
parameter $b$ was variable 36-times at the 3$\sigma$ confidence
level during 0.3--10\,keV IDVs, showing $\Delta
b$=0.14(0.05)--0.31(0.06) within 0.38--23.92\,hr (Figures
\ref{idvext1}--\ref{idvext2}). The fastest instances incorporated a
curvature increase by 0.19(0.06)--0.22(0.06 in $\sim$1\,ks
(Figure\,\ref{idvext1}, at MJD\,58135.93; Figure\,\ref{idvext2}, at
MJD\,57375.4). Moreover, the curvature parameter varied with
2$\sigma$ and 1$\sigma$ significances (capable of causing a
significant flux change) 43 and 22 times, respectively.

\begin{figure*}
  \includegraphics[trim=7.3cm 0.6cm 0cm 0cm, clip=true, scale=0.83]{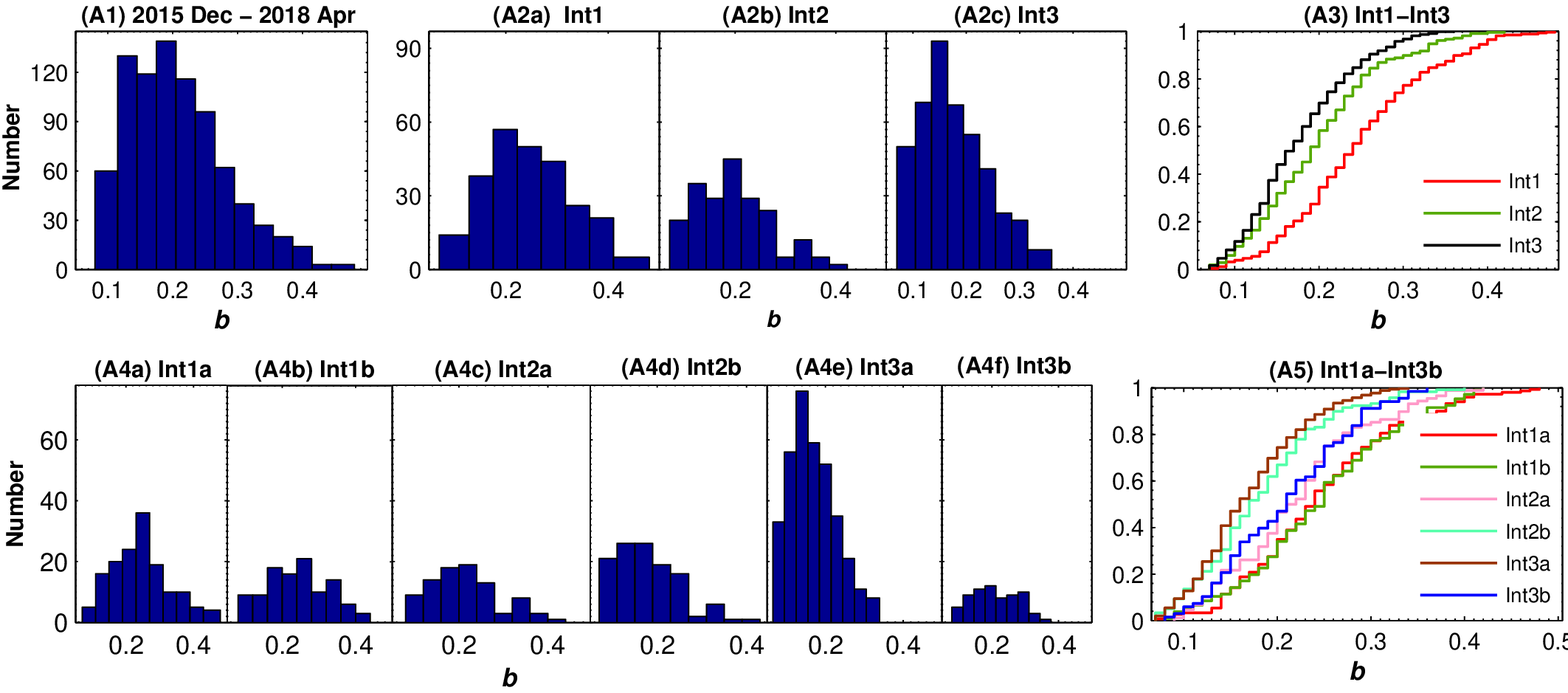}
   \includegraphics[trim=7.3cm 0.3cm 0cm 0cm, clip=true, scale=0.83]{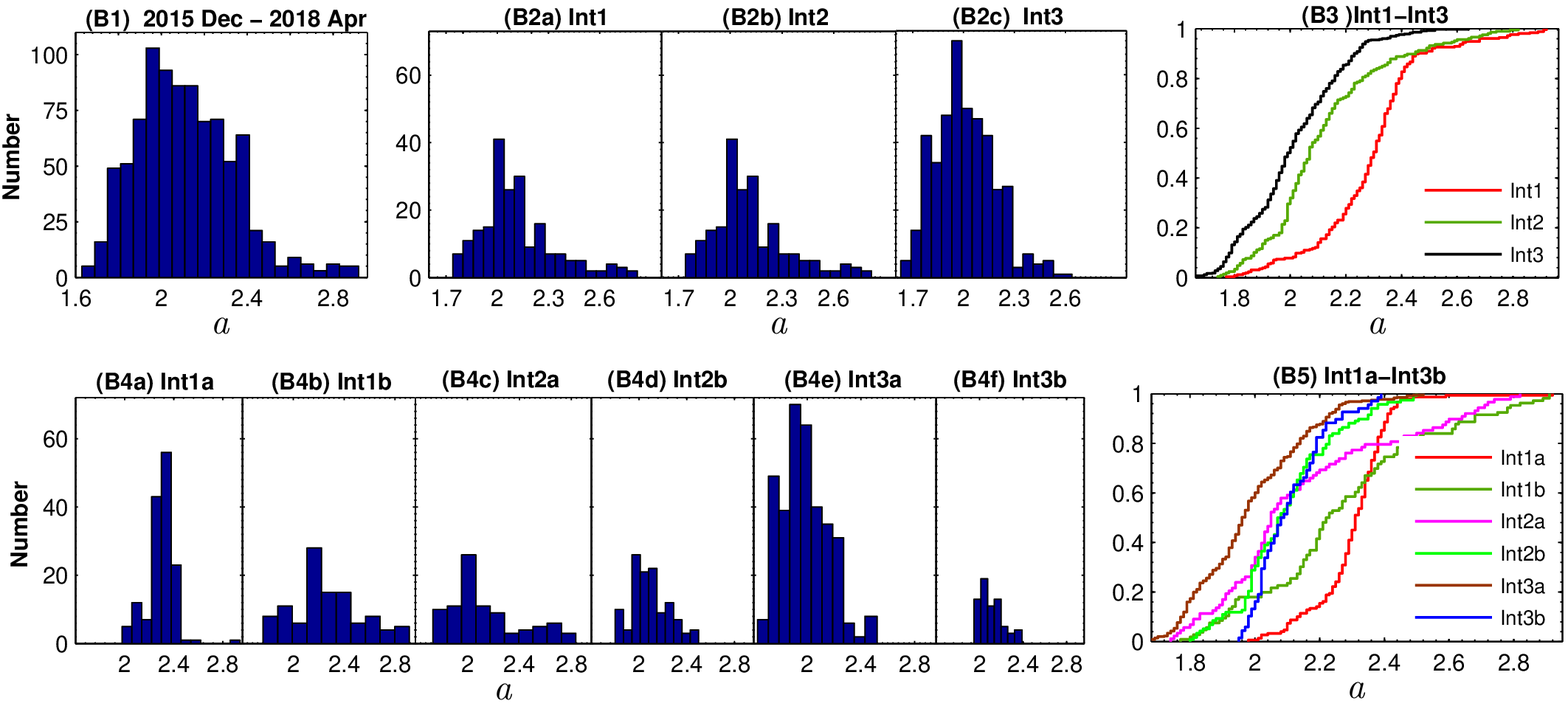}
 \includegraphics[trim=7.3cm 5.4cm 0cm 0cm, clip=true, scale=0.83]{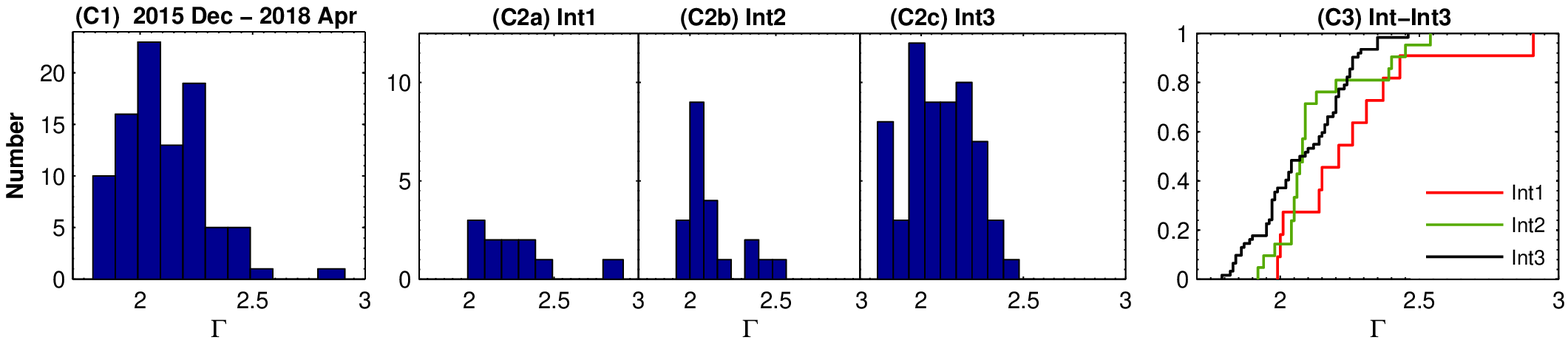}
 \includegraphics[trim=7.3cm 5.9cm 0cm 0cm, clip=true, scale=0.83]{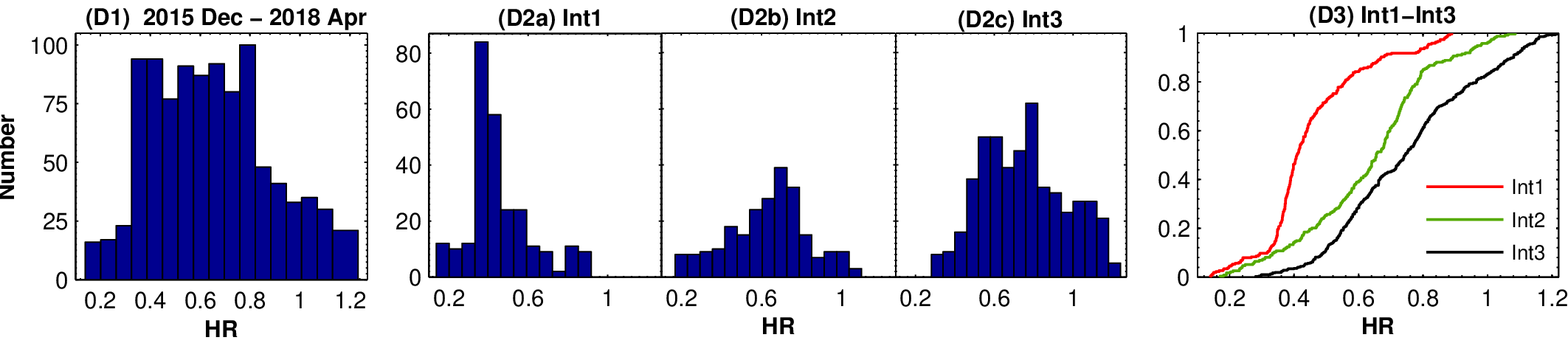}
\vspace{-0.4cm}
 \caption{\label{figdistr} Distribution of the spectral parameter values in different periods: histograms and the corresponding normalized
 cumulative distributions (the last plot in each row). }
  \end{figure*}

 \addtocounter{figure}{-1}
  \begin{figure*}
  \includegraphics[trim=7.3cm 5.8cm 0cm 0cm, clip=true, scale=0.80]{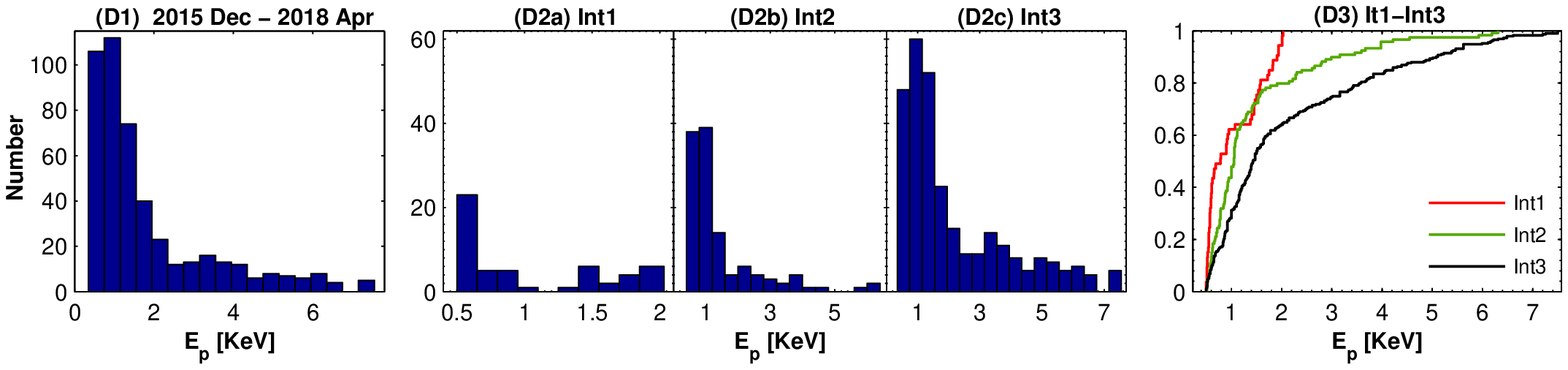}
 \vspace{-0.7cm}
 \caption{\label{figdistr} - Continued. }
 \vspace{-0.1cm}
 \end{figure*}

 \begin{figure*}[ht]
 \includegraphics[trim=7.3cm 5.3cm 0cm 0cm, clip=true, scale=0.82]{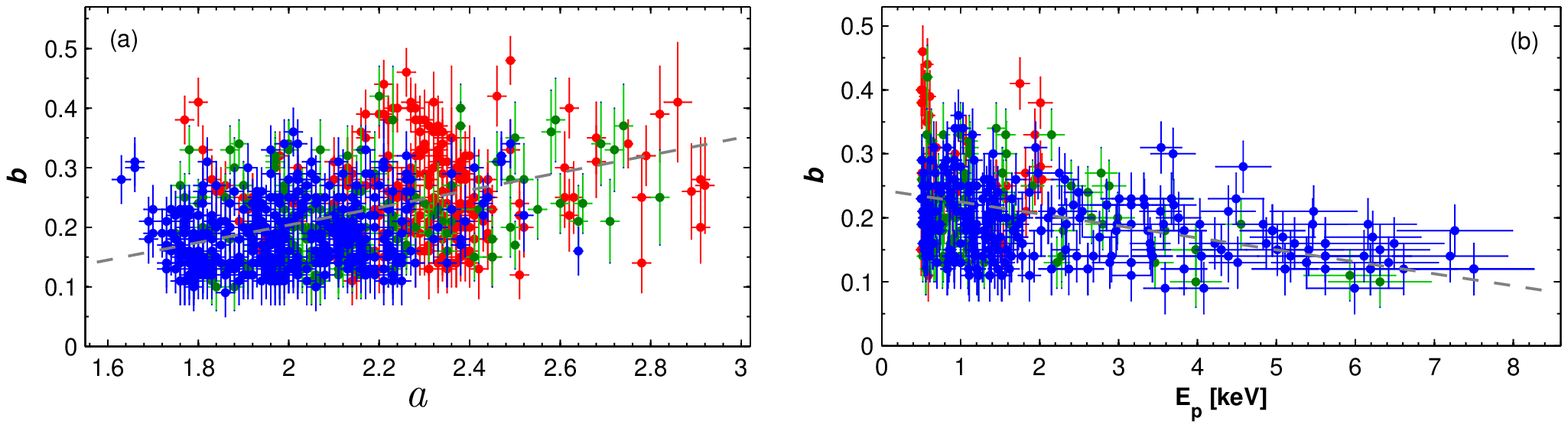}
\includegraphics[trim=7.3cm 5.1cm 0cm 0cm, clip=true, scale=0.82]{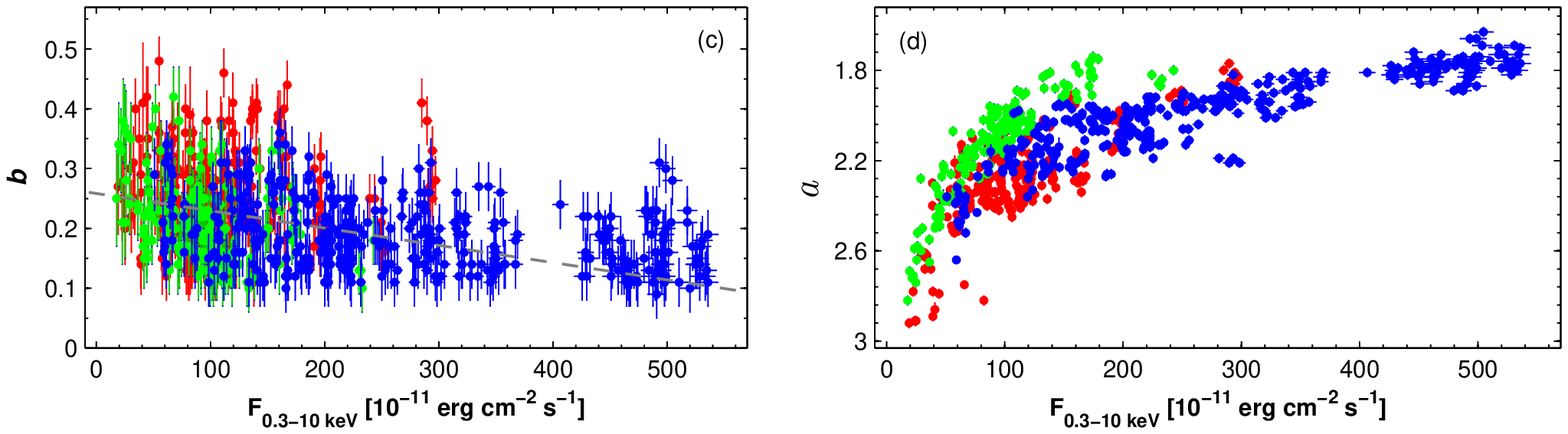}
\includegraphics[trim=7.3cm 5.1cm 0cm 0cm, clip=true, scale=0.82]{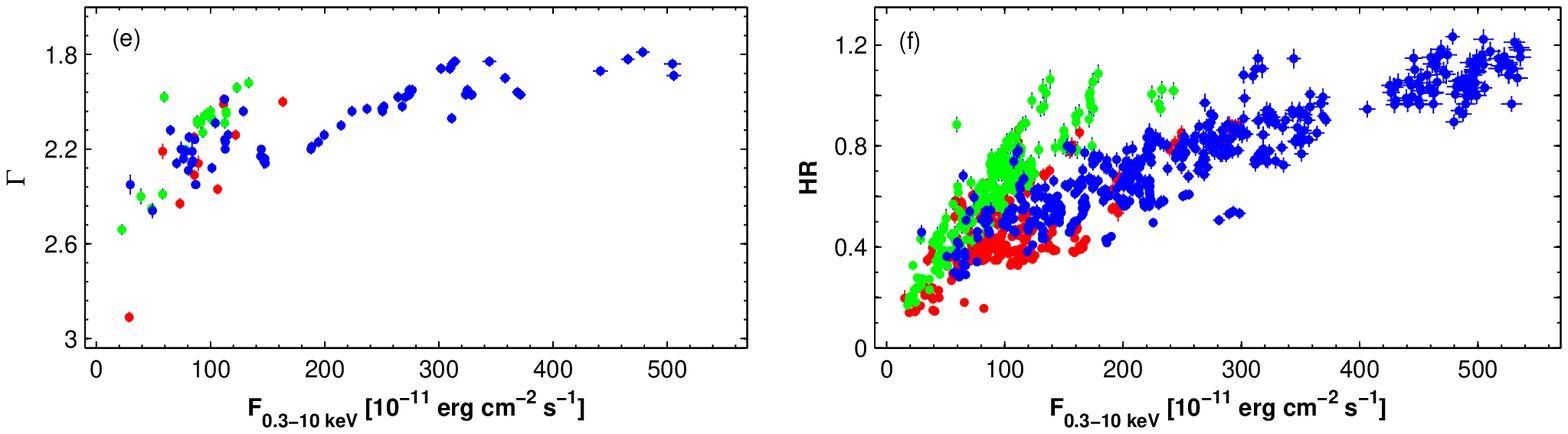}
\includegraphics[trim=7.3cm 5.2cm 0cm 0cm, clip=true, scale=0.82]{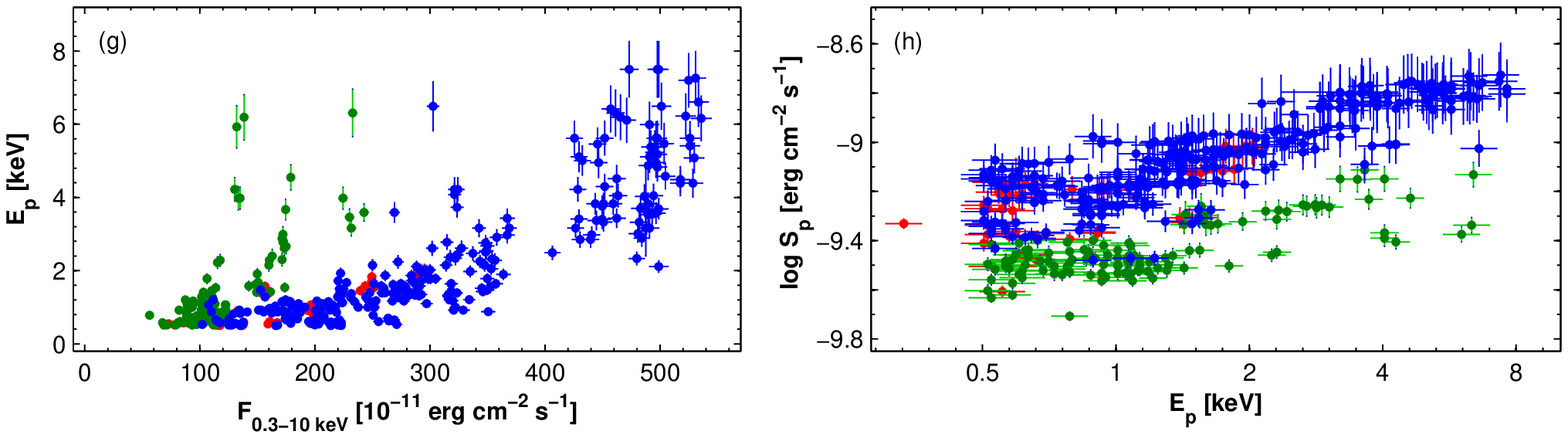}
\vspace{-0.6cm}
 \caption{\label{figcor} Correlation between the spectral parameters and de-absorbed 0.3--10\,keV
 flux. Panels\,(a)--(3): the curvature parameter \emph{b} plotted versus
 the photon index \emph{a}, the position of the synchrotron SED peak $E_{\rm
p}$ and de-absorbed 0.3--10\,keV flux, respectively. The parameters
\emph{a}, $\Gamma$, HR and $E_{\rm p}$ are plotted versus the
de-absorbed 0.3--10\,keV flux in Panels\,(d)--(g). The scatter plot
$E_{\rm p}$--$\log S_{\rm p}$ is provided in the last plot. The
colored points correspond to the different intervals as follows: red
-- Interval\,1; green -- Interval\,2; blue --
 Interval\,3. Gray dashed lines represent linear fits to the scatter plots.}
 \end{figure*}

 \subsubsection{Photon Index and Hardness Ratio}
Although the photon index at 1\,keV showed a very wide range of
values ($\Delta a$=1.29) with the hardest spectrum yielding
$a$=1.63$\pm$0.02 in the period presented here, this range is
narrower compared to that recorded in the time interval 2009--2012
($\Delta a$=1.51). In the latter, the source showed the hardest
($a$=1.51--1.61) and the softest spectra ($a$=2.93--3.02) since the
start of the \emph{Swift }operations. On the other hand, narrower
ranges were found in the time intervals 2005--2008, 2013 April and
2013\,November--2015\,June (1.01, 0.87 and 0.94, respectively).

\begin{figure*}[ht]
\includegraphics[trim=7.3cm 1.0cm -1cm 0cm, clip=true, scale=0.83]{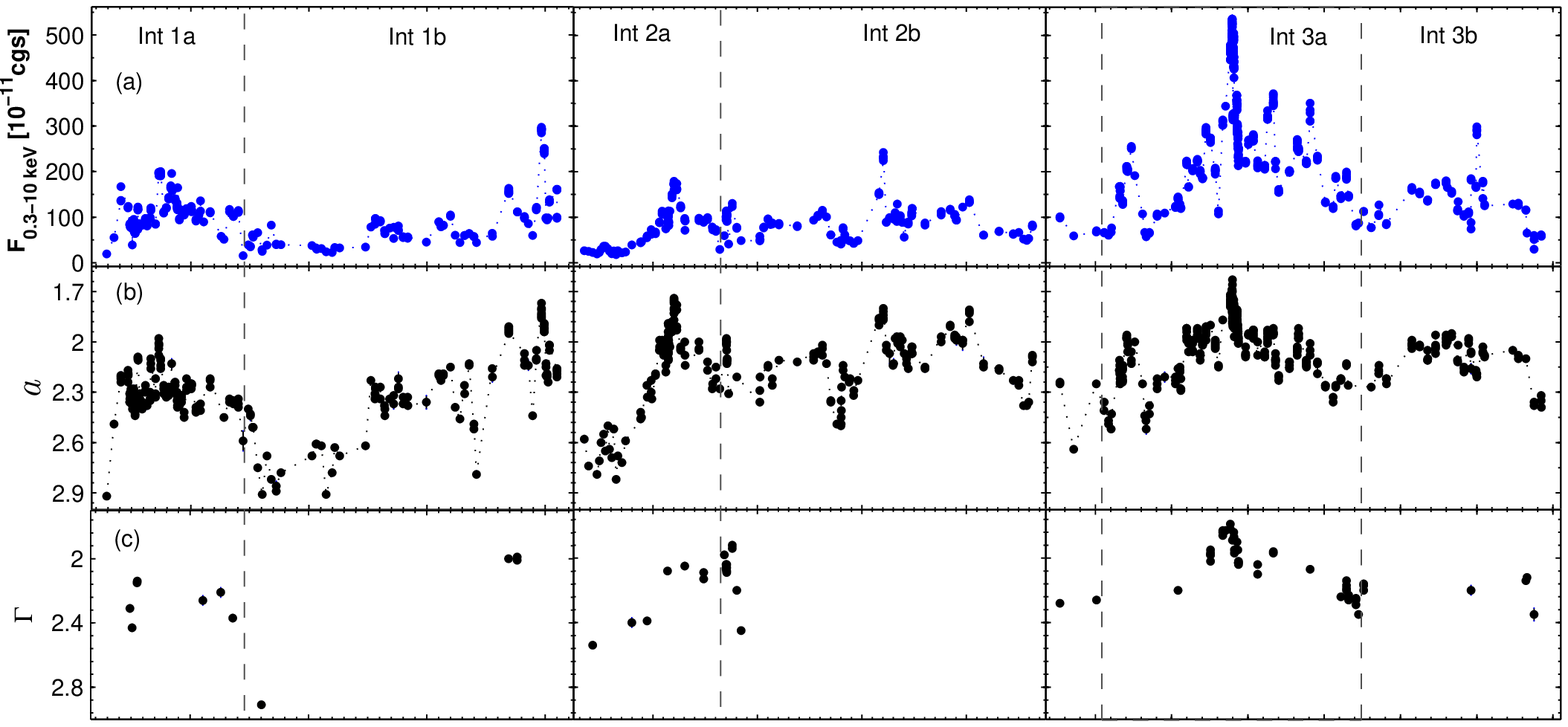}
\includegraphics[trim=7.3cm 0.1cm -1cm 0cm, clip=true, scale=0.83]{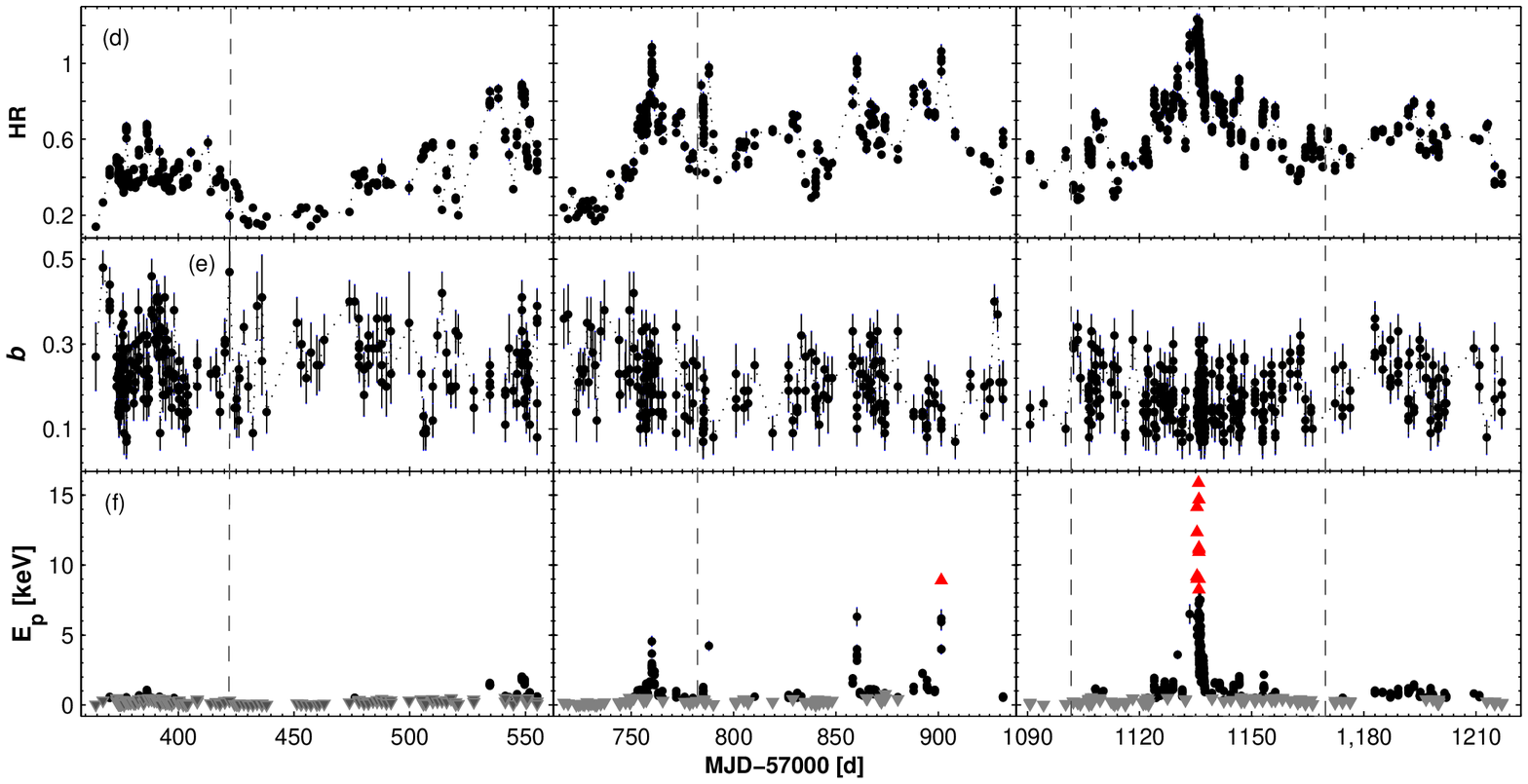}
\vspace{-0.4cm}
 \caption{\label{partiming} De-absorbed 0.3--10\,keV flux and different spectral parameters plotted versus time. The gray and red triangles in the bottom panel indicate the upper and lower limits to the intrinsic position of the synchrotron SED peak, respectively.}
 \end{figure*}

Similar to the previous periods, the source clearly demonstrated a
\textquotedblleft harder-when-brighter\textquotedblright ~spectral
behaviour during the 0.3--10\,keV flares. Figure\,\ref{figcor}d
exhibits a strong anti-correlation between the parameter $a$ and the
de-absorbed 0.3--10\,keV flux (see also Table\,\ref{cortable}). This
trend was evident in Intervals\,1--3 and sub-intervals separately,
although with different strengths and slopes of the corresponding
scatter plot. Moreover, Figure\,\ref{partiming}b shows some short
time intervals when the opposite spectral trend was observed (during
MJD\,57370.42--57374.47, 57426.45--57428.31 etc.).

\begin{table*}
\vspace{-0.3cm}
 \tabcolsep 2pt
 \hspace{-0.9cm}
 \begin{minipage}{185mm}
  \caption{\label{ks}  Results from the KS-test. In Cols\,(2)--(7), the symbol \textquotedblleft
1\textquotedblright~ indicates that data sets are different from
each other, and the corresponding distance between the normalized
cumulative distributions are provided in Cols\,(8)--(13);
\textquotedblleft 0\textquotedblright -  no significant difference
between the data sets.} \vspace{-0.2cm}
  \begin{tabular}{ccccccccccccccccc}
    \hline
   & \multicolumn{6}{c}{KS} & \multicolumn{6}{c}{$D_{\rm KS}$}     \\
  \hline
Quant. & Int1--Int2 &Int1--Int3 & Int2--Int3  & Int1a--Int1b  &Int2a--Int2b & Int3a--Int3b  & Int1--Int2 &Int1--Int3 & Int2--Int3 & Int1a--Int1b  &Int2a--Int2b & Int3a--Int3b  \\
(1) & (2) & (3) & (4) & (5) & (6) & (7) & (8) & (9) & (10) & (11) & (12) & (13)  \\
 \hline
\emph{b} &  1 &  1 &  1&   0  & 1 &  1  & 0.25  &  0.38 &   0.15  &  -  & 0.34  &  0.27 \\
\emph{a} & 1 &  1  & 1 &  1 &  1 &  1 &  0.49 &   0.61 &   0.26  &  0.32  &  0.19  &  0.47\\
$\Gamma$ & 0 &  0 &  0  & -  & - &  - &  -  & - &  - &  -  & -  & - \\
HR &  1  & 1 &  1 &  1 &  0 &  1  & 0.50 &   0.62 &   0.25 &   0.31 &   -&   0.55 \\
$E_{\rm p}$ & 0 &  1 &  1 &  - &  -  & - &  -  & 0.37 &   0.28 &   - &  -  & - \\
 \hline
\end{tabular}
\end{minipage}
\end{table*}

294 spectra were harder than $a$=2 (conventional threshold between
the hard and soft X-ray spectra), amounting to 33.2\% of all
log-parabolic spectra (see Figure\,\ref{figdistr}B1). Note that this
percentage is smaller than in  the  periods 2005--2008, 2009--2012,
2013\,January--May (39\%--46\%)  and higher than those shown in
2013\,November -- 2015\,June (20\%; see
\citealt{k16,k17a,k18a,k18b}). On average, the hardest spectra with
the mean value $\overline{a}$=1.98$\pm$0.01 were observed in
Interval\,3a versus the softest spectra belonging to Interval\,1
($\overline{a}$=2.29$\pm$0.01).
Figures\,\ref{figdistr}B2--\ref{figdistr}B5 and Table\,\ref{ks}
clearly show that the distribution properties of the parameter $a$
varied not only from interval to interval, but also among the
sub-intervals.

Similar to the curvature parameter $b$, the photon index \emph{a}
showed an extreme variability on diverse timescales. It varied at
the 3$\sigma$ confidence level 69 times with $\Delta
a$=0.08(0.03)--0.31(0.02) within 0.27--23.97\,hr along with the
X-ray IDVs (see Col.\,(5) of Table\,\ref{idvtable} and Figures
\ref{idvext1}--\ref{idvext2}). Among them, 5 subhour instances were
recorded: hardenings by $\Delta a$=0.08(0.03)--0.23(0.03) within
0.13--0.28\,hr. On longer timescales, the largest variabilities with
$\Delta a$=0.66--1.07 in 3.1--27.6\,d were observed along with the
strong 0.3--10\,keV flares (Figure\,\ref{partiming}b).

\begin{table}
\tabcolsep 3pt \vspace{-0.2cm} \centering
\begin{minipage}{80mm}
  \caption{\label{cortable} Correlations between spectral parameters and 0.3--10\,keV flux in different periods. In Cols\,(2)--(3), $\rho$ and $p$ stand for
  the Spearman coefficient and the corresponding p-chance, respectively. }
     \vspace{-0.2cm}
    \begin{tabular}{ccc}
  \hline
  Quantities & $\rho$   & $p$ \\
  \hline
 & 2015\,Dec--2018\,Apr  & \\
     \hline
$a$ and $b$ & 0.33(0.08) & $6.23\times10^{-6}$ \\
$b$ and $E_{\rm p}$ & $-$0.34(0.09) & $1.04\times10^{-6}$ \\
$b$ and $F_{\rm 0.3-10\,keV}$ & $-$0.33(0.10) & $3.42\times10^{-6}$ \\
$a$ and $F_{\rm 0.3-10\,keV}$ & $-$0.80(0.03) & $3.08\times10^{-14}$ \\
$\Gamma$ and $F_{\rm 0.3-10\,keV}$ & $-$0.75(0.07) & $7.42\times10^{-13}$ \\
$HR$ and $F_{\rm 0.3-10\,keV}$ & 0.79(0.04) & $4.19\times10^{-14}$ \\
$E_{\rm p}$ and $F_{\rm 0.3-10\,keV}$ & 0.69(0.08) & $4.34\times10^{-12}$ \\
$\log E_{\rm p}$ and $\log S_{\rm p}$ & 0.65(0.08) & $3.39\times10^{-11}$ \\
$\Gamma_{\rm 0.3-2\,GeV}$ and $\Gamma_{\rm 2-300\,GeV}$ & 0.39(0.09) & $1.99\times10^{-8}$ \\
\hline
& Int1 & \\
     \hline
$a$ and $b$ & 0.29(0.09) & $1.00\times10^{-5}$ \\
$a$ and $F_{\rm 0.3-10\,keV}$ & $-$0.70(0.04) & $6.56\times10^{-12}$ \\
$\Gamma$ and $F_{\rm 0.3-10\,keV}$ & $-$0.73(0.907) & $3.04\times10^{-6}$ \\
$HR$ and $F_{\rm 0.3-10\,keV}$ & 0.64(0.06) & $1.24\times10^{-10}$ \\
$E_{\rm p}$ and $F_{\rm 0.3-10\,keV}$ & 0.73(0.08) & $2.51\times10^{-10}$ \\
$\log E_{\rm p}$ and $\log S_{\rm p}$ & 0.64(0.07) & $8.99\times10^{-10}$ \\
\hline
& Int2 & \\
\hline
$a$ and $b$ & 0.23(0.08) & $3.34\times10^{-4}$ \\
$b$ and $E_{\rm p}$ & $-$0.30(0.10) & $8.60\times10^{-5}$ \\
$b$ and $F_{\rm 0.3-10\,keV}$ & $-$0.31(0.10) & $9.11\times10^{-6}$ \\
$a$ and $F_{\rm 0.3-10\,keV}$ & $-$0.88(0.03) & $<10^{-15}$ \\
$\Gamma$ and $F_{\rm 0.3-10\,keV}$ & $-$0.70(0.10) & $7.66\times10^{-5}$ \\
$HR$ and $F_{\rm 0.3-10\,keV}$ & 0.88(0.03) & $<10^{-15}$ \\
$E_{\rm p}$ and $F_{\rm 0.3-10\,keV}$ & 0.69(0.09) & $7.15\times10^{-9}$ \\
$\log E_{\rm p}$ and $\log S_{\rm p}$ & 0.62(0.07) & $6.02\times10^{-10}$ \\
\hline
& Int3 & \\
 \hline
$a$ and $b$ & 0.32(0.08) & $6.23\times10^{-12}$ \\
$b$ and $E_{\rm p}$ & $-$0.32(0.09) & $4.48\times10^{-5}$ \\
$b$ and $F_{\rm 0.3-10\,keV}$ & $-$0.30(0.11) & $8.19\times10^{-5}$ \\
$a$ and $F_{\rm 0.3-10\,keV}$ & $-$0.86(0.03) & $1.03\times10^{-15}$ \\
$\Gamma$ and $F_{\rm 0.3-10\,keV}$ & $-$0.85(0.05) & $2.47\times10^{-14}$ \\
$HR$ and $F_{\rm 0.3-10\,keV}$ & 0.89(0.03) & $<10^{-15}$ \\
$E_{\rm p}$ and $F_{\rm 0.3-10\,keV}$ & 0.85(0.06) & $1.69\times10^{-14}$ \\
$\log E_{\rm p}$ and $\log S_{\rm p}$ & 0.83(0.05) & $5.54\times10^{-14}$ \\
 \hline
\end{tabular}
\end{minipage}
\vspace{0.3cm}
\end{table}

The hardness ratios, derived from the log-parabolic and power-law
spectra, showed a wide range of values ($\Delta HR$=1.09) with
68.7\% of the spectra with $HR>$0.5, and 90 spectra (9.2\%) showing
$HR>$1 (see Figure\,\ref{figdistr}D1 and Table\,\ref{distrtable})
when the de-absorbed 2--10\,keV flux is higher than the 0.3--2\,keV
one. A vast majority of the latter (90\%) belong to Interval\,3a,
characterized by the highest mean value
$\overline{HR}$=0.79$\pm$0.01 (versus $\overline{HR}$=0.42--0.65 in
other sub-intervals; Figures \ref{figdistr}D2--\ref{figdistr}D3). A
positive $F_{\rm 0.3-10 keV}$--HR correlation was observed in all
intervals, demonstrating a dominance of the \textquotedblleft
harder-when-brighter\textquotedblright ~spectral evolution during
X-ray flares, although this trend was significantly weaker in
Interval\,1 (see Figure\,\ref{figcor}f) and Table\,\ref{cortable}).
The long-term behaviour of the hardness ratio followed that of the
parameter $a$: during the largest variability of the photon index,
HR increased by a factor of 3.5--6.4 and showed 75 IDVs by
11\%--88\% per cent (see Col.\,(8) of Table\,\ref{idvtable}).

   \begin{table*}[!ht]
 \vspace{-0.4cm}
\tabcolsep 4pt  \centering
    \begin{minipage}{170mm}
  \caption{\label{pl} The results of the XRT spectral analysis with a simple power-law model (extract; see the corresponding machine-readable table for the full
  version). Col.\,(1) gives the ObsID of the particular observation, along
with the duration of the separate segments (in seconds) used to
extract a spectrum  in parentheses. The acronyms \textquotedblleft
Or\textquotedblright~ and  \textquotedblleft S\textquotedblright
~stand for \textquotedblleft orbit\textquotedblright~ and
\textquotedblleft segment\textquotedblright, respectively.; The
value of the 0.3--10\,keV photon index is provided in Col.\,(2);
Cols (3) and (4) present the norm and the reduced Chi-squared,
respectively; de-absorbed 0.3--2\,keV, 2--10\,keV and 0.3--10\,keV
fluxes (Cols\,(5)--(7)) are given in erg\,cm$^{-2}$\,s$^{-1}$. The
hardness ratio is presented in the last column}.
        \vspace{-0.1cm}
   \begin{tabular}{cccccccc}
  \hline
ObsId & $\Gamma$ & 10$\times K$ &$\chi^2/d.o.f.$ & $ F_{\rm 0.3-2\,keV}$ & $ F_{\rm 2-10\,keV}$ & $ F_{\rm 0.3-10\,keV}$ & HR   \\
(1) & (2) & (3) & (4) & (5) & (6) & (7) & (8) \\
\hline
35014245\,Or5  &  2.31(0.02) & 1.73(0.02) & 0.95/186&57.68(0.66)&28.12(0.83)& 85.90(1.18)& 0.49(0.02)\\
35014246\,Or4  &  2.43(0.02) & 1.51(0.02)&  0.88/182  &  52.60(0.60)&20.70(0.52)& 73.28(0.67)& 0.39(0.01)\\
 34228001\,Or1\,S2\,(410\,s)&2.15(0.02) & 1.64(0.02)&  1.09/229 &   51.88(0.71) &33.81(0.77)&85.70(0.98)& 0.65(0.02)\\
 34228001\,Or2\,S2\,(285\,s)& 2.14(0.02)&2.32(0.02)& 1.08/233  &  73.45(0.84)& 48.42(1.10)& 121.90(1.40)&0.66(0.02)\\
 34228023 &2.26(0.03)&  1.77(0.03) & 0.91/139 &58.34(1.06)&31.05(1.26) &89.33(1.43)&0.53(0.02)\\
 \hline
\end{tabular}
\end{minipage}
\vspace{0.1cm}
\end{table*}

\subsubsection{The Position of the Synchrotron SED Peak}
During the period 2015\,December--2018\,April, 463 spectra (52.3\%
of those showing a curvature) are characterized by 0.5$\leqslant$$
E_{\rm p}$$\leqslant$8\,keV when the position of the synchrotron SED
peak is well-constrained by the XRT data (see \citealt{k18b}). In
the case of 407 spectra (46\%), $E_{\rm p}$$<$0.5\,keV when the
synchrotron SED peak position, derived via the X-ray spectral
analysis, should be assumed as an upper limit to the intrinsic peak
position (not used by us for the construction of the scatter plots
and distributions). Note that such instances amounted to 86.6\%  and
68.9\% of all curved spectra from Intervals 1a and 1b, respectively
(versus 25.4\% in Interval\,3a). Moreover, the majority of the
spectra with $E_{\rm p}$$<$0.1\,keV (when the synchrotron SED peak
is situated in the UV energy range) belonged to these sub-intervals.

On the other hand, for the spectra with $E_{\rm p}$$>$8\,keV, the
synchrotron SED peak is poorly constrained by the observational data
and  such $E_{\rm p}$ values should be considered as lower limits to
the intrinsic position (see \citealt{k18b}). During the period
2018\,December--2018\,April, the source showed 14 spectra with
8.25$\leqslant$$E_{\rm p}$$\leqslant$15.85\,keV, mostly from the
observations corresponding to the highest X-ray states in
Interval\,3a.

The $E_{\rm p}$ values from the range 0.5--8\,keV mainly belong to
Interval\,3 (62.9\%), and their mean is significantly higher than
those from Intervals\,1--2 (2.16\,keV versus 1.02--1.46\,keV in
Intervals\,1--2; see Figures\,\ref{figdistr}D2 and
Table\,\ref{distrtable}). Note that the K-S test and related Monte
Carlo simulations did not show a significant difference between the
distributions corresponding to Intervals 1 and 2 (see
Table\,\ref{ks}). For the entire 2015\,December--2018\,April period,
the parameter $E_{\rm p}$ showed a positive correlation with $F_{\rm
0.3-10 keV}$, which was the strongest in Interval\,3
(Figure\,\ref{figcor}g and Table\,\ref{cortable}. Moreover, a
positive correlation between $E_{\rm p}$ and $S_{\rm p}$ (the height
of the synchrotron SED peak) was detected in all three time
intervals (see Figure\,\ref{figcor}h, Table\,\ref{cortable} and
Section\,4.3 for the corresponding physical implication). Note that
the latter quantity was calculated for each spectrum as \citep{m04}
\vspace{-0.1cm}
\begin{equation}
S_{\rm p}=1.6\times 10^{-9}K10^{(2-a)^2/4b} ~~~ \mbox{ erg cm$^{-2}$
s$^{-1}$}. \vspace{-0.1cm}
\end{equation}

 \begin{figure*}[ht]
\includegraphics[trim=7.2cm 2.5cm -1cm 0cm, clip=true, scale=0.82]{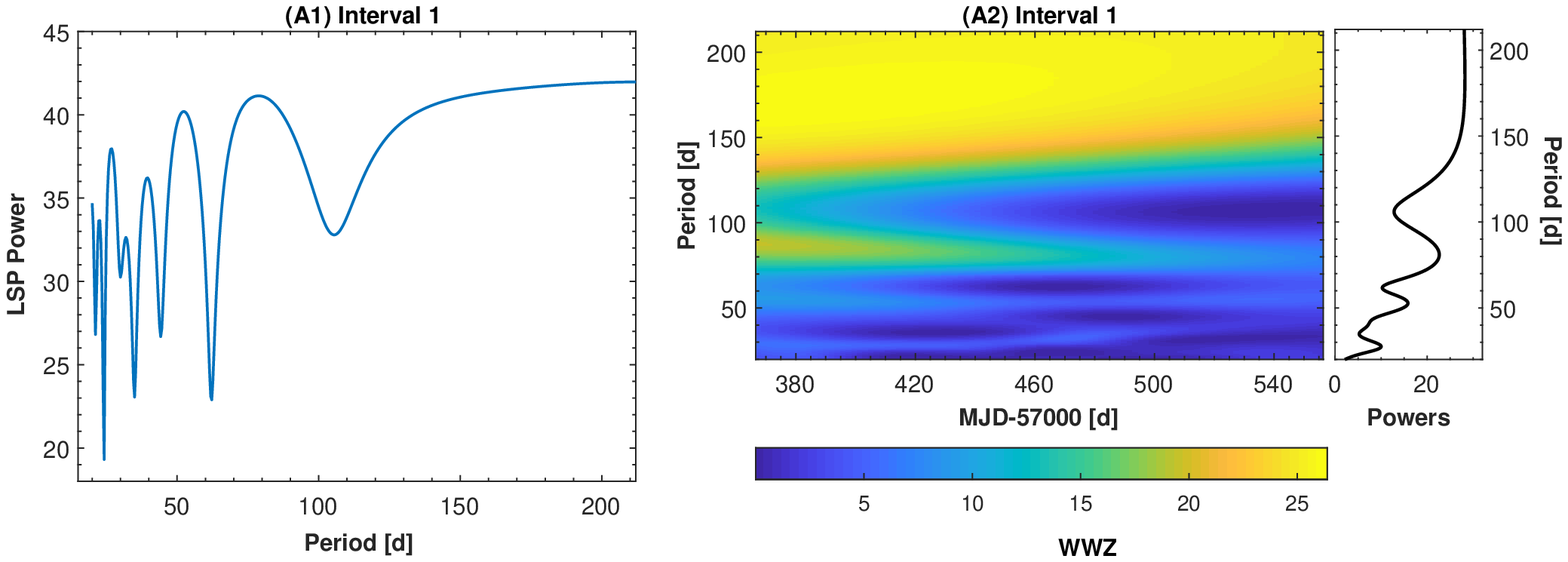}
\includegraphics[trim=7.2cm 2.5cm -1cm 0cm, clip=true, scale=0.82]{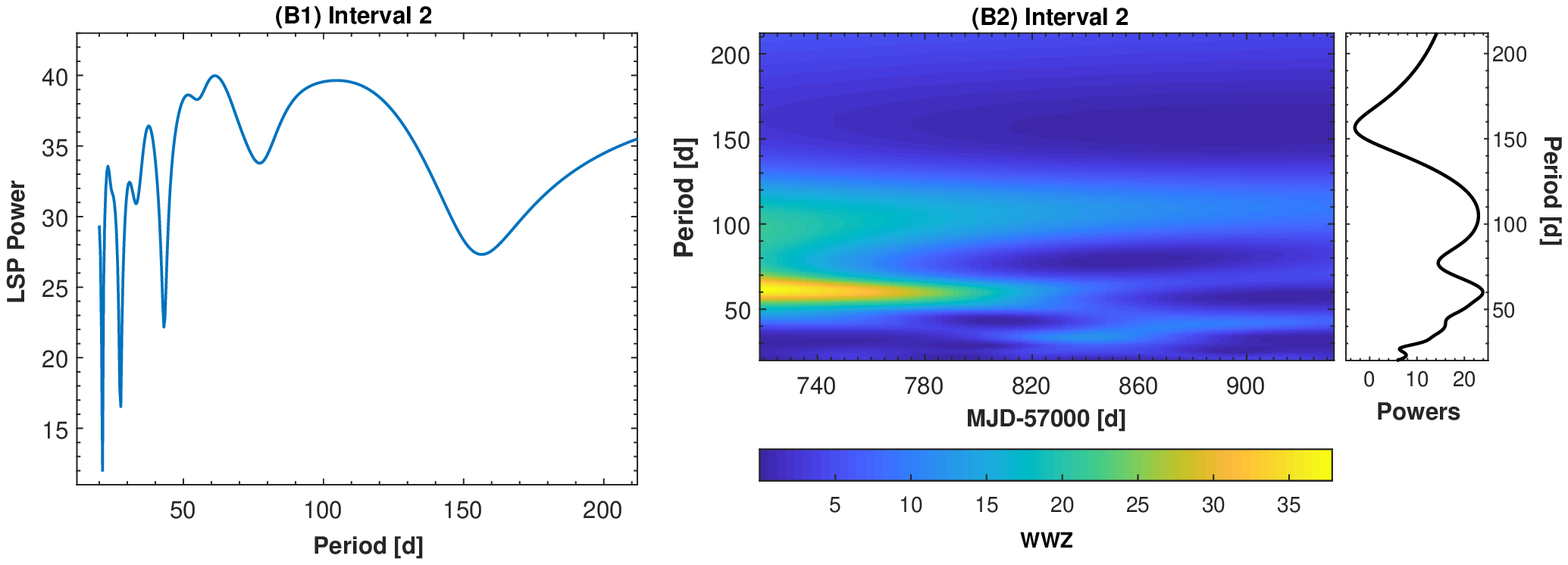}
\includegraphics[trim=7.2cm 3.1cm -1cm 0cm, clip=true, scale=0.82]{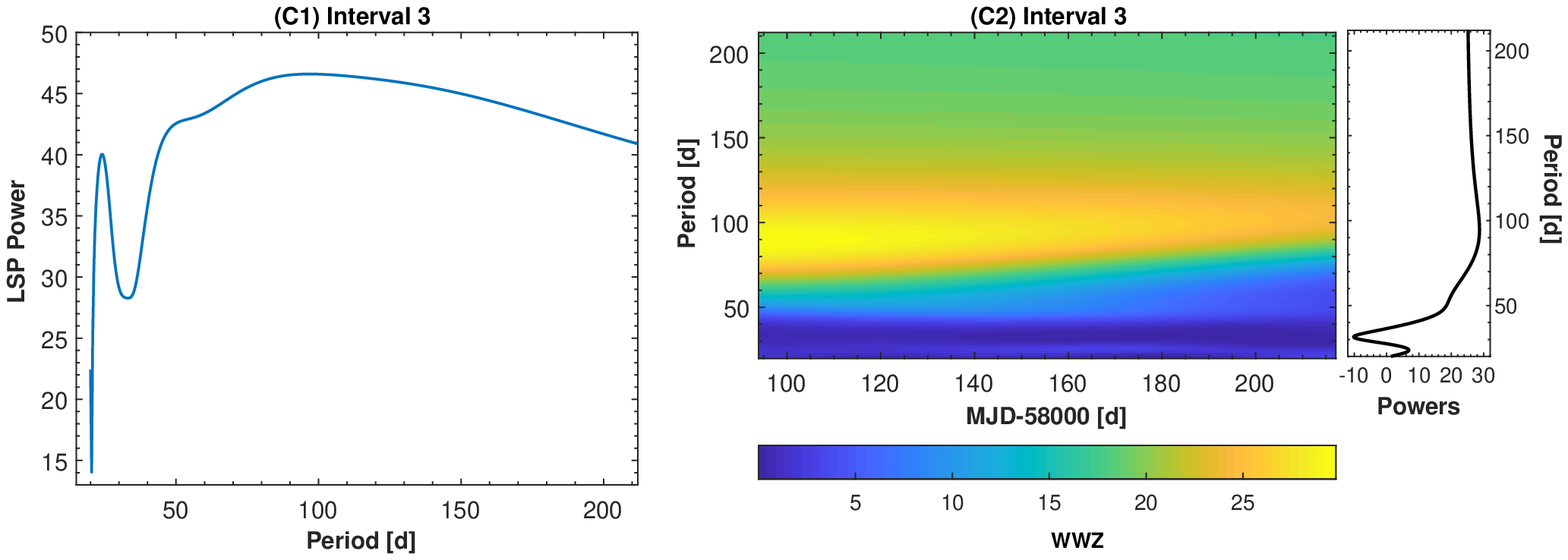}
 \vspace{-0.4cm}
 \caption{\label{lsp} The LSP and WWZ plots from the XRT observations of Mrk\,421 in different intervals.}
 \end{figure*}

On intraday timescales, the parameter $E_{\rm p}$ varied 56 times at
the 3$\sigma$ confidence level, observed during the 0.3--10\,keV
IDVs (see Col.\,(7) of Table\,\ref{idvtable}). The most dramatic
changes were observed during the extreme flare in
2018\,January\,14--30 (see the bottom panels of Figures
\ref{idvext1} and \ref{idv1ks}A1--\ref{idv1ks}A2): $E_{\rm p}$
sometimes showed shifts by several keV  within 0.1--9.5\,hr to
higher energies and moved back in comparable timescales.

\begin{figure*}
 \centering
   \vspace{-0.2cm}
\includegraphics[trim=7.1cm 5.0cm 0.2cm 0cm, clip=true, scale=0.83]{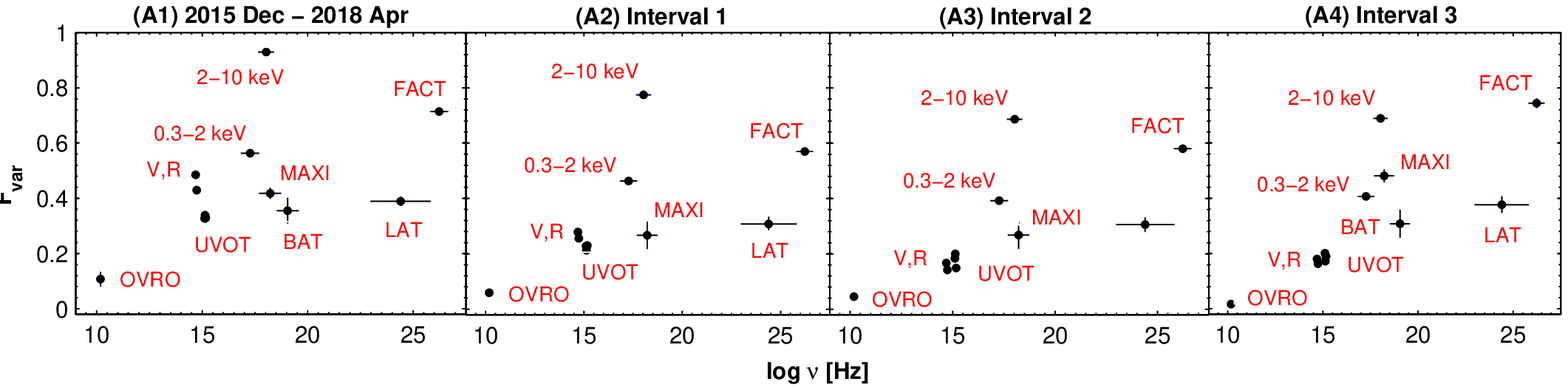}
\includegraphics[trim=7.4cm 4.9cm 0.2cm 0cm, clip=true, scale=0.85]{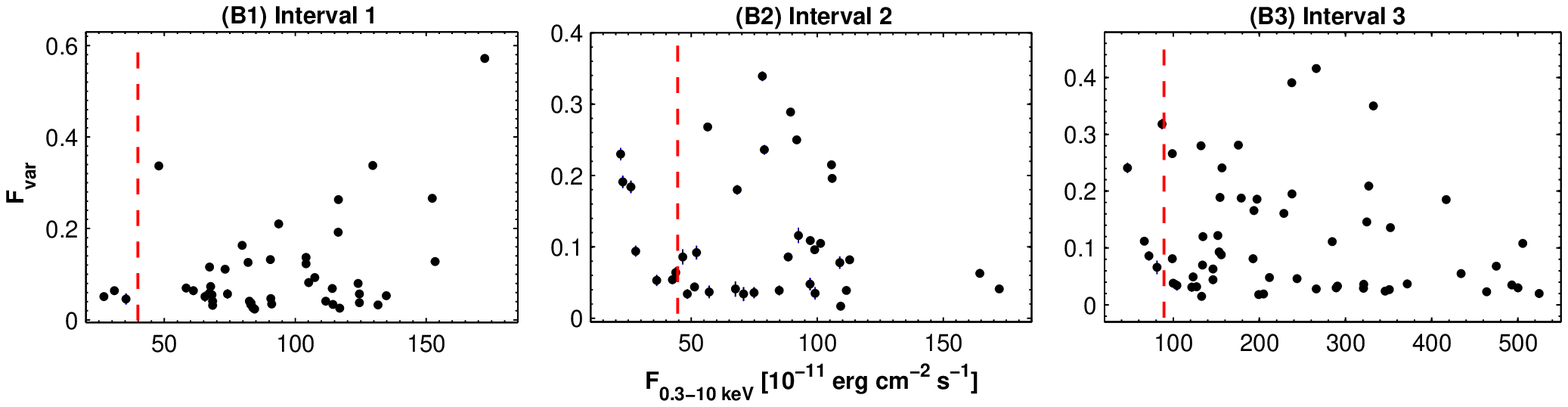}
   \vspace{-0.5cm}
\caption{\label{idvs} Panels (A1)--(A4): multi-instrument fractional
amplitude $F_{var}$ as a function of the energy in different
periods; panels (B1)--(B3): $F_{var}$ of the 0.3--10\,keV IDVs  as a
function of the flux in different periods. In each plane, a
vertical, red dashed line represents the threshold below which the
low X-ray states were observed (see the text for the quantitative
definition)}.
    \end{figure*}

\vspace{0.5cm}
\section{CONCLUSIONS  AND DISCUSSION}

\subsection{X-Ray and MWL Flux Variability}
\subsubsection{Variability Character}
The 0.3--10\,keV brightness of the source reached the highest level
in the time interval 2018\,January--February and Mrk 421 was
brighter only during the giant outburst in 2013\,April. A similar
situation was observed in the VHE energy range:  the highest VHE
states were recorded in 2018\,January (coinciding  those in the XRT
band), while the strongest VHE flare was observed in 2013\,April.
The TeV-band variability mostly showed a good correlation with the
X-ray one, although there were some exceptions (see
Figure\,\ref{subper} and Section\,4.2). Conversely, there were
significantly fewer detections with 5$\sigma$ significance and/or
lower fluxes in the BAT-band, compared to the periods
2005\,December--July, 2008\,March--July, 2009\,October--November,
2010\,January--May, 2011\,September (see a more detailed discussion
in Section\,4.3.5).

In other spectral ranges, Mrk\,421 exhibited a relatively different
behaviour. Namely, the highest 0.3--\textbf{300}\,GeV flux from the
weekly-binned LAT data was
(2.2$\pm$1.8)$\times$10$^{-7}$\,ph\,cm$^{-2}$\,cm$^{-1}$ (in
2016\,February, not coinciding with the highest X-ray states).
Contrary to the X-ray and VHE observations, significantly higher
levels were recorded in 2012\,July--August (with the highest
historical MeV--GeV level) and the comparable states - in 2013 March
and 2014\,April. There was a frequent absence or weakness of the
correlation between the LAT-band and X-ay variability
\citep{k16,k17a}.

In the UVW1--UVW2 bands, the highest states, corresponding to the
de-reddened  and host-subtracted fluxes of 23.3--29.8\,mJy, were
observed during 2016\,January--February, and they were significantly
lower 2\,yr later when the source showed its highest X--ray
activity. The higher UV states were observed in
2010\,June--2011\,April, 2012\,April--May,
2012\,December--2013\,April (the highest historical UV brightness,
preceding the giant X-ray outburst), 2013\,November--2014\,April,
2015\,February--June. A nearly-similar situation was found in the
optical \emph{V--R} and OVRO bands.

We checked the MWL data of Mrk\,421 for periodicity during
2015\,December--2018 April. As an example, Figure\,\ref{lsp}
presents the LSP and WWZ plots from the XRT observations performed
in Intervals\,1--3. No clearly expressed quasi-periodic variations
are found in this energy range, similar to the radio--UV and
GeV--TeV ranges. Periodic  brightness variations have not been found
also by different authors (see, e.g., \citealt{c17,s17}).

As in past years, the source showed a double-humped behaviour in the
plane $\log\nu$--$F_{\rm var}$ during the entire
2015\,December--2018\,April period and its separate intervals (with
$F_{\rm var}$, calculated using the entire data set obtained in the
given spectral band during the particular period; see
Figures\,\ref{idvs}A1--\ref{idvs}A4 and
\citealt{k16,k17a,k18a,k18b}). We used the 1-d binned XRT, BAT,
FACT, OVRO and optical-UV data in our study, while the 3-d binned
LAT was used in the 0.3--300\,GeV energy range (similar to the light
curves provided in Figure\,\ref{subper}). Although the latter
binning and the cuts at 3$\sigma$--5$\sigma$ detection significances
can lead to some undersampling in the corresponding $F_{\rm var}$
values, a two-humped shape with the synchrotron and higher-energy
peaks situated at the X-ray and VHE frequencies, respectively, seems
to be inherent for HBLs and are frequently reported for these
sources. For our target, a similar result was reported by various
authors from various MWL campaigns \citep{a15a,a15b,a16,abey17},
favouring the one-zone SSC model (predicting the correlated
X-ray--VHE variability) for Mk\,421 in different shorter-term
periods and indicating that the electron energy distribution is most
variable at the highest energies \citep{a15b}.

Although the optical data points in
Figures\,\ref{idvs}A1--\ref{idvs}A2 violate the general trend of
increasing variability power from radio to hard X-ray frequencies,
this result can be related to the data sampling: some \emph{V} and
\emph{R}-band observations were carried out at Arizona Observatory
in those time intervals when the source was not targeted by UVOT,
although it was showing a strong variability. The presence of lower
VHE peaks in Figures\,\ref{idvs}A1--\ref{idvs}A3, compared to the
peak in the synchrotron frequency, is difficult to explain via the
upscattering of synchrotron photons in the Thomson regime (when a
squared relation is expected). However, this result can be related
to the use of the FACT excess rates instead the linear fluxes in our
study.

 \begin{figure*}[ht!]
         \hspace{-0.5cm}
  \includegraphics[trim=6.6cm 5.3cm 0.8cm 0cm, clip=true, scale=0.83]{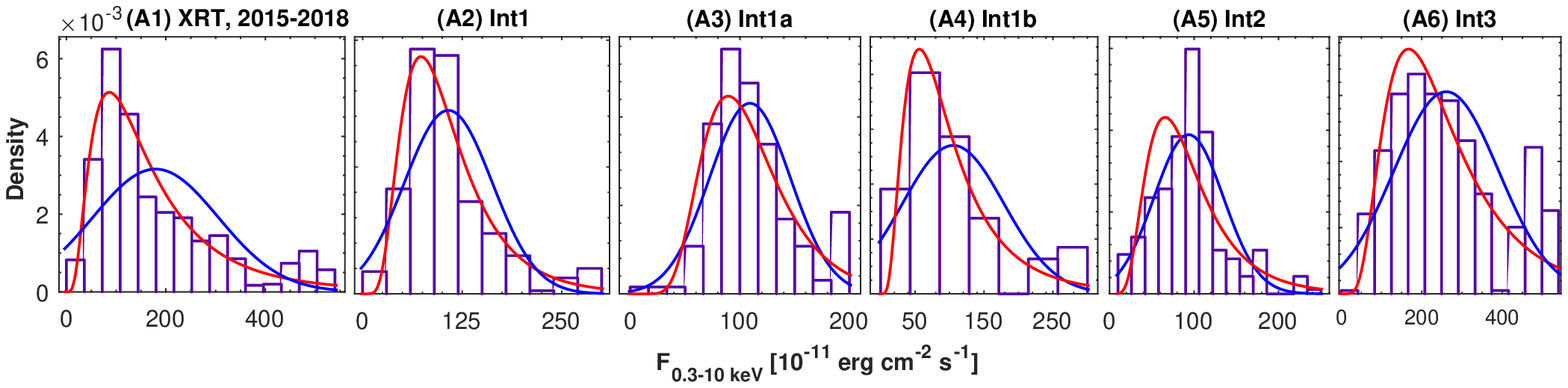}
  \includegraphics[trim=6.9cm 5.3cm 0.8cm 0cm, clip=true, scale=0.82]{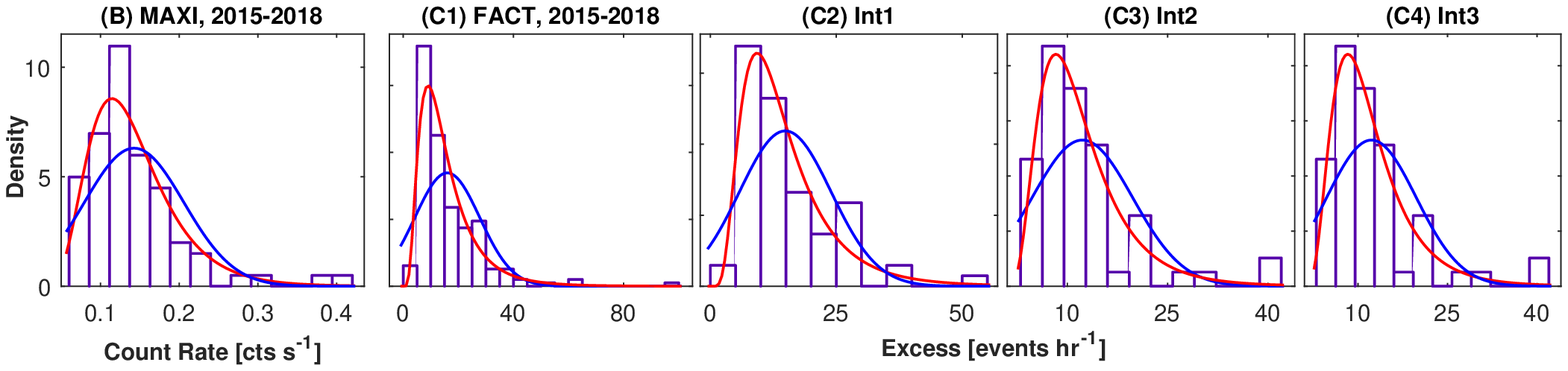}
\includegraphics[trim=6.9cm 5.3cm 0.8cm 0cm, clip=true, scale=0.82]{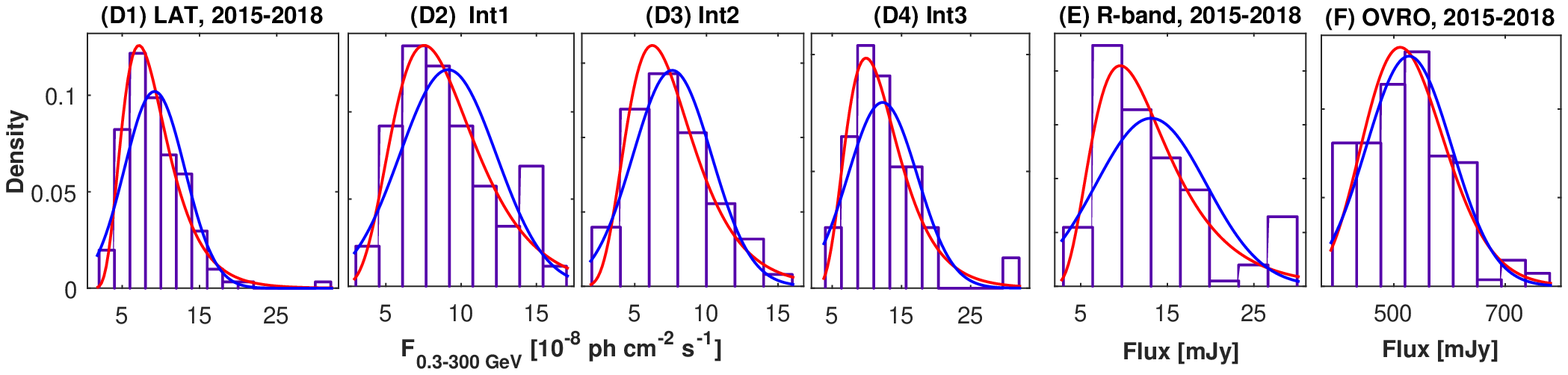}
\includegraphics[trim=6.9cm 5.8cm 0.8cm 0cm, clip=true, scale=0.82]{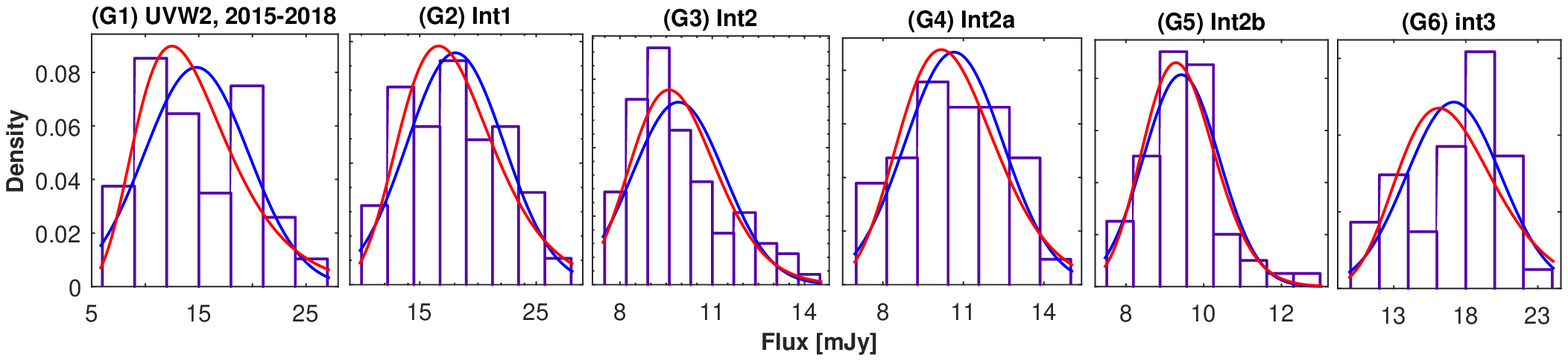}
\includegraphics[trim=6.9cm 5.4cm 0.8cm 0cm, clip=true, scale=0.82]{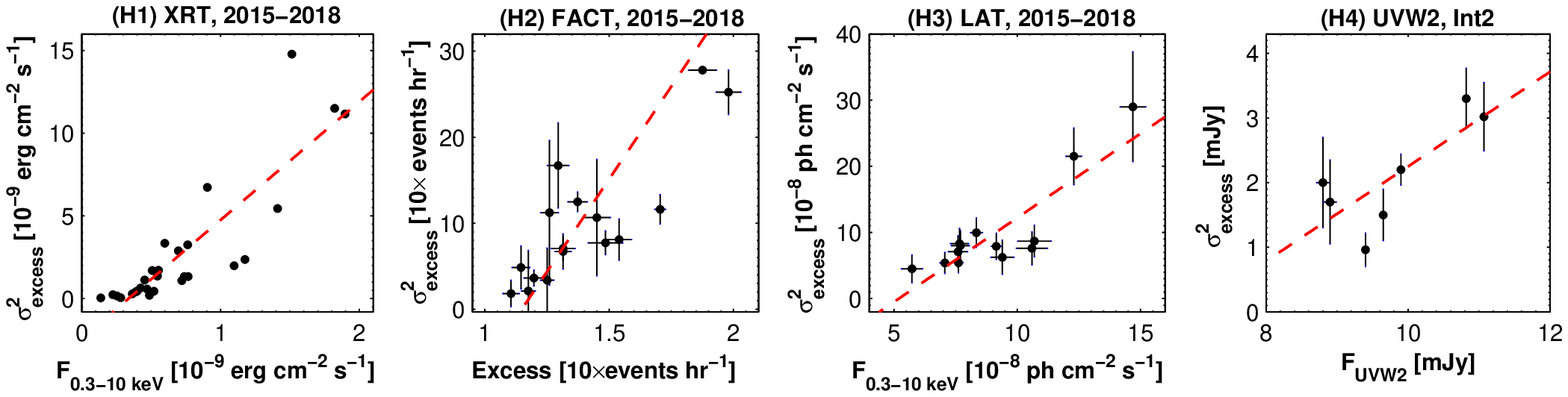}
 \vspace{-0.3cm}
 \caption{\label{logn} Histograms of the MWL fluxes. In each plot, red and blue lines correspond to the lognormal and Gaussian  fits, respectively. For the space
 reason, the period 2015\,December--2018\,April is denoted as \textquotedblleft 2015--2018\textquotedblright~ in the plot titles. The bottom row: scatter plots with $\sigma^2_{\rm excess}$ and the average flux for the MWL
 data. A linear fit is shown in red dashed line.}
\end{figure*}

\subsubsection{Flux Lognormality}
We also checked whether the X-ray and MWL fluxes of Mrk\,421,
observed in the 2015 December-2018 April time interval, showed
lognormal distributions. According to \cite{mc08}, a lognormal flux
behavior in blazars can be indicative of the variability imprint of
the accretion disk onto the jet. Moreover, the lognormal fluxes have
fluctuations, that are, on average, proportional to the flux itself,
and are indicative of an underlying multiplicative, rather than
additive physical process. Consequently, the excess variance
$\sigma^2_{\rm excess}$=$\sqrt{S^2-\overline{\sigma_{\rm err}}^2}$
(with the quantities $S$ and $\overline{\sigma_{\rm err}^2}$ defined
in Equation\,(3)) plotted versus the mean flux for all flares,
should show an increasing linear trend \citep{ch19}.

For BLLs, the lognormality in different spectral ranges and time
intervals were reported for PKS\,2155-304, BL Lac and 1ES\,1011+496
\citep{ch19,g09,si17}. In the case of Mrk\,421, the lognormality was
studied by \cite{si17} using the radio, optical--UV and LAT-band
data obtained during 2009--2015. The lognormal fit to the histograms
were clearly preferred for most of the bands, leading to the
suggestion that the flux variability in the source can be mainly
attributed to changes in the particle spectrum rather than to the
variability of the jet physical parameters such as the magnetic
field or Doppler factor (see \citealt{si16}).

In order to investigate lognormality, we fit the histograms of the
MWL fluxes with the Gaussian and lognormal functions (similar to the
aforementioned studies). Figure\,\ref{logn}A1 demonstrates, that the
distribution of the de-absorbed 0.3--10\,keV flux from the entire
2015\,December--2018\,April period is closer to the lognormal shape
than the Gaussian one (similar to the MAXI observations;
Figure\,\ref{logn}B). A lognormal behaviour is confirmed by the
corresponding scatter plot in the $\sigma^2_{\rm
excess}$--$\overline{F_{\rm 0.3-10 keV}}$ plane where the data
points, corresponding to the XRT-band flares during
2015\,December--2018\,April, show an increasing linear trend with
higher mean flux (Figure\,\ref{logn}H1). Note that this result is
mainly due to the observations performed in Intervals 1 and 3, while
the data from Interval\,2 are closer to the Gaussian function
(Figures\,\ref{logn}A2--\ref{logn}A6).

The longer-term flares may result from the propagation and evolution
of relativistic shocks through the jet (see \citealt{s04} and
references therein). The shock appearance can be related to the
instabilities occurring in the accretion disk, which may momentarily
saturate the jet with extremely energetic plasma with much higher
pressure than the steady jet plasma downstream \citep{s04}.
Consequently, a lognormal flaring activity of the source on longer
timescales may indicate a variability imprint of the accretion disk
onto the jet. However, the fluxes, corresponding to \textbf{the}
highest X-ray states in Intervals\, 1 and 3, produce outliers from
the lognormal distributions. These states generally were recorded
during fast flares superimposed on the long-term one. These flares
could be triggered \textbf{by} the shock interaction with the jet
inhomogeneities whose origin was related to the jet instabilities
(e.g., strong turbulent structures; \citealt{m14}). Therefore, no
lognormal distribution of those fluxes is expected in that case,
owing to the absence of the AD variability imprint. Moreover,
Interval\,2 clearly shows a better fit to the Gaussian function and
since this period was characterized, on average, by lower X-ray
states (see Section\,3.1 and Table\,\ref{persum}), this result could
be related to the propagation of weaker shocks through the jet. On
the other hand, a shock weakness possibly was due to the lower AD
variability in that period, which resulted in a fewer imprint onto
the jet of Mrk\,421.

In each interval, the FACT and LAT-band fluxes clearly showed a
better fit of the lognormal function with the corresponding
distributions (Figures\,\ref{logn}C1--\ref{logn}D4). Although the
same is shown by the \emph{R}-band histogram constructed for the
entire 2015\,December--2018\,April period, it is impossible to draw
a firm conclusion related to the lognormal behaviour of the source
in this band, since we could not construct the corresponding
$\overline{F_{\rm R}}$--$\sigma^2_{\rm excess}$ scatter plot due to
the sparse sampling of the optical flares (see
Figure\,\ref{subper}). In contrast, the OVRO data do not show a good
fit between the corresponding histogram and lognormal function. This
result implies the radio-contributions from various emission regions
with different physical conditions. Finally, the UVOT-band
histograms show a better (but not good) fit with the lognormal
function compared to the Gaussian one only in Interval\,2 (see the
bottom row of Figure\,\ref{logn}).

\subsubsection{Intraday Variability}
 In the 2015 December-2018 April period, the source showed three intraday
flux-doubling  and three flux-halving events with $\tau_{\rm
d}$=14.2--17.8\,hr and $\tau_{\rm h}$=4.8--18.9\,hr, respectively.
Note that the flux doubling time can be used for constraining the
upper limit to the emission zone as follows (\citealt{sa13} and
references therein) \vspace{-0.2cm}
\begin{equation}
 R_{\rm em}\leqslant {{c\tau_{\rm d} \Gamma_{\rm
em}}\over{1+z}}, \vspace{-0.2cm}
\end{equation}
with $R_{\rm em}$ and $\Gamma_{\rm em}$, the size and Lorentz-factor
of the emission zone, respectively. Adopting the typical value of
the bulk Lorentz-factor for the emission zone $\Gamma_{\rm em}$=10
\citep{f14}, we obtain upper limits of
5.0$\times$10$^{15}$--1.9$\times$10$^{16}$\,cm for the emission
zones with intraday flux-doubling instances. Mrk\,421 exhibited more
extreme behaviour during the densely-sampled \emph{Swift}-XRT
campaign in 2009\,May\,22--27 (net exposure time of 23.5\,ks, 59
orbits) $\tau_{\rm h}$=1.4--1.5\,hr and  $\tau_{\rm
d}$=6.5--12.4\,hr \citep{k18b}. A series of the brightness halving
and doubling events with $\tau_{\rm d,h}\sim$1.1\,hr were recorded
on 2009\,February\,17, and a similar variability occurred also
during 2017\,February\,2--3 (see Section\,3.1 and
Figure\,\ref{flares}i). Note that the successive large brightness
drop and rise events can be explained as a consequence of a shock
passage through two inhomogenous areas with strong magnetic fields,
which are separated by a region with significantly weaker field and
lower particle density (yielding the generation of fewer X-ray
photons). The most extreme behaviour was observed during the giant
X-ray outburst in 2013\,April, with several events showing
$\tau_{\rm d}$=1.2--7.2\,hr and $\tau_{\rm h}$=1.0--3.5\,hr
\citep{k16}.

The distribution of the 0.3--10\,keV fluxes extracted from those
\emph{Swift}-XRT pointings which showed IDVs in the period
2015\,December--2018\,April is not well fitted with the lognormal
function, in contrast to the longer-term flares. This result hint at
the absence of an AD variability imprint onto these events.
Moreover, Figures\,\ref{idvs}B1--\ref{idvs}B3 show a rare occurrence
of these events in low X-ray states\footnote{The de-absorbed
0.3--10\,keV flux values, corresponding to the low X-ray states of
the source, were below the thresholds of about
4.0$\times$10$^{-10}$\,cgs, 4.5$\times$10$^{-10}$\,cgs and
9.0$\times$10$^{-10}$\,cgs in Intervals 1,2 and 3, respectively.}
when the IDVs caused by the instability in AD or in the inner jet
regions should be more easily detectable: the variable emission from
these regions will not be \textquotedblleft
overwhelmed\textquotedblright ~ by the huge flux produced near the
shock front, propagating through the jet and causing longer-term
flares (see \citealt{mang93} and references therein). On the other
hand, most of these events are detected from short XRT exposures and
the entire cycle (brightness increase and drop) is generally  not
recorded, in contrast to the longer-term flares. Therefore, it is
not possible to draw a firm conclusion about the absence of
lognormality. However, the 0.3--10\,keV flux values from those IDVs
whose  complete variability cycles were observed do not exhibit a
lognormal flux distribution. Therefore, these results favour the
\textquotedblleft shock-in-jet\textquotedblright ~scenario where
IDVs are triggered by the interaction of a shock front with
small-size jet inhomogeneities \citep{s04,m14,miz14}.

\begin{figure*}[ht]
\vspace{-0.2cm}
\includegraphics[trim=7.2cm 4.5cm -1cm 0cm, clip=true, scale=0.83]{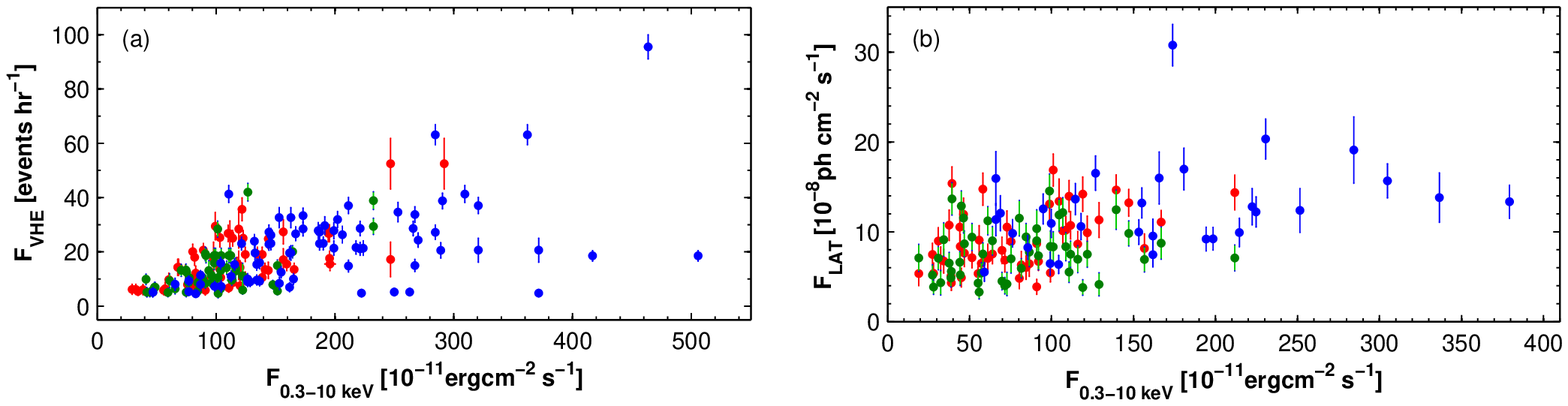}
\includegraphics[trim=7.2cm 4.5cm -1cm 0cm, clip=true, scale=0.83]{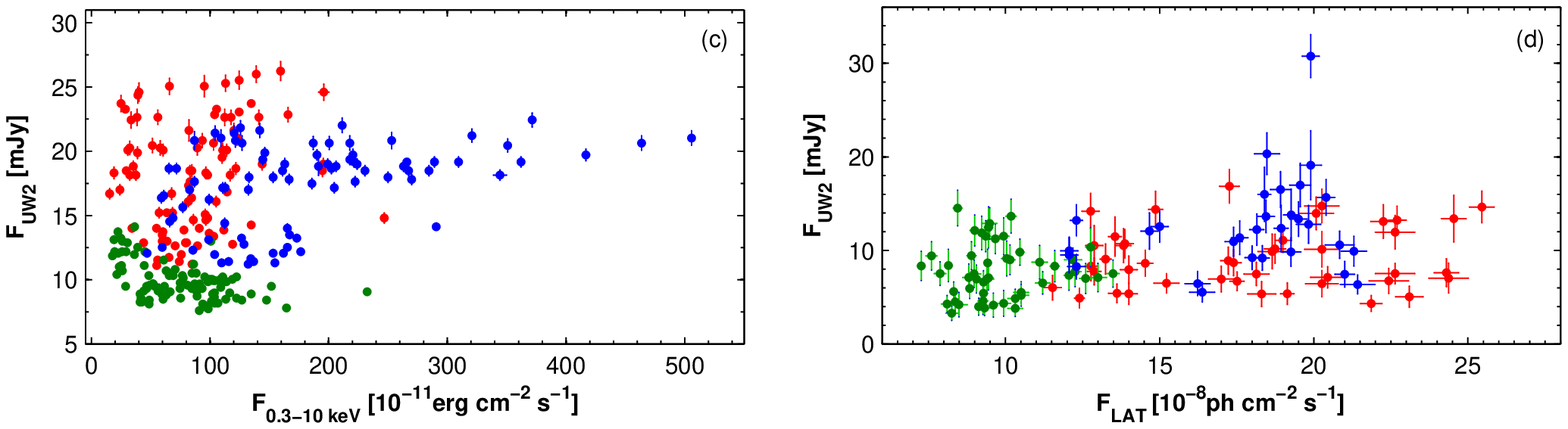}
\includegraphics[trim=7.2cm 4.5cm -1cm 0cm, clip=true, scale=0.83]{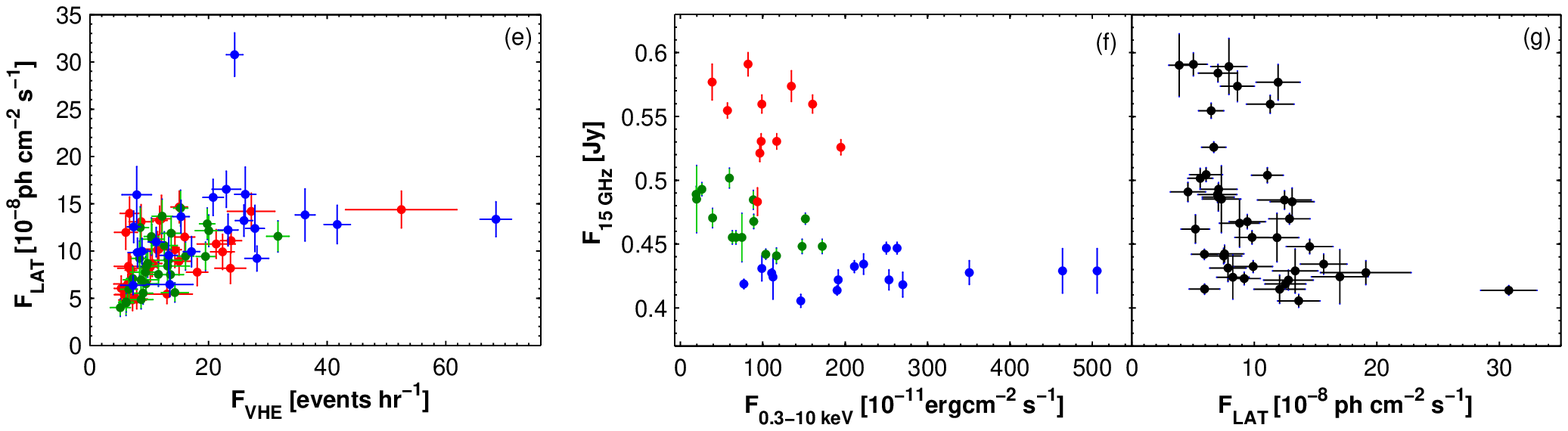}
\includegraphics[trim=7.2cm 4.7cm -1cm 0cm, clip=true, scale=0.83]{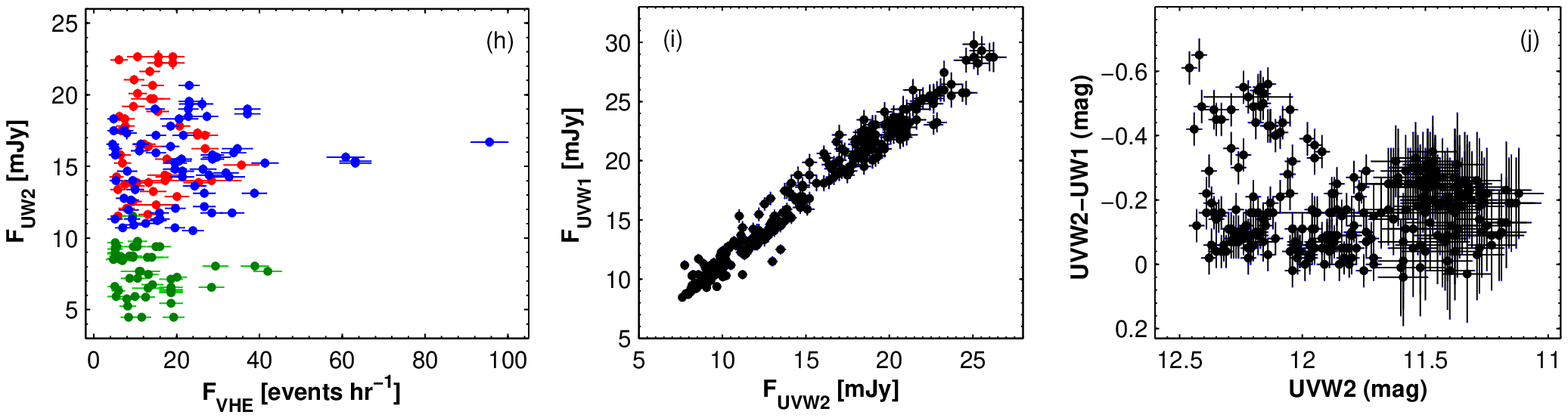}
\vspace{-0.2cm}\vspace{-0.2cm}
 \caption{\label{cor1} Correlations between the MWL fluxes. The coloured points in  Figures\,\ref{cor1}a--\ref{cor1}h correspond
 to the different periods as follows: red -- Interval\,1, green -- Interval\,2, blue -- Interval\,3.}
 \end{figure*}

The duty cycle of the 0.3--10\,keV IDVs (i.e., the fraction of total
observation time during which the object displayed IDVs; see
\citealt{r99} and references therein) amounted to 58.4\%, which is
higher than that shown by the source in 2009--2012 and
2013\,November--2015\,June (43--46 per cent; \citealt{k17a,k18b}),
but significantly lower than compared to the periods 2005--2008 and
2013\,January--May (about 84\%; \citealt{k16,k18a}). Note that this
result can be partially related to significantly more densely
sampled campaigns in some parts of the latter periods. For example,
Mrk\,421 was observed very densely during 2006\,June\,15--25 with a
total exposure of 119\,ks distributed over 124 \emph{Swift}-XRT
orbits.

On the other hand, the source exhibited an unequal 0.3--10\,keV
activity on intraday timescales in different epochs. Note that the
XRT pointings without IDVs were mainly performed during the lower
X-ray states. However, we also discerned some flares with a clear
lack of intraday activity. For example,  such flaring activity was
observed during MJD\,57521--548 with $CR_{\rm
max}$=95.73$\pm$0.32\,cts\,s$^{-1}$, which is a factor of 2.5 higher
than the mean rate during the whole period
2015\,December--2018\,April. This event was preceded by the long
time interval (MJD\,57400-57520) when the source had showed the
lowest mean rate (23.24$\pm$0.01\,cts\,s$^{-1}$) and duty cycle of
IDVs (36\%) for the entire 2015--2018 period. A similar situation
was observed in 2010\,June--July when the source demonstrated low
0.3--10\,keV states with CR$\sim$10\,cts\,s$^{-1}$ and insignificant
variability over 1.5\,months, possibly related to the absence of
strong shock waves, which could cause a long-term flare in different
spectral bands \citep{k18b}. Moreover, the source also showed
significantly slower and weaker IDVs during those densely-sampled
XRT observations presented in Figure\,\ref{idvext2}. These results
lead to the conclusion that the source underwent weaker variability
on intraday timescales during the low X-ray states and some
longer-term flares.

The last column of Table\,\ref{idvtable} provides a list of the
variable spectral factor presumably making the dominant contribution
to the particular 0.3--10\,keV IDV. The most common factor in the
observed fast variability was a change of the photon index (i.e., in
the slope of the Particle Energy  Distribution, PED). Nevertheless,
some events were related to the transition from a log-parabolic PED
into a power-law one and/or vice versa. The most extreme
log-parabolic/power-law/log-parabolic transitions occurred within
the 1-ks exposures during ObsID\,31202174 (MJD\,58135.39), 4-th
orbit of ObsID\,31202175 (MJD\,58136.33) and first orbit of
ObsID\,31202178 (MJD\,58137.11), in the epoch of the strongest X-ay
flaring activity of the source during 2015--2018. Such extreme
events can be related to very fast variability of the magnetic field
properties within the jet regions which are smaller than
$R$=10$^{14}$\,cm (for $\delta$=10): a transition from a magnetic
field characterized by decreasing confinement efficiency with rising
electron's gyroradius or strong turbulence (both yielding a
log-parabolic PED) into a volume without these properties and vice
versa (see below). A number of the IDVs were related also to the
variability of the curvature parameter, the position and height of
the synchrotron SED peak. The changes of the latter two quantities
imply an intraday and even a subhour variability of the physical
parameters (magnetic field, Doppler factor, $\gamma^2_{\rm 3p}$: the
peak of the function $n(\gamma)\gamma^3$) included in the relations
as follows (\citealt{r79} and references therein) \vspace{-0.1cm}
\begin{equation}
E_{\rm p} \varpropto \gamma^2_{\rm 3p}B\delta,~~ S_{\rm p}
\varpropto N\gamma^2_{\rm p}B^2\delta^4.
\end{equation}

\vspace{0.2cm}
\subsection{ Multiwavelength Correlations}

During the time interval presented in this work, the TeV-band
variability showed the strongest correlation with the X-ray one:
strong VHE flares or enhanced activity mostly coincided with strong
X-ray flares (see Section\,3.2). Consequently, Figure\,\ref{cor1}a
shows a positive $F_{\rm 0.3-10 keV}$--$F_{\rm VHE}$ correlation in
each interval, although it was relatively weak in Interval\,3
compared to that in Interval\,1 (the difference between the values
of the coefficient $\rho$ was larger than the corresponding error
ranges; see Table\,\ref{cortable1}). In fact, some data points from
Interval\,3 make the largest exceptions from the general trend. Such
situation was particularly evident during MJD\,58145--153
(Interval\,3a): the VHE flux showed a decline by a factor of $\sim$4
and subsequent low states during the fast X-ray flare with one of
the highest states during the 2015--2018 period (see
Figure\,\ref{subper}a). Two data points in the right lower corner of
the $F_{\rm 0.3-10 keV}$--$F_{\rm VHE}$ plane belong to the X-ray
flare around MJD\,58136 when the 0.3--10\,keV flux reached its
highest values, while the TeV-band state was the highest on the
previous day. This situation is hard to reconcile with one-zone SSC
scenarios \citep{bl05}. However, Fig\,\ref{idvvhe}a demonstrates
that no strictly simultaneous XRT observation was performed during
those 20-min FACT pointings detecting the source in the highest VHE
states. Therefore, we cannot draw any firm conclusion for this case.

The source was not detectable with 3$\sigma$ or showed low VHE
states when it had undergone short-term X-ray flares during
MJD\,(57)413--422 (Figure\,\ref{subper}c), MJD\,(57)425--438 and
MJD\,(57)473--491 (Figure\,\ref{subper}d), MJD\,(57)723--730 and
MJD\,(57)801--838(Figure\,\ref{subper}e). Similar instances were
reported from the FACT observations of Mrk\,421 performed in
previous years \citep{k16,k17a}. Moreover, the MWL campaign in
2002\,December--2003\,January revealed a strong X-ray flare by a
factor of 7 within 3\,d, not accompanied by a comparable
TeV-counterpart \citep{r06}. The TeV flux reached its peak days
before the X-ray flux during the giant flare in 2004 that was
impossible to explain via the standard one-zone SSC model, and
\cite{bl05} suggested this as an instance of the \textquotedblleft
orphan\textquotedblright ~TeV flare. \cite{ac11} also found high
X-ray  states, not accompanied by TeV flaring  and vice versa in
2006--2008. We conclude that the broadband SED can not  always be
modelled using one-zone SSC scenarios, although they were acceptable
for Mrk\,421 during the majority of the X-ray flares (corroborated
by the appearance of a two-hump structure in the $F_{\rm
var}$--log$\nu$ plane, with the peaks at X-ray and TeV frequencies,
respectively).

  \begin{table}[ht!]
   \vspace{-0.2cm}
   \begin{minipage}{85mm}
  \caption{\label{cortable1} Correlations between the MWL fluxes (denoted by \textquotedblleft $F_{i}$\textquotedblright ~for
  the particular $i$-band in Column\,1) during different periods. In Cols\,(2)--(3), $\rho$ and $p$ stand for
  the Spearman coefficient and the corresponding p-chance, respectively. }
     \vspace{-0.2cm}
  \begin{tabular}{ccc}
  \hline
  Quantities & $\rho$ & $p$ \\
  \hline
 & 2015\,Dec--2018\,Apr  & \\
     \hline
$F_{\rm 0.3-2\,keV}$ and $F_{\rm 2-10\,keV}$ &0.93(0.02) & $<10^{-15}$  \\
$F_{\rm 0.3-10\,keV}$ and $ F_{\rm FACT}$ & 0.61(0.08) & $5.44\times10^{-10}$ \\
$F_{\rm 0.3-10\,keV}$ and $ F_{\rm UVW2}$ & 0.30(0.11) & $4.07\times10^{-5}$ \\
$F_{\rm 0.3-10\,keV}$ and $ F_{\rm UVW1}$ & 0.25(0.11) & $8.34\times10^{-4}$ \\
$F_{\rm 0.3-10\,keV}$ and $ F_{\rm LAT}$ & 0.49(0.10) & $1.03\times10^{-8}$ \\
$F_{\rm 0.3-10\,keV}$ and $ F_{\rm 15 GHz}$ & -0.53(0.09) & $8.89\times10^{-9}$ \\
$F_{\rm UVM2}$ and $F_{\rm UVW2}$ & 0.94(0.02) & $<10^{-15}$ \\
$F_{\rm UVW1}$ and $F_{\rm UVW2}$ & 0.96(0.01) & $<10^{-15}$ \\
$F_{\rm UVW2}$ and $F_{\rm LAT}$ & 0.32(0.11) &$7.58\times10^{-5}$ \\
$F_{\rm FACT}$ and $F_{\rm LAT}$ & 0.55(0.09) &$1.24\times10^{-9}$ \\
$F_{\rm LAT} $ and $ F_{\rm 15 GHz}$ & -0.49(0.14) & $3.55\times10^{-4}$ \\
$F_{\rm 0.3-2\,GeV}$ and $ F_{\rm 2-300\,GeV}$ & 0.75(0.06) & $5.02\times10^{-13}$ \\
\hline
 & Int\,1  & \\
  \hline
 $F_{\rm 0.3-2\,keV}$ and $F_{\rm 2-10\,keV}$ & 0.88(0.03) & $<10^{-15}$  \\
$F_{\rm 0.3-10\,keV}$ and $ F_{\rm FACT}$ & 0.62(0.07) & $1.28\times10^{-11}$ \\
$F_{\rm 0.3-10\,keV}$ and $ F_{\rm LAT}$ & 0.44(0.10) & $6.10\times10^{-8}$ \\
$F_{\rm FACT}$ and $F_{\rm LAT}$ & 0.45(0.10) &$3.71\times10^{-8}$ \\
 \hline
 & Int\,1a  & \\
  \hline
 $F_{\rm 0.3-10\,keV}$ and $ F_{\rm UVW2}$ & 0.56(0.11) & $5.01\times10^{-5}$ \\
$F_{\rm 0.3-10\,keV}$ and $ F_{\rm UVW1}$ & 0.48(0.12) & $7.79\times10^{-5}$ \\
 \hline
 & Int\,1b  & \\
  \hline
 $F_{\rm 0.3-10\,keV}$ and $ F_{\rm UVW2}$ & -0.46(0.11) & $6.77\times10^{-5}$ \\
$F_{\rm 0.3-10\,keV}$ and $ F_{\rm UVW1}$ & -0.43(0.11) & $3.40\times10^{-5}$ \\
 \hline
   & Int\,2  & \\
   \hline
$F_{\rm 0.3-2\,keV}$ and $F_{\rm 2-10\,keV}$ & 0.93(0.02) & $<10^{-15}$  \\
$F_{\rm 0.3-10\,keV}$ and $ F_{\rm FACT}$ & 0.51(0.09) & $7.76\times10^{-9}$ \\
$F_{\rm 0.3-10\,keV}$ and $ F_{\rm UVW2}$ & -0.45(0.09) & $1.08\times10^{-6}$ \\
$F_{\rm 0.3-10\,keV}$ and $ F_{\rm UVW1}$ & -0.47(0.09) & $2.03\times10^{-7}$ \\
$F_{\rm 0.3-10\,keV}$ and $ F_{\rm LAT}$ & 0.34(0.12) & $3.17\times10^{-4}$ \\
$F_{\rm 0.3-10\,keV}$ and $ F_{\rm 15 GHz}$ & -0.71(0.09) & $3.33\times10^{-9}$ \\
$F_{\rm FACT}$ and $F_{\rm LAT}$ & 0.64(0.08) &$6.10\times10^{-11}$ \\
 \hline
 & Int\,2a  & \\
  \hline
 $F_{\rm 0.3-10\,keV}$ and $ F_{\rm UVW2}$ & -0.75(0.09) & $1.92\times10^{-8}$ \\
$F_{\rm 0.3-10\,keV}$ and $ F_{\rm UVW1}$ & -0.74(0.09) & $4.56\times10^{-8}$ \\
 \hline
  & Int\,3  & \\
   \hline
 $F_{\rm 0.3-2\,keV}$ and $F_{\rm 2-10\,keV}$ & 0.93(0.03) & $<10^{-15}$  \\
$F_{\rm 0.3-10\,keV}$ and $ F_{\rm FACT}$ & 0.45(0.11) & $9.01\times10^{-8}$ \\
$F_{\rm 0.3-10\,keV}$ and $ F_{\rm UVW2}$ & 0.39(0.10) & $4.07\times10^{-5}$ \\
$F_{\rm 0.3-10\,keV}$ and $ F_{\rm UVW1}$ & 0.38(0.10) & $9.95\times10^{-5}$ \\
$F_{\rm 0.3-10\,keV}$ and $ F_{\rm LAT}$ & 0.39(0.11) & $5.26\times10^{-4}$ \\
 \hline
\end{tabular}
\end{minipage}
\end{table}

The 0.3--10\,keV variability showed a weak positive correlation with
that observed in the MeV--GeV energy range (see Figure\,\ref{cor1}b
and Table\,\ref{cortable1}). The source did not undergo comparable
LAT-band activity or exhibited lower states during the most extreme
X-ray behaviour (MJD\,58115--58160; Figure\,\ref{subper}a).
Moreover, no credible detections or low fluxes were recorded in the
time intervals MJD\,(58)190--215 and (57)364--381
(Figures\,\ref{subper}b--\ref{subper}c), MJD\,(57)731--746
(Figures\,\ref{subper}e). An even weaker $F_{\rm 0.3-10
keV}$--$F_{\rm 0.3-300 GeV}$ correlation was observed during
2009--2012 and 2013\,January--June, while no significant correlation
was found for the period 2013\,November--2015\,June
\citep{k16,k17a,k18b}.

\begin{figure*}[ht!]
         \hspace{-0.5cm}
  \includegraphics[trim=6.9cm 5.5cm 0.8cm 0cm, clip=true, scale=0.83]{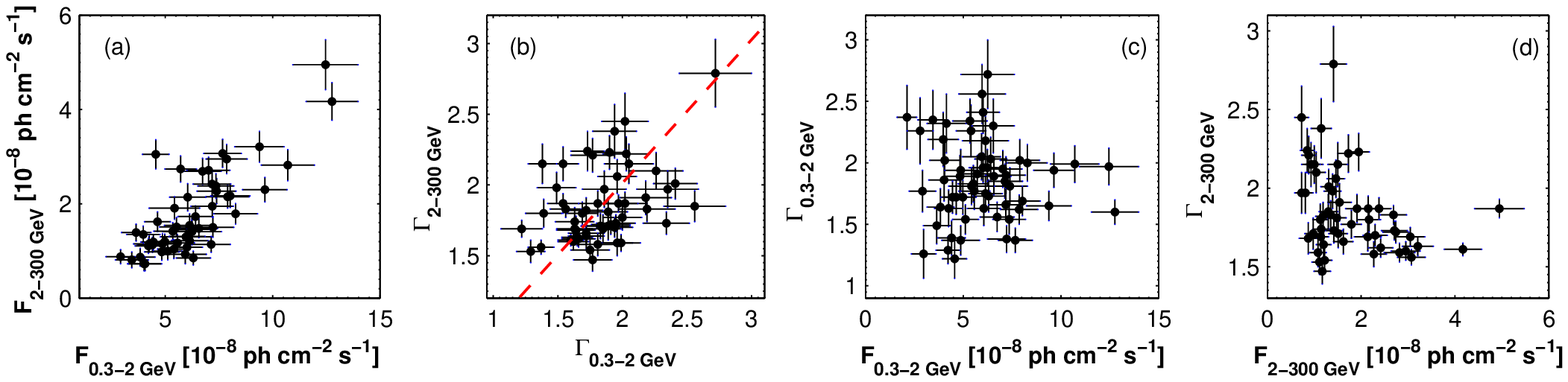}
\includegraphics[trim=7.3cm 6.5cm 0.8cm 0cm, clip=true, scale=0.83]{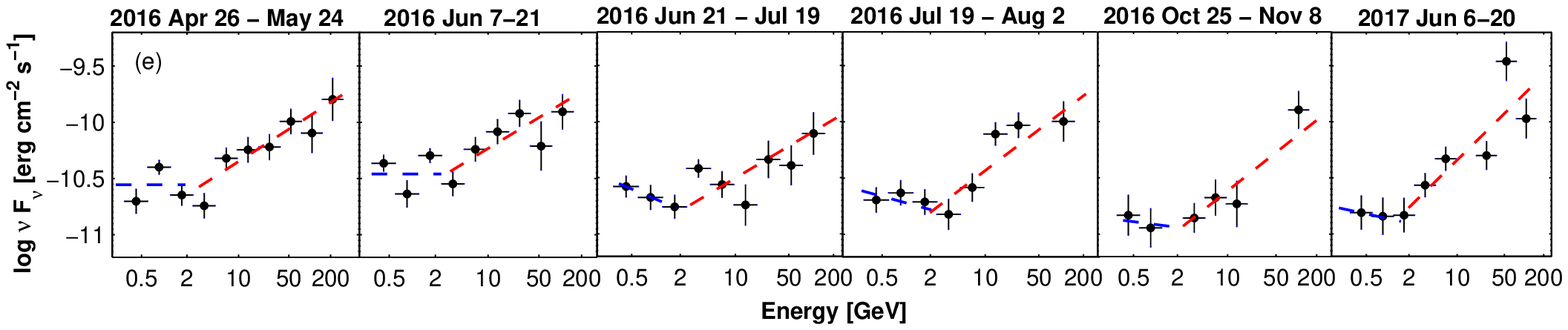}
\includegraphics[trim=7.3cm 6.6cm 0.8cm 0cm, clip=true, scale=0.83]{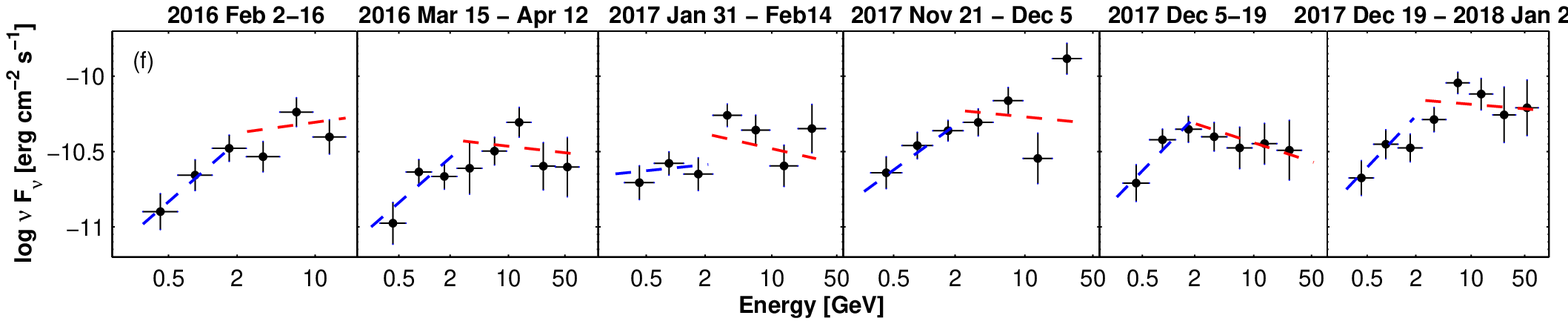}
 \vspace{-0.5cm}
\caption{\label{latgamma}  Results from 2-weekly-binned  LAT data
for the period 2015\,December--2018\,April. The scatter plots:
0.3--2\,GeV and 2--300\,GeV photon fluxes (panel (a)); 0.3--2\,GeV
and 2--300\,GeV photon indices, with the red dashed line
corresponding to $\Gamma_{\rm 0.3-2 GeV}$=$\Gamma_{\rm 2-300 GeV}$
(panel\,(b)); 0.3--2\,GeV photon flux and index (panel\,(c));
2--300\,GeV photon flux and index (panel\,(d)). Panels (e) and (f):
the 0.3-300\,GeV SEDs from the time intervals showing a hardening
and softening in the 2--300\,GeV energy range, respectively. The
blue and red dashed lines represent linear fits to the 0.3--2\,GeV
and 2--300\,GeV SEDs, respectively. }
   \end{figure*}

We extracted the 0.3--2\,GeV and 2--300\,GeV photon fluxes from the
LAT observations to check their cross-correlation and search for
possible contributions from different electron populations to the
LAT-band emission. We used 2-weekly binned observations to ensure
$N_{\rm pred}$$\geqslant$10 (Table\,\ref{lat}).
Figure\,\ref{latgamma}a exhibits a strong $F_{\rm 0.3-2
GeV}$--$F_{\rm 2-300 GeV}$ correlation, demonstrating a
predominantly common origin for the soft and hard LAT-band photons.
This result is in contrast to that obtained for 1ES\,1959$+$650
during the period 2016\,August--2017\,November, with a weak
correlation between the softer and harder LAT-band fluxes that hints
at uncorrelated behaviour and possible contribution from different
particle populations \citep{k18e}.

Note that the LAT-band flux was also correlated weakly with the
UVOT- and FACT-band fluxes, respectively (see
Figures\,\ref{cor1}d--\ref{cor1}e). This result can be explained as
an IC uppscatter of the UV photons to the MeV--GeV energies in the
Thomson regime (a similar relation between the X-ray and VHE
photons). Another source of the LAT-band emission can be an
upscatter of X-ray photons in the Klein-Nishina (K-N) regime. An
upscatter of the BAT-band photons in the same regime could be the
case during those time intervals when the source showed low VHE
states with the absent $F_{\rm 0.3-10 keV}$--$F_{\rm VHE}$
correlation or was not detectable in the TeV energy range (owing to
strong suppression of the $\gamma$-ray emission in this regime; see
\citealt{t09}).

Both LAT and XRT-band fluxes showed an anti-correlation with the
15\,GHz emission which was particularly strong in Interval\,2 in the
case of the X-ray emission (see Figures\,\ref{cor1}f--\ref{cor1}g
and Table\,\ref{cortable1}). A similar situation is evident from the
scatter plot $F_{\rm 0.3-10 keV}$--$F_{\rm UVW2}$ constructed for
Interval\,2 and Interval\,1b (Figure\,\ref{cor1}c and the
aforementioned table). Such a feature was reported also by
\cite{a15b} and explained through a hardening of the electron energy
distribution that can shift the entire synchrotron bump to higher
energies. Consequently, the emission of the synchrotron SED in the
radio--UV energy range is expected to decrease with rising X-ray
brightness. Note that such MWL behaviour is expected for the
stochastic acceleration of electrons within the specific initial
conditions (see Section\,4.3).

No $F_{\rm 0.3-10 keV}$--$F_{\rm UVW2}$ correlation at the 99\%
confidence level was detected in Interval\,1, and a very weak
positive correlation occurred in Interval\,3. An anti-correlation
was observed also during 2009--2012, and uncorrelated variabilities
occurred in 2005--2008 and 2014\,February--2015\,June
\citep{k17a,k18b}. A stronger positive correlation was observed
during 2013\,January--June and 2013\,November--2014\,January
\citep{k16,k17a}. We have not found a correlation between the UV and
VHE fluxes, hinting at an insignificant role for the upscatter of
the UV photons to VHE frequencies in the Thomson regime
(Figure\,\ref{cor1}h). Finally, the UVW1--UVW2 fluxes showed very
strong cross-correlations, as in previous periods, reflected in the
absence of the UV colour variability and demonstrating the
generation of these photons by the same electron population via the
synchrotron mechanism in each period
(Figures\,\ref{cor1}i--\ref{cor1}j).

\subsection{Spectral Properties and Particle Acceleration Processes}

\subsubsection{First-Order Fermi Acceleration}

A positive $a$--$b$ correlation (see Figure 9a), detected in
Intervals\,1--3, shows an importance of the first-order Fermi
acceleration since this correlation was predicted for those jet
regions where particles are confined by a magnetic field at the
shock front, whose confinement efficiency is declining with
increasing gyration radius (i.e., particle's energy; so-called EDAP:
energy-dependent acceleration probability process; \citealt{m04}).
Consequently, the probability $p_i$ that a particle undergoes an
acceleration step $i$, with the corresponding energy $\gamma^{q}_i$
and energy gain $\varepsilon$, is given by $p_i$=$g/\gamma^{q}_i$,
where $g$ and $q$ are positive constants. Consequently, the
probability of the particle's acceleration is lower when its energy
increases,  and the differential energy spectrum is given by
$N(\gamma)\sim{\gamma/\gamma_0}^{-s-r\log{\gamma/\gamma_0}}$, with a
linear relationship between the spectral index and curvature terms
(\emph{s} and \emph{r}, respectively)
$s=-r(2/q)\log{g/\gamma_0}-(q-2)/2$. The synchrotron emission
produced by this distribution is given by \vspace{-0.1cm}
\begin{equation}
P_S (\nu)\varpropto (\nu/\nu_0)^{-(a+b\log(\nu/\nu_0))},
\vspace{-0.1cm}
\end{equation}
with $a=(s-1)/2$ and $b=r/4$ \citep{m04}. However, the detected
$a$--$b$ correlation was weak in each interval. This result can be
explained due to the sub-samples having different slopes in the
\emph{a}--\emph{b} plane, leading to a large scatter of the data
points during the entire period 2015\,December--2018\,April. Note
that some sub-samples showed even a negative \emph{a}--\emph{b}
trend (e.g., those corresponding to the short-term flares recorded
during MJD\,57364--375, 57395--57408, 57527-57544 in Interval\,1),
which is expected when $g>\gamma_0$, i.e., there were electron
populations with a very low initial energy $\gamma_0$ in the
emission zone. Moreover, the co-existence of stochastic
(second-order Fermi) acceleration also could weaken the
aforementioned correlation, since the latter is not expected within
the stochastic mechanism. The Monte Carlo simulations of \cite{k06}
revealed that electrons can be accelerated at the shock front via
EDAP and continue gaining energy via the stochastic mechanism into
the shock downstream region (after escaping the shock front). After
some time, a particle will be able to re-enter the shock
acceleration region and repeat the acceleration cycle. Consequently,
such combined acceleration will not result in the strong  $a$--$b$
correlation.

A stronger positive a--b correlation was found for our target during
the XRT observations in 2005--2008 and  2015\,February
($\rho$=0.41--0.57; \citealt{k17a,k18a}). An even stronger
correlation was reported by \cite{m04} from the\emph{BeppoSAX}
observations in 1997--2000. On the other hand, this correlation was
weaker during 2009--2012 ($\rho$=0.21$\pm$0.07) and absent in
2013\,January--2014\,June \citep{k16,k17a}.

\begin{table*} [ht!]
 \tabcolsep 5pt
 \vspace{-0.4cm}
 \begin{minipage}{180mm}
\caption{\label{lat}  The results from 2-weekly-binned LAT data in
the 0.3--2\,GeV and 2--300\,GeV bands. Cols (2) and (6) present the
test-statistics corresponding to Mrk\,421 in each band; Cols (3) and
(7) -- number of the model-predicted photons. The photon fluxes
(Cols (4) and (8)) are provided in units of
10$^{-8}$\,ph\,cm$^{-2}$\,s$^{-1}$.  Cols (5) and (9) -- the
0.3--2\,GeV and 2--300\,GeV photon indices, respectively.  The last
column provides a remark related to the interplay between the
0.3--2\,GeV and 2--300\,GeV spectral hardnesses:  1 -- increasing
hardness with energy; 2 -- harder spectra in the 0.3--2\,GeV energy
range; 3 -- no significant difference between the higher- and
lower-energy photon indices. In Col.\,(8), an asterisk stands  for
the upper limit to the 2--300\,GeV flux.} \vspace{-0.1cm} \centering
  \begin{tabular}{cccccccccc}
    \hline
   &\multicolumn{4}{c}{0.3--2\,GeV} & \multicolumn{4}{c}{2--300\,GeV}&     \\
  \hline
Dates & TS & $N_{pred}$ & Flux & $\Gamma$ &TS & $N_{\rm pred}$ & Flux & $\Gamma$ &  Remark \\
(1) & (2) & (3) & (4) & (5) & (6) & (7) & (8) & (9)& (10) \\
 \hline
2015-12-08$-$08-22 &  55.0  &  143.5 &  6.05(0.85)&  1.63(0.14) & 175.8&24.6&2.14(0.29)&  1.69(0.09)& 3 \\
2015-12-22$-$2016-01-05 & 113.1 &  380.1 &  7.67(0.88)&  1.37(0.10)&639.9&55.9  &  3.07(0.30)&  1.56(0.05)& 2 \\
 2016-01-05$-$01-19  & 80.5 &274.8&9.39(1.15) & 1.65(0.12) & 481.4 &  40.4  &  3.21(0.33)& 1.63(0.06)&3 \\
2016-01-19$-$02-02&   80.5  &  256.1&   7.86(0.96)& 1.62(0.12)&400.8&37.6  &  2.95(0.31)&  1.60(0.06)& 3 \\
2016-02-02$-$02-16  & 62.7  &  183.7  & 4.56(0.62)&  1.22(0.16)&482.6&47.0  &  3.05(0.31) & 1.69(0.06)& 2\\
 2016-02-16$-$03-01 &  103.0 &  280.5&8.28(1.02) & 2.00(0.15) & 257.0 &  28.6  &  1.79(0.22)&  1.77(0.09)&3  \\
2016-03-01$-$03-15&   51.4 &   143.9  & 4.64(0.60) & 1.72(0.14)&247.3&22.4 &   1.62(0.21) & 1.66(0.08)&3 \\
2016-03-15$-$03-29  & 43.2  &  108.5 &  3.65(0.69)&  1.49(0.15)&178.0&20.1  &  1.39(0.18) & 1.98(0.11)& 2\\
 2016-03-29$-$04-12  & 57.7  &  136.5&4.41(0.71)&  1.39(0.14) & 153.8 &  18.0 &   1.13(0.16) & 1.80(0.10)& 2 \\
2016-04-12$-$04-26 &  61.9 &   144.0 &  5.40(0.76) & 2.26(0.21) &95.4&16.2  &  1.04(0.20)&  2.10(0.13)& 2\\
2016-04-26$-$05-10  & 58.6  &  166.9 &  6.02(0.72)&  1.96(0.12)&126.2&13.4  &  1.08(0.18) & 1.59(0.09)&1 \\
 2016-05-10$-$05-24  & 60.5  &  179.6&7.39(0.91) & 1.81(0.15)&  274.3 &  23.3 &   2.27(0.27) & 1.58(0.08)& 3 \\
2016-05-24$-$06-07 &  80.4  &  208 &6.41(0.82)&  2.03(0.18) & 216.3&28.1&1.73(0.23) & 2.22(0.12)&  3 \\
2016-06-07$-$06-21 &  110.6  & 353.7 &  9.64(1.04)&  1.94(0.14)&334.2&33.4 &   2.30(0.26) & 1.70(0.07)&1 \\
 2016-06-21$-$07-05 &  44.3  &  101.1&6.14(1.18) & 2.18(0.20) & 111.0 &  14.3  &  1.54(0.25)&  1.91(0.12)&  1\\
2016-07-05$-$07-19 &  46.7  &  99.8  &  5.97(1.20)&  2.56(0.24)&144.9&13.5  &  1.30(0.20)&  1.85(0.11)&1 \\
2016-07-19$-$08-02 &  64.8  &  156.6 &  5.57(0.80)&  1.77(0.15)&177.9&17.2 &   1.17(0.18)&  1.47(0.08)&1 \\
 2016-08-02$-$08-16 &  37.7  &  106.2&6.54(1.08) & 2.30(0.22) & 66.4 &   7.7 &1.54$^{\ast}$ &  -&   \\
2016-08-16$-$08-30& 39.7& 109.5&   3.84(0.69)&  1.64(0.15)&  86.7&11.2 &0.87(0.19)&1.68(0.10)& 3 \\
2016-08-30$-$09-13 &  36.7 &   70.2 &   2.92(0.68) & 1.77(0.17)&118.2&13.9  &  0.88(0.15)&  2.21(0.12)& 2\\
 2016-09-13$-$09-27  & 25.9  &  31.5&2.12(0.50) & 2.37(0.26) & 38.4  &  6.7 &1.32$^{\ast}$ &   -&    \\
2016-09-27$-$10-11 &28.0&56.7 &   4.15(1.00) & 2.32(0.24) & 6.1 &3.2 &1.56$^{\ast}$ &   -&  \\
2016-10-11$-$10-25 &  22.9 &   75.7 &   2.98(0.63)&  1.26(0.20) &65.5& 8.3&1.61$^{\ast}$&    -&  \\
2016-10-25$-$11-08 &  32.1  &  55.3 &   3.45(0.77)& 2.35(0.24)& 81.6&9.9& 0.81(0.17)&  1.97(0.13)&1 \\
 2016-11-08$-$11-22& 67.0 &   168.4&5.52(0.78) & 1.83(0.15)&  223.4  & 23.2 &   1.51(0.20)& 1.71(0.09)&3 \\
2016-11-22$-$12-06 &  55.0 &   201.8 &  5.12(0.67)&  1.54(0.12)&103.4&13.4 &   1.00(0.19)&  2.15(0.12)& 2\\
 2016-12-06$-$12-20 &  35.1 &   69.5&6.25(1.37) & 2.72(0.28)&  87.6  &  10.5 &   1.41(0.28)&  2.79(0.24)&3 \\
2016-12-20$-$2017-01-03&  75.1  &  187.2 &  6.57(0.87)&  1.89(0.15)&175.1&21.6 &   1.48(0.20) & 1.81(0.10)&3 \\
2017-01-03$-$01-17  & 94.8 &   246.7 &  7.90(1.01)&  2.02(0.17)&317.0&33.0  &  2.15(0.24)&  1.87(0.08)& 3\\
 2017-01-17$-$01-31&   61.7 &   186.3&6.14(0.86) & 1.75(0.15) & 155.7  & 15.4  &  1.22(0.16) & 1.54(0.08)&3 \\
2017-01-31$-$02-14 &  89.7  &  205.9 &  7.19(0.99)&  1.90(0.15)&262.1& 29.9  &  1.95(0.23) & 2.23(0.12)& 2\\
2017-02-14$-$02-28 &   69.3  &   217.0 &   5.72(0.74) &  1.91(0.15)&393.7&43.7  &   2.74(0.28)&   1.72(0.07)& 3\\
 2017-02-28$-$03--14 &  46.2  &   111.8&4.02(0.78)&   1.86(0.17)&   72.3  &   10.7  &   0.73(0.15)&   1.97(0.13)&3 \\
2017-03-14$-$03-28&    63.6  &   195.6 &   6.31(0.80)&   1.73(0.13)&90.2&10.9 &    0.85(0.15) &  2.24(0.14)& \\
 2017-03-28--04-11 &   73.2   &  210.9&4.45(0.58)&   1.72(0.13)&   212.2 &   24.4 &    1.19(0.16) &  1.82(0.09)& 2\\
2017-04-11$-$04-25 &  100.5 &  301.6 &  8.03(0.92) & 1.69(0.12)&331.6&34.1&    2.17(0.25) & 1.80(0.08)& 3\\
 2017-04-25$-$05-09 &  58.4 &   164.5&4.98(0.69) & 1.72(0.14)&  176.1 &  17.6 &   1.20(0.16) & 1.64(0.09)& 2\\
2017-05-09$-$09-23  & 46.1  &  112.5 &  4.25(0.72)&  1.63(0.15)&123.7&15.5 &   1.15(0.18) & 1.74(0.10)&3 \\
2017-05-23$-$06-06 &  48.5  &  111.2 &  5.94(0.98) & 2.05(0.19) &70.0& 9.8&0.93(0.20)&  2.15(0.14)& 3\\
 2017-06-06$-$06-20  & 50.5  &  90.2& 3.98(0.78)&2.19(0.22)&  216.3 &  22.3  &  1.35(0.18)&  1.83(0.09)& 2\\
2017-06-20$-$07-04&26.7  &  39.5  &  2.80(0.66)&  2.26(0.27) & 35.9& 5.2 &1.42$^{\ast}$  &  -& \\
2017-07-04$-$07-18 &  31.1 &   98.6 &   4.87(0.97)& 1.37(0.12)&71.3&6.4& 1.82$^{\ast}$ &   -& \\
2017-07-18$-$08-01 &  75.8  &  171.7 &  6.24(0.83) & 1.96(0.15)&194.0&23.2  &  1.46(0.19) & 2.06(0.12)& 3\\
 2017-08-01$-$08-15 &  49.6 &   107.7&4.06(0.78)&  2.02(0.18) & 76.6  &  11.4  &  0.73(0.15)&  2.45(0.20)&2 \\
2017-08-15$-$08-29& 49.6 &   121.7  & 4.84(0.80) & 1.89(0.17)&  112&12.7& 0.98(0.18) & 1.71(0.10)&3 \\
2017-08-29$-$09-12 &  52.1&    141.5& 4.22(0.62)&1.29(0.11) & 149.4 &  16.6  &  1.11(0.16)&  1.53(0.08)&2 \\
2017-09-12$-$09-26  & 83.8  &  195.2 &  7.14(0.90) & 1.85(0.14)&126.4&17.0   & 1.14(0.17) & 1.69(0.10)& 3\\
 2017-09-26$-$10-10 &  49.0   & 114.7&5.35(0.88)&  2.34(0.18) & 223.0 &  23.1&    1.42(0.18) & 1.73(0.08)&1 \\
2017-10-10$-$10-24 &  58.0  &  134.7  & 4.87(0.80) & 1.94(0.17)&98.5&13.8  &  1.15(0.22) & 2.38(0.19)& 2\\
 2017-10-24$-$11-07 &  67.4 &   173.9&5.43(1.18)&  1.81(0.15) & 280.7&   29.9   & 1.91(0.24) & 1.87(0.09)&3 \\
2017-11-07$-$11-21  & 84.1  &  147.2 &  6.02(0.84) & 2.41(0.17)&216.2&24.2 &   1.31(0.17) & 2.01(0.11)& 1\\
 2017-11-21$-$12-05  & 96.1 &   284.1&7.36(0.84) & 1.54(0.12) & 353.6 &  38.2  &  2.38(0.26)&  1.87(0.07)& 2\\
2017-12-05$-$12-19  & 63.3  &  248.0 &  7.22(0.98)&  1.38(0.11)&152.5&15.8  &  1.50(0.21)&  2.15(0.14)&2 \\
 2017-12-19$-$2018-01-02&  73.8 &206.6&6.74(0.87) & 1.56(0.12)&  387.7 &  36.2&    2.69(0.28) & 1.83(0.06)&2 \\
2018-01-02$-$01-16 &  98.8 &   264.9 &  7.18(0.88)&  1.66(0.13)&360.5&41.6 &   2.42(0.26) & 1.62(0.06)& 3\\
 2018-01-16$-$01-30 &  86.6 &   279.7&10.71(1.24)& 1.99(0.15)&  321.5  & 29.0  &  2.82(0.33) & 1.59(0.06)&1 \\
2018-01-30$-$01-13  &  73.8  &  255.8 &  12.46(1.52)& 1.97(0.15)&386.0&37.3  &  4.95(0.53)&  1.87(0.06)& 3\\
2018-02-13$-$02-27 &  161.5 &  587.7 &  12.77(1.21)& 1.60(0.10)&748.9&65.5  &  4.17(0.40) & 1.61(0.04)& 3 \\
2018-02-27$-$03-13&   86.4 &   208.6&7.02(1.01) & 1.95(0.15)&  434.1 &  42.5  &  2.71(0.28) & 1.73(0.06)& 1\\
  \hline
  \end{tabular}
  \end{minipage}
  \end{table*}

\subsubsection{Stochastic Acceleration and Turbulence}
As in previous years, the source mostly showed low curvatures
($b\sim$0.3 or smaller for a vast majority of the log-parabolic
spectra; see Section\,3.3.1), i.e., wider synchrotron SED, expected
in the case of efficient stochastic acceleration \citep{m11b}. This
result is related to the inverse proportionality of the PED
curvature $r$ to the diffusion coefficient \emph{D} in the
Fokker-Plank kinetic equation: $r\propto D^{-1}$. On the other hand,
$r\propto \varepsilon/(n_{\rm s}\sigma^2_{\varepsilon})$
\citep{m04}, with $n_{\rm s}$, the number of the acceleration steps;
$\sigma^2_{\varepsilon}$, the variance of the energy gain
$\varepsilon$. Consequently, the detection of  low spectral
curvatures implies higher values of $n_{\rm s}$ and diffusion
coefficient, which needs strongly developed turbulence in  smaller
acceleration region.

In fact, the relativistic magnetohydrodynamic simulations of
\cite{miz14} showed that shock propagation can strongly amplify the
turbulence in the shocked jet material due to its interaction with
higher-density inhomogeneities existing in the pre-shock medium; the
more frequent detection of the 0.3--10\,keV IDVs in flaring X-ray
states (compared to quiescence periods; see Section\,4.1.3)
demonstrates the viability of this scenario for Mrk\,421 during the
time interval presented in this work. Moreover, our detection of the
anti-correlation $F_{\rm 0.3-10 keV}$--$b$ (see Figure 9c), i.e., a
dominance of lower curvatures in higher X-ray states, favours the
shock-in-jet scenario and strongly-developed turbulence during those
states.

The simulations of \citealt{miz14} also demonstrated that the
higher-energy photons (including those having 0.3--10\,keV energies)
are expected to originate in the smallest jet regions, which contain
the strongest magnetic field and yield the most rapid time
variability. In fact, the fastest IDVs, occurring within a few
hundred seconds, were observed mostly in the highest X-ray states in
Interval\,3a, and such instances can be related to the interaction
of the relativistic shock front with the smallest-scale turbulent
regions, embodying stronger magnetic fields (according to
light-travel argument). Note also that this sub-interval was
remarkable for the lowest mean curvature observed in the whole
2015-2018 time interval, implying the existence of the most
efficient stochastic acceleration in that sub-interval.

Along with the flux, the 0.3--10\,keV spectral parameters varied on
intraday timescales. Sometimes, these changes were extremely fast,
within 1-ks observational runs: curvature risings by 0.19--0.22;
hardenings by $\Delta a$=0.08--0.23; shifts of the synchrotron SED
peak by several keV to higher or lower energies; transitions from
the log-parabolic PED into the power-law one and/or vice versa. Such
behaviour could be related to the passage of a shock front through
the regions with spatial scales $l$$\lesssim$10$^{14}$\,cm and
significantly stronger turbulence, separated by the region with less
extreme physical properties (magnetic field strength, particle
number density, bulk Lorentz factor etc.). Our detections show the
viability of the simulations of \cite{miz14}, yielding a strong
turbulence amplification on extremely small spatial scales in the
shocked jet medium.

Stochastic acceleration scenarios predict the presence of the
$E_{\rm p}$--$b$ anti-correlation \citep{t09}. However, the latter
was not detected at the 99\% confidence level in Interval\,1 and was
weak during Intervals\,1--3 (see Table\,\ref{cortable} and
Figure\,\ref{figcor}b). This correlation is deduced from the
relation $ \ln E\propto 2\ln \gamma_{\rm p}+3/(5b)$ \citep{t09},
with $\gamma_{\rm p}$, the PED peak energy. Note that the
significant difference in the $\gamma_{\rm p}$ values, corresponding
to the different X-ray flares, may result in a large scatter of the
data points in the $E_{\rm p}$--$b$ plane or even a destruction of
the anti-correlation. Furthermore, this correlation is also
predicted for EDAP through $ \log E\sim 3/(10b)$ \citep{c14}, and
since this relation shows a different slope compared to the
stochastic case, the joint operation of these mechanisms can yield a
large scatter in the $E_{\rm p}$--$b$ plane and weaken the
anti-correlation.

We found a positive $F_{\rm 0.3-10 keV}$--$E_{\rm p}$ correlation
during Intervals\,1--3, i.e., a trend of shifting the synchrotron
SED peak to higher energies with rising X-ray flux (see
Figure\,\ref{figcor}g). \cite{t09} demonstrated that as the peak
energy of the emission increases, the cooling timescale shortens and
can compete with the acceleration timescales. It is then possible to
observe a bias in the $E_{\rm p}$--$b$ relation (weakening the
anticorrelation), since the cooling timescale is shorter than that
of EDAP or stochastic acceleration.

Note that the anti-correlation between the 0.3--10\,keV and radio
variabilities, discussed above and explained as resulting from the
shifting of the PED peak with a rising X-ray flux (corroborated by
our finding of the positive $F_{\rm 0.3-10 keV}$--$E_{\rm p}$
correlation), is expected in the framework of stochastic
acceleration  with a narrow initial energy distribution, having the
mean energy significantly higher than the equilibrium energy
\citep{k06}. Presumably, such a physical condition was not the case
for some X-ray flares when no declining radio brightness was
observed. Moreover, since there were some flares with a negative
$a$--$b$ trend, this result hints at the low initial energies of the
accelerating particles during those events.

In the case of the low spectral curvature, the electron volume
density $n_{\rm e}$ is expected to be higher, yielding a brighter IC
peak within the SSC scenario \citep{m11b}. Since the source
generally shows its IC peak at the TeV frequencies (see, e.g.,
\citealt{ac11}), lower VHE states are expected along with the high
spectral curvatures. In fact, during the majority of the XRT
observations with $b$$>$0.4, Mrk\,421 was not detectable with
3$\sigma$ significance with FACT or showed excess rates lower than
the mean value during 2015\,December--2018\,April. However, there
were two exceptions showing higher VHE states along with $b$$>$0.4
(MJD\,57388 and 57548). On the other hand, the source was not
detectable or showed low VHE states during some time intervals with
predominantly low curvatures (e.g., MJD\,57408-57486, 58147--58162).
These instances demonstrate that the one-zone SSC scenario was not
always acceptable for our target in the here-presented period.

\subsubsection{Spectral Loops}

\begin{figure*}
   \centering
   \vspace{-0.2cm}
   \includegraphics[trim=7.2cm 4cm 0.2cm 0cm, clip=true, scale=0.83]{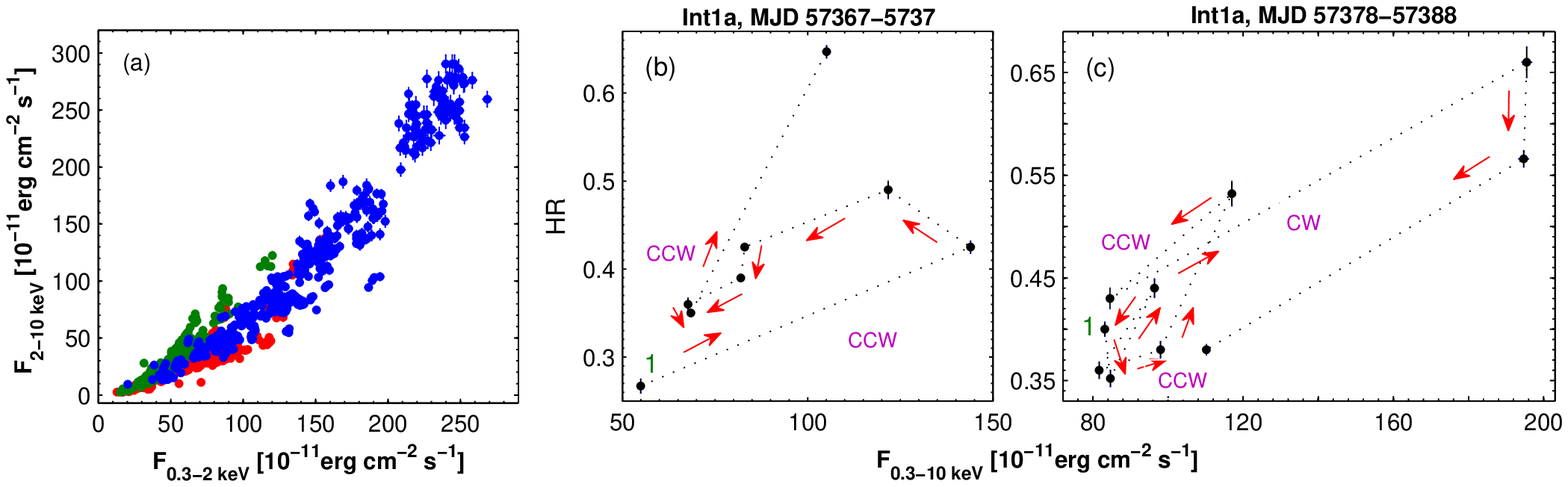}
   \vspace{-0.7cm}
\caption{\label{hyster} Panel (a): scatter plot with de-absorbed
0.3--2\,keV and 2--10\,keV fluxes. The subsequent panels: spectral
hysteresis patterns in different sub-intervals (extract; available
in its entirety in machine-readable form). In each
   HR-flux plane, the start point is denoted by \textquotedblleft 1\textquotedblright. }
    \end{figure*}

In the Bohm limit, the first-order Fermi mechanism yields an
electron acceleration timescale $\tau_{\rm FI}$$\approx
10200(c/v^2_{\rm sh})\sqrt{\gamma^2-1}(B/1G)^{-1}$ , with $v_{\rm
sh}$ -- the shock speed \citep{tam09}. For a 1-G field, relativistic
shock ($v_{\rm sh}\rightarrow c$) and $\gamma$$\lesssim$10$^4$, this
timescale will be a few milli-seconds or shorter. In that case, the
acceleration and injection of electrons into the emission zone will
be instantaneous and a clockwise (CW) evolution of the X-ray flare
in the plane $F_{\rm 0.3-10\,keV}$--HR is expected, making the
spectrum progressively harder in the brightening phase of the
source, due to the emergence of a flaring component starting at hard
X-rays \citep{t09}. Although the de-absorbed soft 0.3--2\,keV and
hard 2--10\,keV fluxes showed strong or very strong
cross-correlations during Intervals\,1--3 (Figure\,\ref{hyster}a and
Table\,\ref{cortable1}), the latter underwent a higher variability
in each interval (see Table\,\ref{persum} for the corresponding
$F_{\rm var}$ and $\Re$ values) and the hysteresis patterns were
clearly evident in the $F_{\rm 0.3-10\,keV}$--HR plane. The CW
loops, expected in the case of EDAP, are evident in Figures
\ref{hyster}c--\ref{hyster}n, \ref{hyster}p,
\ref{hyster}r--\ref{hyster}u during the various short and
longer-term flares discussed in Section\,3.1. Two CW-loops were
detected also on intraday timescales (Figures \ref{hyster}w
and\ref{hyster}y).

However, EDAP can not be considered as instantaneous in the case of
significantly weaker magnetic fields, frequently inferred from the
one or multi-zone SSC modelling of Mrk\,421 ($B\lesssim$0.05\,G;
see, e.g., \citealt{abey17,a12}). Furthermore, no instantaneous
injection is expected for the hadronic content in the emission zone
whose acceleration time-scales are $\sim$1000-times longer than for
electrons \citep{tam09}. Note that the latter is more naturally
compatible with the hard $\gamma$-ray spectra characterized by the
photon index $\Gamma\lesssim1.8$ \citep{sh16,m93}, and such spectra
frequently were recorded during the LAT observations in the
here-presented period (as hard as $\Gamma_{\rm min}$=1.26$\pm$0.13).
Consequently, EDAP will be more gradual than instantaneous, X-ray
flares propagate from low energies to high energies and
counter-clockwise (CCW) spectral evolution should be observed
\citep{tam09}. Such behaviour was also  frequently observed in
Intervals\,1--3 (see Figures\,\ref{hyster}b--\ref{hyster}e,
\ref{hyster}g--\ref{hyster}j, \ref{hyster}o--\ref{hyster}t,
\ref{hyster}v). Note that the slow, gradual acceleration and
CCW-loops are also expected during the stochastic acceleration in
the jet region with low magnetic field and high matter density
\citep{vv05}. On the contrary, this time-scale can be much shorter
and even instantaneous  in the purely or mainly lepton plasma if the
matter density is low  and high magnetic fields are presented. In
such situation, CW-type loops can develop, although this requires
quite ideal turbulence conditions with particle-scattering waves
moving in opposite directions over a sufficiently long length-scale
\citep{tam09}.

\vspace{0.2cm}
\subsubsection{Power-law Spectra}

Along with EDAP and stochastic mechanisms, there could be other
\textquotedblleft competing \textquotedblright ~processes acting in
the emission zone and weakening the observed $E_{\rm p}$--$b$
correlation. Namely, the first-order Fermi process can yield a
power-law PED when the magnetic field properties are variable and
its confinement efficiency becomes independent of the particle's
energy for some time intervals. Note that 9.6\% of the 0.3--10\,keV
spectra showed a simple power-law distribution of the photons with
frequency, and they were observed mostly in higher X-ray states.
This percentage was unprecedented high during 2005--2008 (27.5\%)
and the power-law spectra were observed most frequently during the
densest XRT campaign in 2006\,June\,15--25, although they were
recorded in any brightness states shown by the source in that period
\citep{k18a}. A higher percentage (13\%) and no clear trend with
brightness was observed also during 2013\,January--May \citep{k16}.
On the contrary, the periods 2009--2012 and
2013\,November--2015\,June were characterized by significantly fewer
occurrence of power-law spectra (4.9\%--6.7\%; \citealt{k17a,k18b}).

Hard power-law PEDs with slopes $p$$<$2 can be established by
relativistic magnetic reconnection, expected to operate efficiently
in highly-magnetized plasma  with the magnetization parameter
$\sigma$$\gtrsim$10 \citep{sir14}. However, the simulated broadband
SEDs, obtained by \cite{p16} for the cases $\sigma$=10--50, differ
significantly from those of Mrk\,421 constructed using the data
obtained during the different MWL campaigns
(\citealt{abey17,b16,a12} etc.), particularly, in the MeV--TeV
energy range.

\vspace{0.2cm}
\subsubsection{The Position of the Synchrotron SED Peak and Variable Turbulence Spectrum}

During 2015\,December--2018\,April, only 15\% of the spectra showed
$E_{\rm p}$$>$2\,keV, i.e., peaking at the hard X-ray frequencies
(taking into account the error ranges). This percentage is
significantly lower than that shown by Mrk\,421 during 2005--2008
(24\%; \citealt{k18a}) and a higher occurrence of such spectra were
recorded also during 2009--2012 (17\%; \citealt{k18b}). Note that
the spectra with $E_{\rm p}>$2\,keV were mostly concentrated in the
sub-periods 2006\,April--July, 2008\,March--June, 2009\,November,
2010\,January--May. Due to the position of the synchrotron SED peak
at higher frequencies, BAT detected the source with 5$\sigma$
confidence frequently in the aforementioned time intervals: the
BAT-band photons are generally of synchrotron origin in the HBL
sources and no significant contribution from the IC photons are
found, in contrast to the low-energy-peaking BLLs (LBLs; e.g.,
OJ\,287; see \citealt{k18c}).

On the other hand, the periods 2013\,January--May and
2013\,November--2015\,June were characterized by a significantly
lower occurrence of hard X-ray peaking spectra (2\% and 5\%,
respectively; see \citealt{k16,k17a}). Note that the percentage of
the spectra with $E_{\rm p}$$>$2\,keV was significantly higher for
1ES\,1959+650 in 2016\,January--August (48\%) and
2016\,August--2017\,November (28\%) \citep{k18d,k18e}. The highest
value of this parameter was 12.80$\pm$0.86\,keV. However, more
extreme case with 94\% spectra with the synchrotron peaks in X-rays
was recorded for Mrk\,501 during the extended X-ray flaring activity
in 2014\,March--October \citep{k17b}. In that period, the maximum
value $E^{max}_{\rm p}$=20.96$\pm$2.81\,keV and unprecedented
spectral behaviour when the synchrotron SED peak position underwent
a shift by at least two orders of frequency and moved beyond
100\,keV \citep{tav01}. For our target, the most extreme SED
position was observed on 2006\,April\,22 with $E^{max}_{\rm
p}$=26$^{+19}_{-8}$\,keV, obtained by \cite{t09} from the joint fit
of the log-parabolic model with the XRT and BAT spectra. Although
the same authors reported more extreme cases $E_{\rm p}$$>$100\,keV
from the 2006\,April--June observations, our thorough analysis of
these spectra showed insignificant spectral curvature and a good fit
with a simple power-law \citep{k18a}.

Fig.\,\ref{figcor}h demonstrates a positive $\log E_{\rm p}$--$\log
S_{\rm p}$ correlation with a slope of  0.63$\pm$0.08 -- the value
of the exponent $\alpha$ in the relation $S_{\rm p} \varpropto
E^{\alpha}_{\rm p}$. This relation was predicted by the simulations
of \cite{t11} corresponding to the case when the momentum-diffusion
coefficient $D$ is variable during stochastic acceleration of the
X-ray emitting electrons. Consequently, there should be a transition
from the Kraichnan spectrum of the turbulence with the exponent
$Q$=3/2  into the \textquotedblleft hard sphere\textquotedblright
spectrum ($Q$=2). In the latter regime, the scattering and
acceleration timescales are independent of the particle energy.
During the transition, the synchrotron SED follows the expectation
of a lower curvature for the harder turbulence spectra \citep{t11}.
On average, the lowest curvatures were observe in Interval\,3 which
shows the strongest $\log E_{\rm p}$--$\log S_{\rm p}$ correlation
during 2015\,December--2018\,April (see Table\,\ref{cortable}). This
results can serve as another confirmation of the stochastic
acceleration of particles in that period.

\subsubsection{LAT-Band Spectral Properties and Possible Jet-Star Interaction}

 The photon indices corresponding to the softer $\Gamma_{\rm 0.3-2 GeV}$
 and harder $\Gamma_{\rm 2-300 GeV}$ LAT-bands showed a weak cross-correlation during 2015\,December--2018\,April (see
Figure\,\ref{latgamma}b and Table\,\ref{cortable}). On some
occasions, the index $\Gamma_{\rm 2-300 GeV}$ was lower than the
0.3--2\,GeV one, which can be related to the soft gamma ray excess
at energies of several hundred MeV. One of the possible explanations
consists in a star-jet interaction, expected in the blazars hosted
by elliptical galaxies (including Mrk\,421; see \citealt{s00}).
These galaxies may have a population of red giants surrounding the
blazar jet and  can carry large wind-blown bubbles into the jet,
leading to gamma-ray emission through bubble-jet interactions
\citep{t19}. Note that those instances, characterized by a spectral
hardening with energy, are mostly observed in the period
2016\,April--August, while no opposite spectral trend was observed
during that time (see Figure\,\ref{latgamma}e and Table\,\ref{lat}).
The simulations of \cite{t19} have shown that the IC emission,
resulting from the jet-bubble interaction, is negligible ($L_{\rm
IC}\sim$10$^{40}$\,erg\,s$^{-1}$), while that generated by the
synchrotron mechanism can make a significant contribution to the MeV
energy budget ($L_{\rm IC}\sim$10$^{44}$\,erg\,s$^{-1}$; see
\citealt{a12} for comparison). In the latter case, the equipartition
value of the magnetic field and acceleration efficiency
$\xi\gtrsim0.1$ are required.

In 17 cases, the index $\Gamma_{\rm 2-300 GeV}$ was higher than its
lower-energy \textquotedblleft counterpart\textquotedblright ~(the
data points situated below the red dashed line in
Figure\,\ref{latgamma}b; see Figure\,\ref{latgamma}f for the
corresponding SEDs). These cases could be related to the up-scatter
of X-ray photons to the 2--300\,GeV energy range in  the K--N
regime, yielding a steepening of the corresponding photon spectrum
with respect to that established in the 0.3--2\,GeV range by means
of the Thompson up-scattering of the optical--UV photons
\citep{k18e}. Finally, 8 out 23 LAT-band SEDs, where the difference
between the $\Gamma_{\rm 0.3-2 GeV}$ and $\Gamma_{\rm 2-300 GeV}$
photon indices did not exceed the error ranges, were very hard and
their origin could be related to the hadronic contribution to the
0.3--300\,GeV energy range (as suggested by \citealt{sh16}).

 \section{Summary}
In this paper, we have presented the spectral and timing results
obtained during the intensive \emph{Swift}-XRT and MWL observations
of Mrk\,421 in 2015\,December--2018\,April. The main results of our
study are as follows:
\begin{itemize}
\vspace{-0.2cm}
\item Similar to the previous years, the source exhibited strong and erratic
X-ray variability (without any quasi-periodicity). The most extreme
behaviour was recorded during 2017\,December--2018\,February when a
long-term flare, lasting more than 2\,months, was superimposed by
short-term ones during which the 0.3--10\,keV flux exceeded a level
of 5$\times$10$^{-9}$erg\,cm$^{-2}$s$^{-1}$, similar to that
recorded on 2008\,June\,12, and even higher states were observed
during the giant outburst in 2013\,April. This period was also
characterized by several intraday flux doubling and halving events
with $\tau_{\rm d}$=4.8--18.9\,hr, as well as by numerous
lower-amplitude 0.3--10\,keV IDVs with $F_{\rm var}$=0.20--0.42,
including extremely fast brightness fluctuations by 5\%--18\% within
180--600 seconds. Six another long-term flares of comparable
durations but with lower amplitudes were evident during other parts
of the period presented in this work. \vspace{-0.2cm}

\item The highest VHE states were recorded in 2018\,January (coinciding  with those in
the XRT band) and the TeV-band variability mostly showed a good
correlation with the X-ray one, although there were several
exceptions when the VHE flux showed a decline or low states during
the fast X-ray flare, or the X-ray and VHE peaks were separated by
the time interval of $\sim$1\,d or longer, posing problems for
one-zone SSC scenarios. In other spectral ranges, Mrk\,421 exhibited
a relatively different behaviour: there was only a weak positive
$F_{\rm XRT}$--$F_{\rm LAT}$ correlation and the highest
0.3--300\,GeV states were recorded about 2-yr earlier than the X-ray
ones, while the source exhibited only a moderate LAT-band flaring
activity along with the strongest X-ray flares recorded in
Interval\,3a. Similarly, the highest optical--UV states were
observed during 2016\,January--February, and they were significantly
lower 2 \,yr later when the source showed its highest X--ray
activity. Consequently, the latter was anti-correlated with the
UVOT-band variability, and a similar $F_{\rm 0.3-10 keV}$--$F_{\rm
15 GHz}$ relation was observed during the entire
2015\,December--2018\,April period. Such MWL variability favours
some earlier simulations of the second-order Fermi process, when a
population of the accelerating electrons are characterized by a
narrow initial distribution of energy, having a mean value
significantly higher than the equilibrium energy. \vspace{-0.2cm}

\item During Intervals 1 and 3, the distributions of the de-absorbed 0.3--10\,keV
flux showed lognormality features, which could be indicative of the
variability imprint of AD on the jet. However, the data from
Interval\,2 and the highest X-ray states did not show the same
property. Since Interval\,2 clearly shows a better fit with the
Gaussian function and since this period was characterized, on
average, by lower X-ray states (see Section\,3.1 and
Table\,\ref{persum}), a lack of the lognormality could be related to
weaker shocks through the jet compared to other periods (possibly,
due to weaker AD instabilities). The FACT and LAT-band fluxes showed
lognormality features in all intervals, in contrast to the radio-UV
observations. The 0.3--10\,keV IDVs were observed significantly more
frequently during higher X-ray states and did not exhibit a
lognormality. This result favours the \textquotedblleft
shock-in-jet\textquotedblright ~scenario. The longer-term flares may
result from the propagation and evolution of relativistic shocks
through the jet. The shock appearance could be related to an abrupt
increase of the collimation rate at the jet base owing to some
processes in the accretion disc, yielding a lognormal flaring
behaviour on longer timescales. \vspace{-0.2cm}

\item Along with the strong flux variability, the source also exhibited an
extreme spectral behaviour. The 0.3--10\,keV spectra generally
showed their best fits with the log-parabolic model, yielding wide
ranges of the curvature parameter $b$=0.07(0.05)--0.48(0.04) and
photon index at 1\,keV $a$=1.63(0.03)--2.92(0.02). The position of
the synchrotron SED peak underwent extreme variability on various
timescales between the energies $E_{\rm p}$$<$0.1\,keV (the UV
frequencies) and $E_{\rm p}$$>$15\,keV, with 15\% of the spectra
peaking at hard X-rays. The synchrotron SED showed a positive
correlation with the 0.3--10\,keV flux: it shifted by several keV to
higher energies during the flaring phases and moved back along with
brightness drops, exhibiting the most violent intraday variability
by several keV during the strongest X-ray flares. 33\% of the
spectra were harder than $a$=2, and the energy spectral shape
generally followed a \textquotedblleft
harder-when-brighter\textquotedblright ~trend (except for some short
time intervals with the opposite trend, explained by the emergence
of a new soft X-ray component in the emission zone). The photon
index varied on diverse timescales with variations from $\Delta
a$=0.08--0.23 within 0.13--0.28\,hr to $\Delta a$=0.66--1.07 in
3--27\,d. 9.6\% of the spectra were fitted well with a simple
power-law, with photon indices $\Gamma$=1.79-2.91 and strongly
following the \textquotedblleft
harder-when-brighter\textquotedblright ~trend. The source mostly
showed a low spectral curvature ($b\sim$0.1--0.3) and an
anti-correlation $E_{\rm p}$--\emph{b}, as predicted for the
efficient stochastic acceleration of X-ray emitting electrons by the
magnetic turbulence. Moreover, the source showed a positive $a$--$b$
correlation, expected within the EDAP scenario, although it was
weak, possibly due to the \textquotedblleft
competition\textquotedblright ~with other types of the acceleration
mechanisms and cooling processes, not displaying the same
correlation. \vspace{-0.1cm}

\item The 0.3--10\,keV spectra showed a relation $S_{\rm
p}$$\varpropto E^{\alpha}_{\rm p}$, with $\alpha$$\sim$0.6 which
demonstrates a transition from the Kraichnan-type turbulence
spectrum into the \textquotedblleft hard sphere\textquotedblright~
one, due the variability of the momentum-diffusion coefficient. This
result corroborates the importance of stochastic acceleration in the
here-presented period. Our study of the spectral hysteresis patterns
in the flux--HR plane shows the patterns of both the instantaneous
injection and the gradual acceleration of X-ray emitting electrons,
owing to first and second-order Fermi processes.
 \vspace{-0.2cm}
\item The source frequently showed very hard 0.3--300\,GeV spectra,
predicted for a hadronic contribution to the HE emission. On some
occasions, the corresponding SED showed a soft $\gamma$-ray excess,
possibly owing to the jet interaction with a wind-blown bubble from
a nearby red giant. This suggestion is corroborated by the fact that
the MeV-excess SEDs mostly belong to the period 2016\,April--August.
On the contrary, there was a softening in the 2--300\,GeV energy
range compared to the 0.3--2\,GeV spectrum, possibly due to the
upscatter of X-ray photons in the 2--300\,GeV energy range in  the
K-N regime. This may yield a steepening of the corresponding photon
spectrum with respect to the 0.3--2\,GeV range, corresponding  to
the Thompson up-scatter of the optical--UV photons.
\end{itemize}

 \bigskip
PR acknowledges the contract ASI-INAF I/004/11/0. We acknowledge the
use of public data from the \emph{Swift} data archive. This research
has made use of the \texttt{XRTDAS} software, developed under the
responsibility of the ASDC, Italy, and the data from the OVRO 40-m
monitoring program which is supported in part by NASA grants
NNX08AW31G and NNX11A043G, and NSF grants AST-0808050 and
AST-1109911.   We thank the FACT collaboration for making their
analysis results publicly available. Finally, we thank the anonymous
referee for his/her useful comments and suggestions that helped to
improve the quality of the paper.


\begin{thebibliography}{}
\bibitem[Abdo et al.(2011)]{ab11} Abdo, A. A., et al. 2011, \apj, 736, 131
\bibitem[Abeysekara et al.(2017)]{abey17} Abeysekara, A. U., et al. 2017, \apj, 834, 2
\bibitem[Acciari et al. (2011)]{ac11} Acciari, V. A., et al. 2011, \apj, 738, 25
\bibitem[Acero et al.(2015)]{ace15} Acero, F., et al. 2015, \apjs, 218, 23
\bibitem[Ahnen et al.(2016)]{a16} Ahnen, M. L., et al. 2016, \aap, 593, 91
\bibitem[Aleksic  et al.(2012)]{a12} Alecsic, J., et al. 2012, \aap, 542,100
\bibitem[Aleksic et al.(2015a)]{a15a} Alecsic, J., et al. 2015a, \aap, 578, 22
\bibitem[Aleksic et al.(2015b)]{a15b} Alecsic, J., et al. 2015b, \aap, 576, 176
\bibitem[Anderhub et al.(2013)]{a13} Anderhub, H. et al. 2013, Journ. of Instr., 8, article id. P06008
\bibitem[Atwood et al.(2009)]{at09} Atwood, W. B., et al. 2009, \apj, 697, 1071
\bibitem[Balokovic et al.(2016)]{b16} Balocovic, M., et al. 2016, \apj, 819, 156
\bibitem[Begelman et al.(2008)]{b08} Begelman, M.C., Fabian, A C., \& Rees, M. J. 2008, \mnras, 384, L19
\bibitem[Barthelmy et al.(2005)]{ba05} Barthelmy, S. D., et al. 2005, \ssr, 120, 143
\bibitem[Blazejowski et al.(2005)]{bl05} Blazejowski, M., et al. 2005, \apj, 630, 130
\bibitem[Bessel (1979)]{b79} Bessel, M. S. 1979, \pasp, 91, 589
\bibitem[B\"{o}ttcher \& Dermer(2010)]{b10} B\"{o}ttcher, M., \& Dermer, C. 2010, \apj, 711, 445
\bibitem[Breeveld et al. (2011)] {b11} Breeveld, A. A. et al. 2011, AIPC, 1358, 373
\bibitem[Burrows et al.(2005)]{b05} Burrows, D. N., et al. 2005, \ssr, 120, 165
\bibitem[Carnerero et al.(2017)]{c17} Carnerero, M. I., et al. 2017, \mnras, 472, 3789
\bibitem[Celotti \& Ghisellini(2008)]{c08} Celotti, A., \& Ghisellini, G. 2008, \mnras, 385, 283
\bibitem[Cesarini (2012)]{c12} Cesarini, A., PhD  Thesis, Nat. Univ. of Ireland Galway, 2013
\bibitem[Chen (2014)]{c14} Chen, L. 2014, \apj, 788,179
\bibitem[Chevalier et al.(2019)]{ch19} Chevalier, J. et al. 2019, \mnras, 484, 749
\bibitem[Cui(2004)]{c04} Cui, W. 2004, \apj, 605, 662
\bibitem[Dermer et al.(1992)]{d92} Dermer, C. D., Schlickeiser, R., \& Mastichiadis, A., 1992, \aap, 256, 27
\bibitem[Dorner et al.(2015)]{d15} Dorner, D. et al. 2015, astro-ph/1502.02582
\bibitem[Falomo et al.(2014)]{f14} Falomo, R., et al. 2014, \aapr, 2014, 22, 37
\bibitem[Fitzpatrick \& Massa (2007)]{fi07} Fitzpartick, E. L., \& Messa, D. 2007, \apj, 663, 320
\bibitem[Foster(1996)]{f96} Foster, G. 1996, \aj, 112, 1709
\bibitem[Fukugita et al. (1995)]{f95} Fukugita, M., et al. 1995, \pasp, 107, 945
\bibitem[Gehrels et al.(2004)]{g04} Gehrels, N., et al. 2004, \apj, 611, 1005
\bibitem[Giebels et al.(2002)]{g02} Giebels, B., et al. 2002, \apj, 571, 763
\bibitem[Giebels \& Degrange (2009)]{g09} Giebels, B., \& Degrange, B. 2009, \aap, 503, 797
\bibitem[Grossmann \& Morlet(1984)]{g84} Grossmann, A., \& Morlet, J. 1984, SIAM J. Math. Anal., 15, 723
\bibitem[Kapanadze et al.(2016)]{k16} Kapanadze, B., et al. 2016, \apj, 831, 102
\bibitem[Kapanadze et al.(2017a)]{k17a} Kapanadze, B., et al. 2017a, \apj, 848, 103
\bibitem[\protect\citeauthoryear{Kapanadze et al.}{2017b}]{k17b} Kapanadze, B., et al. 2017b, \mnras, 469, 1655
\bibitem[Kapanadze et al.(2018a)]{k18a} Kapanadze, B., et al. 2018a, \apj, 854, 66
\bibitem[Kapanadze et al.(2018b)]{k18b} Kapanadze, B., et al. 2018b, \apj, 858, 68
\bibitem[Kapanadze et al.(2018c)]{k18c} Kapanadze, B., et al. 2018c, \mnras, 480, 407
\bibitem[Kapanadze et al.(2018d)]{k18d} Kapanadze, B., et al. 2018d, \mnras, 473, 2542
\bibitem[Kapanadze et al.(2018e)]{k18e} Kapanadze, B., et al. 2018e, \apjs, 238, 13
\bibitem[Kalberla et al.(2005)]{k05} Kalberla, P. M. W., et  al. 2005, \aap, 440, 775
\bibitem[Katarzynski et al.(2006)]{k06} Katarzynski, K., et al. 2006, \aap, 453, 47
\bibitem[Lomb(1976)]{l76} Lomb, N. R. 1976, \apss, 39, 447
\bibitem[Macomb et al.(1995)]{m95} Macomb, N., et al. 1995, \apj, 449, L99
\bibitem[Mangalam \& Wiita(1993)]{mang93} Mangalam, A. V., \& Wiita P. J. 1993, \apj, 406, 420
\bibitem[Mannheim(1993)]{m93} Mannheim, K. 1993, \aap, 269, 60
\bibitem[Marscher \& Gear(1985)]{m85} Marscher, A. P., \& Gear, W. K. 1985, \apj, 298, 114
\bibitem[Marscher(2014)]{m14} Marscher, A. P. 2014, \apj, 780, 87
\bibitem[Massaro et al.(2004)]{m04} Massaro, E., et al. 2004, \aap, 413, 489
\bibitem[Massaro et al.(2011a)]{m11a} Massaro, F., et al.  2011a, \apjs, 197, 24
\bibitem[Massaro et al.(2011b)]{m11b} Massaro, F., et al. 2011b, \apj, 742, L32
\bibitem[Matsuoka et al.(2009)]{m09} Matsuoka, M., et al. 2009, \pasj, 61, 999
\bibitem[Mattox et al.(1996)]{m96} Mattox, J. R., et al. 1996, \apj, 461, 396
\bibitem[McHardy(2008)]{mc08} McHardy, I. 2008, in Blazar Variability across the Electromagnetic Spectrum,
Workshop Proc. (Trieste: PoS), 14
\bibitem[Mizuno et al.(2014)]{miz14} Mizuno, Y., et al. 2014, \mnras, 439, 3490
\bibitem[Moretti et al.(2005)]{m05} Moretti, A., et al. 2005, in Siegmund O., ed., Proc. SPIE Vol. 5898, In-flight
Calibration of the Swift XRT Point Spread Function. SPIE, Bellingham, p. 360
\bibitem[Nilsson et al. (2007)]{n07} Nilsson, K., et al. 2007, \aap, 475, 199
\bibitem[Padovani \& Giommi(1995)]{p95} Padovani, P., \& Giommi, P. 1995, \apj, 444, 567
\bibitem[Page et al.(2013)]{pa13} Page, M., et al. 2013, \mnras, 436, 1684
\bibitem[Petropoulou et al.(2016)]{p16} Petropoulou, M., et al. 2016, \mnras, 462, 3325
\bibitem[Pian et al.(2014)]{p14} Pian, E., et al. 2014, \aap, 570, 77
\bibitem[Poole et al.(2008)]{p08} Poole, T. S., et al. 2008, \mnras, 383, 627
\bibitem[Rebillot et al. (2006)]{r06} Rebillot, P. F., et al. 2006, \apj, 641, 740
\bibitem[Richards et al.(2011)]{r11} Richards, J. L., et al. 2011, \apj, 194, 209
\bibitem[Romano et al. (2006)]{ro06} Romano, P., et al. 2006, \aap, 456, 917
\bibitem[Romero et al. (1999)]{r99} Romero, G. E., et al. 1999, \aaps, 135, 477
\bibitem[Roming et al.(2005)]{r05} Roming, P. W. A., et al. 2005, \ssr, 120, 95
\bibitem[Rybicki \& Lightman (1979)]{r79} Rybicki, G. B., \& Lightman, A. P. 1979, in Radiative processes in astrophysics
(New York: Wiley-Interscience), 393
\bibitem[Saito et al.(2013)]{sa13} Saito, S., et al. 2013, \apj, 766, L11
\bibitem[Sandrinelli et al.(2017)]{s17} Sandrinelli, A., et al. 2017, \aap, 600, 132
\bibitem[Scargle(1982)]{s82} Scargle, J. D. 1982, \apj, 263, 835
\bibitem[Scarpa et al.(2000)]{s00} Scarpa, R., et al. 2000, \apj, 532, 740
\bibitem[Shukla et al.(2016)]{sh16} Shukla, A., et al. 2016, \aap, 591, 83
\bibitem[Sinha et al.(2016)]{si16} Sinha, A., et al. 2016, \aap, 591, 81
\bibitem[Sinha et al.(2017)]{si17} Sinha, A., et al. 2017, \apj, 836, 83
\bibitem[Sironi \& Spitkovsky (2014)]{sir14} Sironi, L., \& Spitkovsky, A. 2014, \apj, 783, L21
\bibitem[Smith et al.(2009)]{s09} Smith, P. S., et al. 2009, eConf Proc. C091122
\bibitem[Sokolov et al.(2004)]{s04} Sokolov, A., et al. 2004, \apj, 613, 725
\bibitem[Tavecchio et al.(2001)]{tav01} Tavecchio, F., et al. 2001, \apj, 554, 725
\bibitem[Tammy \& Duffy (2009)]{tam09} Tammy, J., \& Fuffy, P. 2009, \mnras, 393, 1063
\bibitem[Torres-Alba \& Bosch-Ramon (2019)]{t19} Torres-Alba, N., \& Bosch-Ramon, V. 2019, \aap, 623, 91
\bibitem[Tramacere et al.(2009)]{t09} Tramacere, A., et al. 2009, \aap, 501, 879
\bibitem[Tramacere et al.(2011)]{t11} Tramacere, A., et al. 2011, \apj, 739, 66
\bibitem[VanderPlas(2018)]{v18} VanderPlas, J. T. 2018, \apjs, 236, 16
\bibitem[Vaughan et al.(2003)]{v03}  Vaughan,  S., et al. 2003, \mnras, 345, 1271
\bibitem[Virtanen \& Vainio(2005)]{vv05} Virtanen, J. J. P., \& Vainio R. 2005, \apj, 621, 313
\end{thebibliography}
\end{document}